\documentclass[12pt, leqno]{amsart}
\usepackage[utf8]{inputenc}
\usepackage{geometry}
\usepackage{amsmath, appendix}
\usepackage{amssymb}
\usepackage{amsthm}
\usepackage{soul}
\usepackage{booktabs}
\usepackage{multicol}
\usepackage[countmax]{subfloat}
\usepackage{float}

\newtheorem{corr}{{\bf\sc Corrolary}}
\newtheorem{example}{{\bf\sc Example}}
\usepackage{subfig}
\usepackage[rightbars, color]{changebar}
\cbcolor{blue}
\usepackage{ctable} 

\newtheorem{thm}{{\bf \sc Theorem}}[section]
\newtheorem{lem}{{\bf \sc Lemma}}[section]

\newtheorem{prop}{{\bf\sc Proposition}}[section]
\newtheorem{ass}{{\bf\sc Assumption}}[section]

\usepackage{algorithm2e,algcompatible,amsmath}
\RestyleAlgo{boxruled}
\usepackage{graphicx}
\graphicspath{{Figures/}}
\usepackage{comment}
\usepackage{mathtools}  

\usepackage{amsthm,amsmath,natbib}
\usepackage[colorlinks=true,linkcolor=blue, citecolor = blue]{hyperref}%

\providecommand{\Prob}{\mathbb{P}}

\renewcommand{\Pr}{\Prob}

\usepackage{mwe}
\usepackage{caption}

\usepackage{booktabs}
\usepackage{siunitx}

\title[Identifying latent geometry]{Identifying the latent space geometry of network models through analysis of curvature}
\author[Lubold]{Shane Lubold${}^\S$}
\author[Chandrasekhar]{Arun G. Chandrasekhar${}^\ddagger$}
\author[McCormick]{Tyler H. McCormick${}^{\S.\star}$}

\thanks{We thank Eric Auerbach, Abhijit Banerjee, St\'{e}phane Bonhomme, Emily Breza, Jacob Burchard, Catherine Calder, Gabriel Carroll, James Evans, Bailey Fosdick, Jeremy Fox, Paul Goldsmith-Pinkham, Ben Golub,  Matthew Grant, Alan Griffith, Fang Han, Rachel Heath, Yunmi Kong,  Mengjie Pan, Mallesh Pai, Adrian Raftery, Abel Rodriguez, Anna Smith, Xun Tang, Matt Thirkettle, Aravindan Vijayaraghavan, Jon Wellner, and Steve Wilkins-Reeves. 
We thank Sumhith Veda Aradhyula, Vasu Chaudhary, Alex Philip, and Adam Visokay for excellent research assistance. Chandrasekhar is grateful for support from the Alfred P. Sloan Foundation. McCormick is supported by the National Institute Of Mental Health of the National Institutes of Health under Award Number DP2MH122405.  Partial support for this research came from a Eunice Kennedy Shriver National Institute of Child Health and Human Development research infrastructure grant, P2C HD042828, to the Center for Studies in Demography \& Ecology at the University of Washington. McCormick and Chandrasekhar are supported by award SES-2215369 from the National Science Foundation.}

\thanks{$^{\ddagger}$ Department of Economics, Stanford University, USA; NBER; JPAL}
\thanks{$^{\S}$ Department of Statistics, University of Washington, USA}
\thanks{$^{\star}$ Department of Sociology, University of Washington, USA}

\begin{document}


\maketitle 
\begin{abstract}
A common approach to modeling networks assigns each node to a position on a low-dimensional manifold where distance is inversely proportional to connection likelihood.  More positive manifold curvature encourages more and tighter communities; negative curvature induces repulsion. We consistently estimate manifold type, dimension, and curvature from simply connected, complete Riemannian manifolds of constant curvature.  We represent the graph as a noisy distance matrix based on the ties between cliques, then develop hypothesis tests to determine whether the observed distances could plausibly be embedded isometrically in each of the candidate geometries.  We apply our approach to data-sets from economics and neuroscience.~
\end{abstract}




\section{Introduction}\label{sec:Introduction}

Social, economic, biological, and technological networks play a crucial role in a myriad of environments. Job referrals \citep{granovetter1973,calvo2004,beaman2012,heath2018firms},  neurological function \citep{leung2008caenorhabditis}, epidemics \citep{hoffrh2002,bansal2010dynamic,sewell2015latent},  social media \citep{romero2011differences,myers2014bursty,cho2016latent}, informal insurance \citep{ambrusms2012,cai2017interfirm}, education decisions \citep{calvo2009peer}, sexual health \citep{handcock2004likelihood},  financial contagion \citep{gai2010contagion,elliott2014financial,acemoglu2015systemic}, international trade \citep{chaney2014network},  and politics \citep{diprete2011segregation} are among the many settings in which  networks play a major role.   Modeling network formation is, therefore, essential for both descriptive and counterfactual analyses.  

Constructing such models is 
 challenging from a statistical perspective since networks typically feature higher-order dependence between the connections.  Phenomena such as transitivity 
  are common 
  and mean that standard regression 
  approaches, which assume independence across connections, are not appropriate.  A 
  common approach for modeling this dependence structure is the latent space model, introduced by \cite*{hoffrh2002}.  
 One estimates a probability distribution over graphs that is consistent with the single, observed 
  graph. 
The model 
  assigns each node in the network to a position on a low-dimensional manifold.  Likelihood of a connection is inversely proportional to distance between actors on a manifold with a pre-specified dimension and geometry. Connections are assumed independent conditional on the latent positions. Standard practice in this area assumes a manifold class beforehand.

The choice of latent manifold 
is extremely consequential for both the interpretation  
 and its theoretical properties.  In enumerating these properties, it is first critical to distinguish between the intrinsic and extrinsic geometries.  Our focus is on the former, which embodies the fundamental properties of the geometric space.  Modeling in a geometry 
 manifold and inheriting the distance measure of that geometry (e.g., using a 2-dimensional sphere 
 means points exist on the surface of the sphere and distances are measured by arc length).

Moving now to the connection between geometry and networks, we first note that, holding constant the distribution of points in the latent space, the choice of the geometry in particular determines the nature of network  structure captured by the latent space.  For a simple example, consider a two dimensional Euclidean space (a plane).  
Here it is not possible to place four nodes in such a way that they are equidistant from one another, meaning that it isn't possible to represent groups of four such that, holding constant node effects, each node has the same likelihood of interacting with any other. Another way to see the impact of geometry is through triangles.  Since a sphere has bounded area, there is an upper bound to how far apart nodes can be from one another before they start getting closer together.  Positive curvature 
also encourages the formation of triangles and communities. 
Additionally, certain networks, such as a network of neurons or a network exchange built along a supply chain, may have a tree-like structure.  Trees are difficult to embed in spherical or Euclidean space but fit more naturally
in hyperbolic space.  Recent work on statistical modeling has also shown the importance of modeling networks using non-Euclidean latent representations. For instance, \citet{mccormick2015latent} model latent space as a sphere and  \cite{krioukov2010hyperbolic}  and  \citet{asta2015geometric}  use hyperbolic space.  \cite{Abel2019} explores both spherical and Euclidean representations in a Bayesian model for spatial voting patterns. \cite{weber2018curvature} examines the relationship between the latent space curvature and graph motifs.
\cite{SWR_THM_2022} propose a test of the assumption that the latent space has constant curvature which uses the clique structure in the graph. Finally, \citet*{smith2019geometry} provides a comprehensive review of the implications and consequences of the choice of geometry.

A second consequence of the choice of geometry arises in the theoretical properties of latent space model estimates. 
Consistency of the estimates of the individual locations on the unobserved manifold 
is the subject of recent work by \citet{shalizi2017consistency}.  Since the distribution of the network formation process depends on the manifold itself, the key open question is whether a researcher can consistently estimate the latent space. After all, the network formation process is sensitive to the geometry 
  inclusive of its curvature and dimension.
  
  It is currently common practice to assume the latent dimension and manifold type.  Our approach provides a data-driven alternative.  It also contrasts with cross-validation based selection procedures that are sometimes used, in particular to estimate the  
 dimension.  These approaches subsample connections and then use either model fit diagnostics or out-of-sample prediction metrics.  Our approach avoids a critical issue with these approaches, namely   subsampling can fundamentally alter graph properties in unpredictable ways \citep{chandrasekharl2016}, calling into question the relevance of the subsampled distribution.  
We  approach the question from a fundamentally different perspective than currently available alternatives that use the likelihood or penalized likelihood to estimate model fit.  First, rather than characterizing fit or predictive accuracy with a particular dataset, our approach takes a more classical hypothesis testing perspective.  Comparing (for example) an information metric across a model fit with a spherical or hyperbolic latent space is fundamentally characterizing the congruence between the embedding for a given dataset and the spaces under consideration.  Uncertainty in this framework arises from sampling, but also from potential model mis-specification.  A likelihood based metric for a spherical space with small curvature will likely perform quite well for a graph generated from Euclidean embeddings, for example.  
 We isolate uncertainty to only sampling error by using a test for isometric embeddings of distances into the space under consideration.  We conceptualize variability in the observed distance matrix as representing expected noise due to sampling realizations of a graph of a given size.  Second, we isolate the test to uniquely distance, rather than to the model as a whole, as would be the case with a likelihood-based measure.  A likelihood ratio test for whether or not the curvature of the space is zero, for example, may seem to be an appealing alternative to our approach. Such a test would, however, confound changes in the latent geometry with changes in the fixed effects.  To see this, recall that the surface area of the sphere changes as a function of the curvature.  To preserve the overall density of the graph, therefore, the individual effects must change when the curvature changes.  
 We sidestep this issue by leveraging the structure of the network formation model to isolate the test as specific to the latent geometry. Constructing an appropriate likelihood-based test, in contrast, would require marginalizing over the individual effects, which would be computationally intensive and require specifying distributions for the individual effects (which we do not require). 
 
We address the question of how to choose the manifold $\mathcal{M}^{p^\star}(\kappa^\star)$, meaning the manifold class ($\mathcal{M}^\star)$, curvature ($\kappa^\star$), and dimension ($p^\star$) of the   latent space.  We present a hypothesis testing framework which connects distances in the latent space to feasible embeddings on  simply connected, complete  Riemannian manifolds with constant curvature.  Rather than relying on likelihood or cross-validation techniques, we directly leverage geometric results that provide necessary and sufficient conditions to embed points into particular manifolds, given pairwise distances between the points.

Our main insight is as follows. The manifolds we consider in this work all come equipped with a metric---an inner product which is used in calculation of distances between points. The metric uniquely identifies the manifold. Given a distance matrix $D$ between $K$ points, even without knowing where the points are located, we can check if the points can be isometrically embedded in the manifold. It can be embedded if and only if the distance matrix is compatible with the candidate manifold's metric. Specifically, given $D$, we can construct a test matrix $W_{\kappa^\star}(D)$ and check that the eigenvalue spectrum has the same signature as the manifold's metric. The manifold detection problem is therefore reduced to testing the spectrum of $W_{\kappa^\star}(D)$.

We make two contributions. Our first contribution is in statistical geometry (Theorem \ref{thm: intro_thm}). We show how to estimate the manifold consistently when observing a noisy estimate $\hat D$ of the true distances. For this, we must estimate the signature of the spectrum of $\hat W_{\hat \kappa}(\hat D)$, the empirical test matrix. The  logic relies on Weyl's inequality, which places bounds on the change in eigenvalues due to perturbations of a matrix. This  provides an avenue for consistent estimation   of $\mathcal{M}^{p^\star}(\kappa^\star)$.

Our second contribution is to then extend the argument to the latent space network model. The main idea is that we take the observed graph $G$ on $n$ nodes and construct some distance matrix $\hat D(G)$ among $K<<n$ points and then apply our statistical geometry result.  We define distance based on interaction rates between $K$  \emph{groups} of nodes. Using the observed graph, we define the distance between two cliques based on the probability of an edge between a node in clique $C_1$ and a node in clique $C_2$, which we can calculate using the definition of the latent space model plus the 
 fraction of realized links between cliques.   We estimate the matrix of cross-clique edge probabilities, and then use the latent space model to estimate the pairwise distances between cliques. We then leverage our statistical geometric result to estimate the manifold.
 
In the remainder of this section, we formally define the two problems we address and provide an overview of our approach.  Specifically, we address (i) the general problem of estimating geometry from a noisy distance matrix and (ii) estimating network geometry from latent space models using cliques.  Using cliques represents the most challenging case for our method since we expect that in many settings cliques may be relatively small.
 Next, in Section \ref{sec: Geometry_Overview} we review key geometric concepts crucial for our testing procedure. 
 Section \ref{sec:geometry-general} covers the general geometry problem in detail 
 and develops our general method of estimating geometry from a noisy distance matrix. We turn to studying the estimation of the latent space in general using cliques 
 in Section \ref{sec: network_geometry_identification}. 
  In Section \ref{sec: Simulations} we present simulation experiments that explore the efficacy of our approach.  We apply our results to two empirical examples in Section \ref{sec: Empirics}.   The first empirical example considers data from 75 Indian village social networks, comprised of informal finance, information, and social links. We study the financial flows by geometry and also how the introduction of microfinance impacts geometry. The second example focuses on the neural network of the \emph{C. Elegans} worm. Section \ref{sec: Conclusion} concludes. All proofs are in Appendix \ref{sec:proofs} unless otherwise noted.

\subsection{Statistical Geometry Problem}
\label{sec: stat_geom_prob}
 The estimation methods we propose are   quite general. They apply broadly to a large set of problems (Section \ref{sec: examples_dist} provides examples) in which the researcher observes a noisy distance matrix $\hat D$ and wishes to estimate the properties of the underlying space. We make the following assumption about the latent space $\mathcal{M}^p(\kappa)$. After that we provide a broadly applicable classification theorem of $\mathcal{M}^p(\kappa)$ from an estimator $\hat D$ of distances computed between points in $\mathcal{M}^p(\kappa).$

\begin{ass} 
\label{ass:manifold}
    $\mathcal{M}^{p^\star}(\kappa^\star)$ is a simply connected, complete Riemannian manifold of constant sectional curvature $\kappa^\star$, with $p^\star \in \mathbb{Z}$ with known upper bound and  $\kappa^\star \in [-b,-a] \cup \{0\} \cup [a,b]$ with $a>0,b>0$.
\end{ass}

The technical geometric definition of simply connected, complete Riemannian manifolds is provided in a self-contained manner in  Appendix \ref{sec: sec_curvature}. By \citet{Killing1891}, Assumption~\ref{ass:manifold} means that the manifold must be Euclidean, spherical, or hyperbolic with a bounded dimension and curvature value in some compact set. We  emphasize that $a > 0$  means that in the curved cases, the geometry is not arbitrarily close to a Euclidean space ($\kappa = 0$). All such Riemannian manifolds are locally Euclidean, by definition, so this is required to be meaningful. We also emphasize that by dimension $p^\star$ we mean the minimum such dimension, as one can clearly embed $\mathcal{M}^{p^\star}$ in $\mathcal{M}^{p^\star+h}$ for $h>0$.

Algorithm \ref{alg: statistical_geometry} takes in $\hat D$ and returns consistent estimates of the manifold type, curvature, and dimension. We use $\circ$ to denote the Hadamard product and $1_K$ denotes a vector of ones of length $K$.

\begin{algorithm}[H]

{\small Input: noisy $K \times K$ distance matrix  $\hat D$.
\begin{enumerate}
\item Estimate the curvature of the manifold for  each of the curved cases $(\hat \kappa_S, \hat \kappa_H)$, from (\ref{eq: curvature_estimator}).
\item For each of the three candidate geometries, calculate the test matrix $W_\kappa$ using the corresponding curvature estimate from step (1), with projection $J = I_K - 1_K 1_K^T / K$:
\[
W_{\kappa}\left(\hat D\right):=\begin{cases}
\frac{1}{\kappa}\cos\left(\sqrt{\kappa}\hat D\right) & \text{ for }\kappa\neq0\\
-\frac{1}{2}J\hat D\circ \hat DJ & \text{ for }\kappa=0.
\end{cases}
\]
\item Construct $\hat{\mathcal{M}}$, the estimate  the manifold class, using Proposition \ref{prop: ordered_test}.
\item Given $\hat{\mathcal{M}}$, select $\hat p$, the estimate of its dimension using Proposition \ref{prop: consistent_rank}.
\end{enumerate}
}
 \caption{Consistent Estimation with a Noisy Distance Matrix}
 \label{alg: statistical_geometry}
\end{algorithm}

\begin{thm}[Consistent estimation with a noisy distance matrix]
\label{thm: intro_thm}
Let $\mathcal{M}^{p^\star}(\kappa^\star)$ be a latent space of unknown manifold class, dimension, and curvature that satisfies Assumption \ref{ass:manifold}. Fix a set of $K > p^\star$ locations on $\mathcal{M}^{p^\star}(\kappa^\star)$
and suppose they uniquely identify the latent space. Suppose 
there is a sequence of $K \times K$ matrices $\hat D_T$, indexed by 
$T \in \mathbb{N}$, such that $\hat D_T \overset{p}{\rightarrow}  D$ as $T \rightarrow \infty$. Under these assumptions, the estimators produced by Algorithm \ref{alg: statistical_geometry} 
are consistent as $T \rightarrow \infty.$ That is,
\(
\Pr( \hat{\mathcal{M}}^{\hat p} \neq \mathcal{M}^{p^\star} ) = o(1) \text{ and } \hat \kappa - \kappa^\star = o_P(1).
\)
\end{thm}

\vspace{0.1in}
\subsection{Latent Space Model}
Having proposed a consistent classification method of the manifold class, 
dimension, and curvature from noisy distances (see Theorem~\ref{thm: intro_thm}), we now turn to the latent space model. 
Consider a graph $G=(V,E)$ where $V$ are nodes and $E$ are edges (also called links or connections), with $|V| = n$.  For simplicity, we assume throughout that the graph is un-directed, all connections are symmetric, and unweighted, all connections are either present or absent.  Our methods readily extend to the weighted and directed case, though it increases the complexity in terms of both notation and exposition.    
We assume that edges in $G$ are drawn independently according to
\begin{equation}
\Pr\left\{G_{ij}=1 \mid \nu^\star, z^\star, X^\star_{ij}, \mathcal{M}^{p^\star}(\kappa^\star) \right\} 
= \Lambda\left(\nu_i^\star  + \nu_j^\star -   d_{\mathcal{M}^{p^\star}(\kappa^\star)}(z^\star_i,z^\star_j) \right) \;,
\label{eq:main_model}
\end{equation}
for some increasing, invertible link function $\Lambda$.  We represent $\mathbf{\nu} = (\nu_1, \dotsc, \nu_n)$ as the vector of individual effects, restricted to lie in some set to ensure \eqref{eq:main_model} is a probability value in $[0, 1]$.\footnote{One way to model a directed graph is to allow each node to have two different fixed effects, $\nu_i$ and $\chi_i$, one playing a role when $i$ is the sender of the link ($G_{ij}$) and the other when $i$ is the receiver ($G_{ji}$).}  These are independent effects that encode individual gregariousness, and are related to the total number of connections \citep{chatterjeeds2010,graham2014econometric}.  The $d_{\mathcal{M}^p(\kappa )}(z_i,z_j)$ terms represent the distance on the manifold $\mathcal{M}^p(\kappa)$, with dimension $p$ and curvature $\kappa$, between locations $z_i$ and $z_j$.   {Most of our analysis is done on the model above in (\ref{eq:main_model}) using an exponential link function, so $\Lambda(x) = \exp(x)$, but this is mostly out of convenience. In Appendix \ref{sec: other_graph_models}, we show how to handle other common link functions
(e.g., logistic link, \citep{hoffrh2002}) or how to handle node- and pair-level covariates effects \citep{graham2014econometric}.}

\begin{figure}
    \centering
    \subfloat[Flat $\kappa^\star = 0$]{
    \includegraphics[scale = 0.15]{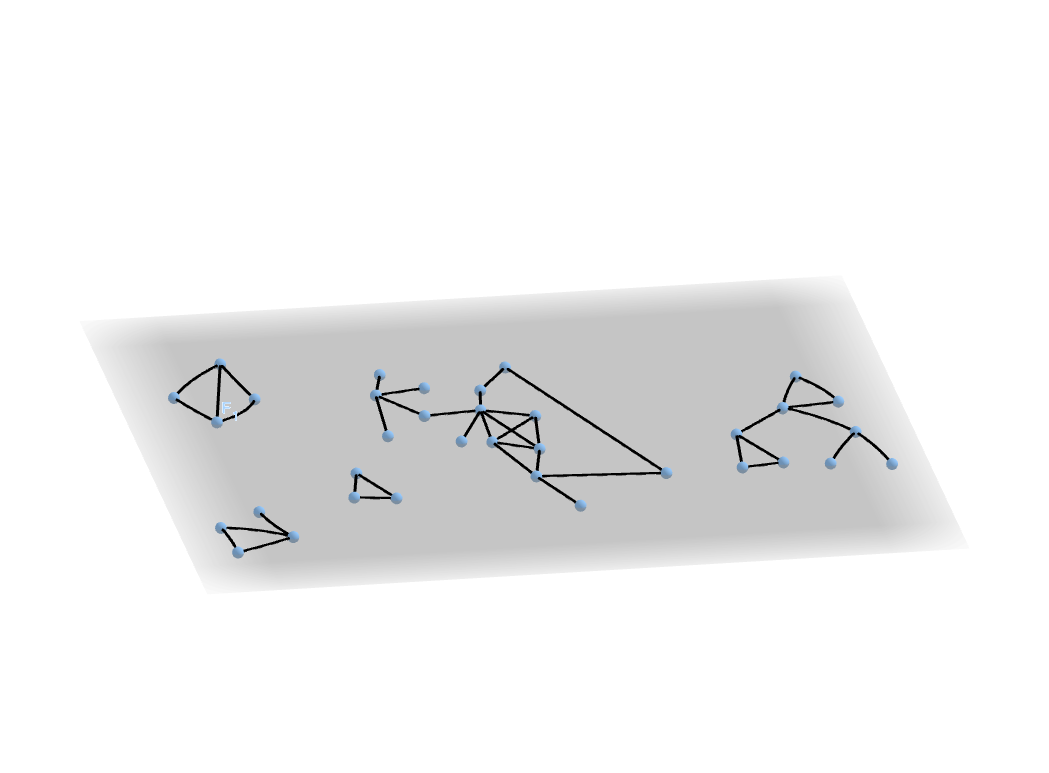}
    }
    \subfloat[Positive $\kappa^\star > 0$]{
    \includegraphics[scale = 0.15]{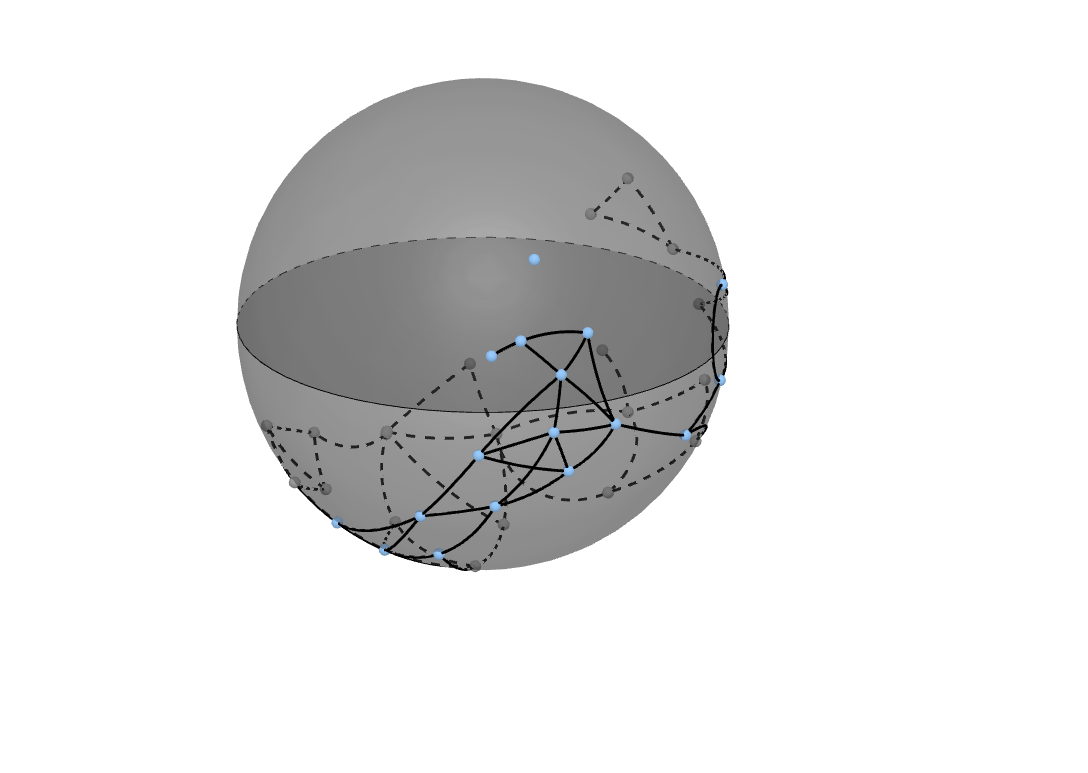}
    }
    \subfloat[Negative $\kappa^\star < 0$]{
    \includegraphics[scale = 0.15]{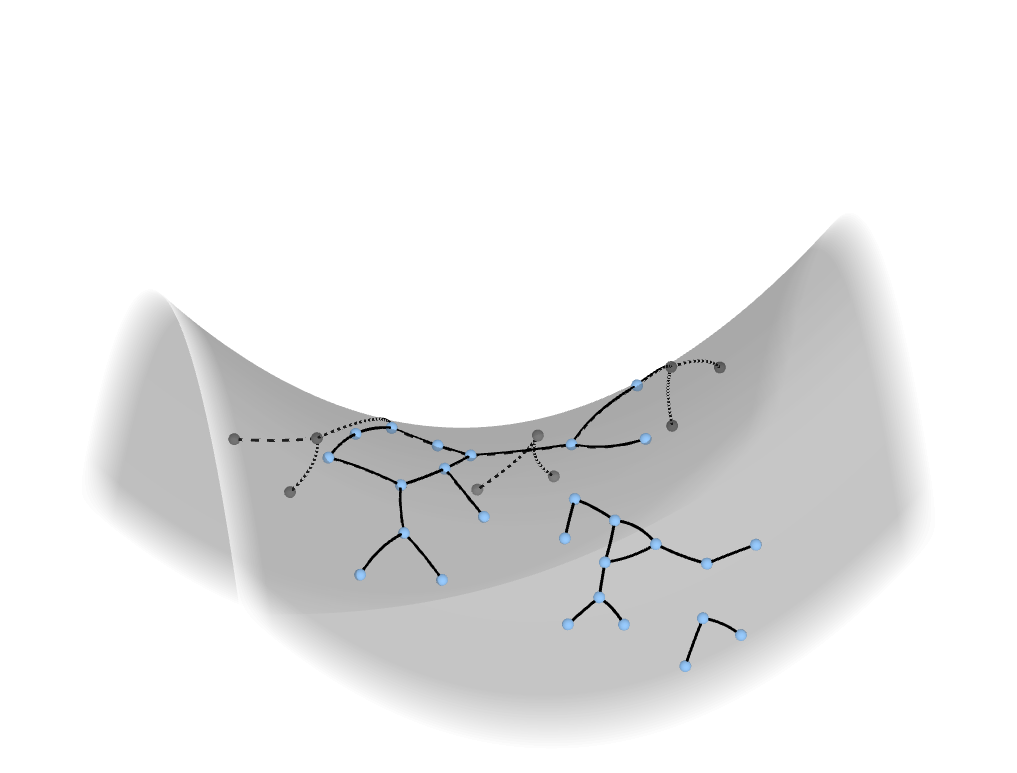}
    }
    \caption{\footnotesize{How curved geometries affect network embeddings where each displayed graph has 36 nodes.}}
    \label{fig:curvature}
\end{figure}

We now discuss how Assumption \ref{ass:manifold} applies to our network problem.  Simple connectedness and completeness are innocuous and constant curvature  provides a place to start and nests all manifolds used in the literature, but rule out inhomogeneous latent spaces entirely such as those with ``structural holes'' in the manifold, like the torus. Nevertheless, these three types of manifolds span a large and usable set of empirically relevant networks. With zero curvature, we model networks that allow for many paths where following them along nodes takes one increasingly far from nodes in other directions, while preserving local clustering. So while there is clustering, a flat space models a sort of vastness. Meanwhile, a sphere which has constant positive curvature does force such behavior. Following friends of friends of friends and so forth typically leads to encountering some distant friends in common at a much higher rate. Therefore there is a sort of cloistering in addition to clustering. Finally, hyperbolic spaces in contrast naturally embed trees or hierarchical networks or any context where expansiveness is a key feature.  Intuitively this is because any set of initially parallel lines spread apart. Figure \ref{fig:curvature} presents intuitions. \cite{smith2019geometry} provides a comprehensive discussion on the relationship between network properties and the latent space.

 Assumption~\ref{ass:nu} is a mild assumption that ensures that (\ref{eq:main_model}) produces probabilities.\footnote{Equivalently if one defined $\theta_i = \exp(\nu_i)$, then these fixed effects are simply multiplicative factors on the linking probability due to distances.}

\begin{ass}\label{ass:nu}
    Every node $i$ has a fixed-effect $\nu_i^\star$ i.i.d. from a distribution $F_\nu$. The support of $F_\nu$ is required to be such that (\ref{eq:main_model}) always returns values in $[0,1].$
\end{ass}
If the link function is exponential, Assumption \ref{ass:nu} requires that $\text{support}(F_\nu) \subseteq (-\infty, 0]$, but for the logistic link function Assumption \ref{ass:nu} is satisfied for any distribution. Recall that our goal is to apply Theorem \ref{thm: intro_thm} to identify the manifold properties from just one network. We 
 need to identify a set of $K$ locations on the manifold and a sequence of estimators $\hat D_T$ that satisfy the assumptions in Theorem~\ref{thm: intro_thm}.  

Our approach, which we study in detail in Section \ref{sec: network_geometry_identification}, exploits the clique structure in the network.  
Because of (\ref{eq:main_model}), nodes in even modest sized cliques (e.g., 5 nodes) are very likely to be close in the latent space. 
 In other words, we can imagine that nodes in a clique are at the same location on the manifold. Finding $K$ disjoint cliques in the network therefore gives us $K$ distinct points in the latent space. By counting the number of edges between pairs of these $K$ cliques, we can therefore estimate the probability that nodes on the latent space connect. Since (\ref{eq:main_model}) relates distances in the latent space and edge probabilities, we can therefore use the estimated probability of connections to estimate distances between these $K$ points.\footnote{This makes clear that if the researcher observed weighted graph data, since $G_{ij}$ is now a smooth function of distance and fixed effects, they can dispense with the clique approach altogether, since they directly observe a transformation of distances between specific points. The problem is easier. } We use the notation $C(\ell)$ to denote a clique of size $\ell$ in a graph, and $C_1(\ell), \dotsc, C_K(\ell)$ refers to a collection of $K$ cliques each of size $\ell$.

\begin{algorithm}
\SetAlgoLined
{\small Input: graph $G$.

\begin{enumerate}
\item Construct $\hat D = \hat D(G)$.
\begin{enumerate}
    \item Identify $K$ disjoint $\ell$-cliques $C_1(\ell), \dotsc, C_K(\ell)$.
    \item Estimate the cross-clique linking probability $ \hat P_{kk'} = \frac{1}{\ell^2} \sum_{i \in C_k(\ell)} \sum_{j \in C_{k'}(\ell)} G_{ij}$.
    \item Calculate $\hat D_{kk'} = -\log(\hat P_{kk'}) + \log( \hat \gamma)$, where $\hat \gamma$ is an estimate of $E\{\exp(\nu)\}^2$.
\end{enumerate}
\item Apply Algorithm \ref{alg: statistical_geometry} to $\hat D$ to construct estimator $\hat{\mathcal{M}}^{\hat p}(\hat \kappa)$.
\end{enumerate}
}
 \caption{Estimating Geometry of Latent Space Network Model using Cliques}
 \label{alg: cliques}
\end{algorithm}

We require an assumption to ensure that observed cliques are likely to be comprised of nodes that are near each other in the latent space. Assumption~\ref{ass:z} sets out a general requirement for $F_z$ which makes explicit the condition that is required for our proofs: 
proportionally most of the cliques in the graph are comprised of nodes that are proximate in latent space. This assumption 
 captures a typical feature of latent space models and empirical data. 

Let $G$ be a graph drawn from (\ref{eq:main_model}). For an arbitrary set of nodes $V_0 \subseteq V$, let $G_{V_0}$ denote the sub-graph induced by these nodes. If $|V_0| = \ell$, we use   $\{G_{V_0} \in C(\ell)\}$ to denote the event that  $G_{V_0}$ is an $\ell$-clique; that is, $G_{V_0}$ is a complete graph on $\ell$ nodes.

\begin{ass}\label{ass:z}
Every node $i$ resides at a location $z_i^\star$ that is drawn independently and are identically distributed from a distribution $F_{z}$ on manifold $\mathcal{M}^{p^\star}(\kappa^\star)$.  The latent location distribution must satisfy two properties: 
\begin{enumerate}
    \item [(a)] \emph{Identifiability:} The support of $F_{z}$ must consist of at least $K > p^\star$ distinct points that uniquely identify the manifold.
    \item [(b)] \emph{Local cliques:} 
    For any collection $V_0$ of $\ell$ nodes   with locations drawn i.i.d. from $F_z$ 
     we have for all $\delta > 0$, 
     $\Pr\left\{\max_{ij \in V_0} d(z_i, z_j) \leq \delta |G_{V_0} \in C(\ell) \right\} \rightarrow 1$,
    as $n, \ell \rightarrow \infty$.\end{enumerate}
\end{ass}
Part (a) states that we need $K$ to be larger than the true dimension and that we need there to be only one latent space in which we can embed these points isometrically (Section \ref{sec: Geometry} contains definitions of these terms). Part (b) states that given a clique of size $\ell$, the probability that nodes in this clique are close to each other goes to 1 as the clique size and graph size grow.  Aside from this condition, we impose no restrictions on the distribution of $F_z$.  This allows for continuous, discrete, or mixed distributions as well as dependence on $n$. In Section \ref{sec: ass_1_3} we provide a discussion on these two conditions in the context of the network model and provide high-level conditions which imply Assumption \ref{ass:z}. We  prove that these high-level conditions  hold in common cases, such as when the node locations are drawn from a lattice model,  Gaussian mixture model, or uniformly over a bounded but expanding region. 

Before continuing, we  emphasize a few key points. First, Assumption \ref{ass:z}(b) is written with $n, \ell \rightarrow \infty$. Let $\ell = \ell(n)$ depend on the graph size. In order for there to be cliques of size $\ell$ as $n \rightarrow \infty$, we need $\ell(n)$ to grow slowly, usually $\ell(n) \propto \log(n)$. Appendix \ref{sec: existence_of_cliques} shows that cliques of size $\log(n)$ exist with high probability as $n$ grows for many common location distributions. 
Second, the existence of cliques is guaranteed by the latent space model under our assumptions as the number of nodes increases.
 The conditional independence relation that is key to the latent space model requires an assumption of exchangeability.\footnote{Large graphs with exchangeable nodes are only dense if the graph is a subgraph of an infinitely exchangeable graph. If the distribution of the graph changes with $n$, the limiting graph need not be dense.}  The Aldous-Hoover Theorem implies that exchangeable sequences of nodes correspond to dense graphs in the limit~\citep{aldous1981representations,orbanz2015bayesian}, which implies that cliques are present in the limit. We also
  examine the existence of cliques 
   using our empirical and simulated examples.  We find that 
  the number and size of cliques in our empirical examples is sufficient to match settings in simulations where the method controls Type 1 error and has high power.

\begin{thm}[Estimating geometry  via cliques]\label{thm: main_cliques}
Let $\mathcal{M}^{p^\star}(\kappa^\star)$ be a latent space of unknown type, dimension, and curvature that satisfies Assumption \ref{ass:manifold}. Consider a sequence of graphs on $n$ nodes drawn from the distribution in \eqref{eq:main_model}, satisfying Assumptions \ref{ass:nu}-\ref{ass:z}. Under these assumptions, the estimators produced by Algorithm \ref{alg: cliques}  are consistent   as $n, \ell \rightarrow \infty$. That is,
\(
\Pr( \hat{\mathcal{M}}^{\hat p} \neq \mathcal{M}^{p^\star} ) = o(1) \text{ and } \hat \kappa - \kappa^\star = o_P(1).
\)
\end{thm}

 \section{Overview of Geometry and embedding conditions}\label{sec: Geometry_Overview}
In this section, we provide a brief overview of the three manifold types we consider in this work: Euclidean, spherical, and hyperbolic space. As noted above, these spaces span a class of empirically relevant manifolds on which much of the latent space network literature focuses \citep{Killing1891}. We then provide necessary and sufficient conditions on an arbitrary distance matrix that ensures the points from which the distances are computed can be embedded isometrically into one of these three geometries.

 \subsection{Candidate Geometries}
\label{sec: Geometry}

In order to study the candidate manifold $\mathcal{M}^p(\kappa)$, we  embed them in $\mathbb{R}^{p+1}$. Clearly the Euclidean case is trivial. In the spherical case we embed it in Euclidean space ($\mathbb{R}^{p+1}$ with the usual metric) and in the hyperbolic case we use Minkowski space ($\mathbb{R}^{1,p}$). Note that the only difference is that the bilinear form of the space, denoted by $Q$ below, varies in signature described below.

The model for each  is constructed by looking at a locus of points in the ambient space in which it is embedded:\footnote{For the hyperboloid $x_0 > 0$ is an additional restriction.}
\[
\mathcal{M}^p(\kappa) := \left\{ x \in \mathbb{R}^{p+1}: \ Q(x,x) = \kappa^{-1} \right\}.
\]
This implies a way of calculating distances between points on the manifold. Specifically, 
\[
d_{{\bf \mathcal{M}}^{p}}\left(x,y\right)=\frac{\arccos\left\{\kappa Q\left(x,y\right)\right\}}{\sqrt{\kappa}}.
\]

\

Let us turn to our candidate cases. The Euclidean space $\mathbb{R}^p$ is the $p$-dimensional Euclidean space with the usual Euclidean metric. In the case of the sphere $\mathbf{S}^p$, we have the usual Euclidean inner product $Q_{\mathbb{R}^{p+1}}\left(x,y\right):=\sum_{i=1}^{p+1}x_{i}y_{i}$.  The locus of points and distances between two points  
$x,y\in\mathbb{R}^{p+1}$ for the embedding is

\[
{\bf S}^{p}\left(\kappa \right):=\left\{ x\in\mathbb{R}^{p+1}:\ Q_{\mathbb{R}^{p+1}}\left(x,x\right)=\kappa^{-1},\ \kappa>0\right\} \text{ and } d_{{\bf S}^{p}}\left(x,y\right)=\frac{\arccos\left\{\kappa Q_{\mathbb{R}^{p+1}}\left(x,y\right)\right\}}{\sqrt{\kappa}}.
\]

\

Hyperbolic space $\mathbf{H}^p$ is embedded in Minkowski space, $\mathbb{R}^{1,p}$ which is $\mathbb{R}^{p+1}$ equipped with the Minkowski bilinear form:   $Q_{\mathbb{R}^{1,p}}\left(x,y\right):=-x_{0}y_{0}+\sum_{i=1}^{p}x_{i}y_{i}$. 
  The important point is that the signature is distinguished from the Euclidean space which  will play  a key  role in distinguishing the geometries. The locus of points and  distances are given by 
\[
{\bf H}^{p}\left(\kappa \right):=\left\{  x = (x_0,x_{1:p})\in\mathbb{R}^{1,p}:\ Q_{\mathbb{R}^{1,p}}\left(x,x\right)=\kappa^{-1},\ x_{0}>0,\ \kappa <0\right\}\]
and  
\[d_{{\bf H}^{p}}\left(x,y\right)=\frac{{\rm arccos}\left\{ \kappa Q_{\mathbb{R}^{1,p}}\left(x,y\right)\right\}}{\sqrt{\kappa}}.
\]

\subsection{Isometric Embedding Conditions}\label{sec:embedding}

Equipped with a notion of how distances are calculated between points in our candidate manifolds, we briefly review the conditions to check if a collection of $K$ points can be isometrically embedded in each of the manifolds.

Let $D$ be a known distance matrix from $K$ points given by $Z = \{z_1,\ldots,z_K\}$. We say that $Z$ can be \emph{isometrically embedded} in manifold  $\mathcal{M}^p(\kappa)$, written as $Z \overset{\text{isom}}{\rightarrow} \mathcal{M}$, if there exists an isometry $\phi$ such that for all $l, l' \in \{1, \dotsc, K\}$, $d_\mathcal{M}(\phi(z_l), \phi(z_{l'})) = d_{ll'}.$ 

Given $D$, we define the $K \times K$ matrix $W_\kappa(D)$ which will allow us to determine whether an isometric embedding is possible in one of the three candidate geometries. To do this, we choose the matrix to correspond to the bilinear form $Q(\cdot,\cdot)$ above. For the Euclidean case we need a matrix $J := I_K - \frac{1}{K}1_K 1_K'$. Our test matrix is given by
\begin{equation}
\label{eq: test_W}
W_{\kappa}\left(D\right):=\begin{cases}
\frac{1}{\kappa}\cos\left(\sqrt{\kappa}D\right) & \text{ for }\kappa\neq0\\
-\frac{1}{2}JD\circ DJ & \text{ for }\kappa=0,
\end{cases}
\end{equation}
where we apply the cosine operation element-wise, as before. We write $W_\kappa  = W_\kappa(D)$, suppressing the dependency on  $D$ unless otherwise noted. By using a Taylor series of $W_\kappa(D)$ around $\kappa = 0$, one can see the relationship between the expression of $W_\kappa(D)$ for $\kappa > 0$ and $W_0(D).\footnote{We would like to thank Gabriel Caroll for pointing this out to us.}$

The following lemma characterizes the conditions for isometric embedding and is a concise restatement of classical results: \cite{Schoenberg35} Theorem 1 (which we include as part (1) of our Lemma \ref{lem: embedding}) and \cite{begelfor2005world} Theorem 1 (which we include as parts (2-3) of our Lemma \ref{lem: embedding}). See \cite{Belton_geometry} for an overview of related topics in distance geometry. 
The \emph{signature} of a square matrix $A$ is a triple $\text{sig}(A)=(a, b, c)$, where $a$, $b$, and  $c$ are respectively the number of positive, zero, and negative eigenvalues of $A$. A positive semi-definite matrix has $c=0$. Throughout the paper, we use the convention that $\lambda_{\max}(A) := \lambda_1(A) \geq \lambda_2(A) \geq \dotsc \geq \lambda_K(A) =: \lambda_{\min}(A)$ are the eigenvalues of the $K \times K$ matrix $A$ sorted in decreasing order.

\begin{lem}\label{lem: embedding}Let  $p_{\min}$ be the minimum dimension for which $Z \overset{\text{isom}}{\rightarrow} \mathcal{M}^p(\kappa)$, and assume $a>0$. 
\begin{enumerate}
    \item $Z \overset{\mathrm{isom}}{\rightarrow} \mathbb{R}^p$ for some $p$ if and only if $\text{sig}(W_0)=(a,K -a,0)$. Further, $p_{\min} = a$.
    \item $Z \overset{\text{isom}}{\rightarrow} \mathbf{S}^p(\kappa)$ if and only if $\mathrm{sig}(W_\kappa) = (a,K -a,0)$.  Further, $p_{\min} = a-1$.
    \item $Z \overset{\text{isom}}{\rightarrow} \mathbf{H}^p(\kappa)$ if and only if $\mathrm{sig}(W_\kappa) = (1,K-a-1, a)$. Further, $p_{\min} = (K-a)-1$.
\end{enumerate}
\end{lem}

The result above tells us, for example, that $\lambda_{\min}(W_0) \geq 0$ when $D$ is computed from points in $\mathbb{R}^p$. This allows us to then phrase the problem of geometry identification, where we do not observe the manifold, into a problem about eigenvalues of $W$, which we do observe. This re-framing of the statistical geometry problem is the main insight behind our geometry classification procedure in Theorem \ref{thm: intro_thm}. The lemma also allows us to estimate the dimension of the latent space through the rank of $W_\kappa$, which provides the basis of our dimension estimation procedure in Section \ref{sec: dim_estimate}.


\section{Testing Geometry from an Arbitrary Distance Matrix Estimate $\hat D$}\label{sec:geometry-general}
This section addresses how to test whether a set of $K$ points can be isometrically embedded into a candidate manifold out of the set we consider, given only a consistent estimator $\hat D$ of the pairwise distance matrix $D$ between them.  We develop the constituent pieces for our main result, Theorem \ref{thm: intro_thm}.

This section, therefore, is not about networks exclusively but rather about a general statistical geometry problem that we outlined in Theorem \ref{thm: intro_thm}.
We show below that the general statistical geometry problem outlined in Section \ref{sec: stat_geom_prob} is applicable in a wide range of relational data examples, extending the scope of our work beyond the binary adjacency matrix case we consider in Section~\ref{sec: network_geometry_identification}. Each of the following examples will produce an estimator $\hat D$ which approximates some unknown matrix $D$, which contains pair-wise distances between $K$ objects (people, nodes, firms, etc.). For the asymptotic results presented below, index $\hat D$ by some  $T$ such that $\hat D = \hat D_T \overset{p}{\rightarrow} D$ as $T \rightarrow \infty$. Informally, $T$ maybe be thought of as the sample size. In our main application to networks, $T$ is a function of network size $n$. However, the general statistical geometry problem may generate an estimator of distances in other ways. For example, longitudinal data where $T$ indexes time in an international trade example or number of samples in a neuroscience example. For convenience, we often drop the notation $\hat D_T$ and instead simply write $\hat D$. 

Because the results of this section are more general
than the network model studied in equation \eqref{eq:main_model}, our main application, we provide
a few examples to develop an intuition for other potential applications. These examples deal with a general set of problems where we observe weighted relationships between pairs of nodes, and these relationships are formed based on distances or dissimilarities between nodes. The latent space model is a special case where the relationship is binary. The geometric results in this
section are not confined to even these applications -- in fact, our estimation procedure in Theorem \ref{thm: intro_thm} only requires a consistent estimate of distances along the latent space.

\subsection{Examples of Relational Data with Distances, $\hat D$}\label{sec: examples_dist} We provide four motivating examples for $\hat D$. 
{
\begin{example}[A Single Large Network]
\label{example: network}
The researcher only observes a single large network, $G$, drawn
from the distribution specified in equation \eqref{eq:main_model}. 
We study it in detail in Section \ref{sec: network_geometry_identification}.
\end{example}
\begin{example}[Relational data, \cite{petersen2015brain,abdelnour2018functional, ginestet2017hypothesis, Hoff_Music, salter2017latent}]
\label{ex: longitudinal}
There are $K$ units, such as individuals, neurons, sensors, or firms.
The researcher observes an outcome of an interaction between two units
$i$ and $j$, given by $f_{ij,t}$ at time $t=1,\ldots,T$ with some disturbance $\epsilon_{ij,t}$. For instance, $f_{ij,t}=\Lambda\left\{d_{\mathcal{M}^{p^{\star}}\left(\kappa^{\star}\right\}}\left(z_{i}^{\star},z_{j}^{\star}\right)+\epsilon_{ij,t}\right)$ where $\Lambda$ is a bijective function. This may be binary or continuous, such as an instance of a signal being transmitted between
neurons or sensors, some financial flows between individuals, or some
transactions between firms. We can then compute $\hat{D}$
with entries $\hat{d}_{ij}=\frac{1}{T}\sum_{t}\Lambda^{-1}\left(f_{ij,t}\right)$
and under regularity conditions (such as differentiability of $\Lambda^{-1}$) our results will follow.

A particularly relevant case is the following. The researcher observes $T$ networks $G_1, \dotsc, G_T$, where $T$ could represent the number of observations of one network with a fixed set of nodes, or it could represent the number of observed networks, each with a potentially distinct set of nodes. Here we write $p_{ij} := \mathbb{P}(g_{ij} = 1 | z_i^\star, z_j^\star)$ is the probability that nodes $i$ and $j$ connect, given their latent space locations. This term depends on $d_{\mathcal{M}}(z_i^\star, z_j^\star)$. We write this as $p_{ij} = H_{ij}\{d_{\mathcal{M}}(z_i^\star, z_j^\star)\}$ for some invertible function $H_{ij}$. Here we suppose that the generative model for the networks is constant across all $T$ networks.  We can then estimate $\hat D_{ij} = H^{-1}(\hat P_{ij})$, where $\hat P_{ij} = T^{-1} \sum_{i = 1}^T G_{ijt}$ is the number of observed edges between nodes $i$ and $j$, normalized by the number of observations $T$. 
\end{example}

\begin{example}[Trait Groups, \cite{killworth1990estimating,mccarty2001comparing}]
\label{ex: trait_groups}
The $n$ nodes each have one of $K$ traits and the number with a
trait $k$ is given by $n_{k}$. Locations in the latent space are determined uniquely by a node's trait, and nodes with the same trait share the same location on the latent space, with $z^\star_{k_i}$ denoting the common location of nodes with trait $k_i$. An interaction between nodes follows
$f_{ij}=\Lambda\left\{d_{\mathcal{M}^{p^{\star}}\left(\kappa^{\star}\right)}\left(z_{k_{i}}^{\star},z_{k_{j}}^{\star}\right)+\epsilon_{ij}\right\}$
where again $\Lambda\left(\cdot\right)$ is bijective. In this example,
the researcher can construct $\hat{D}$ with entries \begin{equation*}
    \hat{D}_{kk'}=\frac{1}{n_{k}n_{k'}}\sum_{i,j}\Lambda^{-1}\left(f_{ij}\cdot{\bf 1}\left\{ k_{i}=k,k_{j}=k'\right\}\right) \;,
\end{equation*}
where $n_k$ is the number of nodes with trait $k$, which can be obtained from census or surveys.
\end{example}
}

\subsection{Perturbation}
In Section \ref{sec:embedding} we saw that the isometric embedding conditions related the manifold class, curvature, and dimension to the spectrum of a test matrix $W_\kappa(D)$. In practice, since we observe $\hat D$, we must construct $\hat W_{\hat \kappa}(\hat D)$ and study its spectrum. The main idea of our approach to estimating the manifold class, curvature, and dimension comes from Weyl's inequality, which says that the eigenvalues of the estimated test matrix, $\hat W_{\hat \kappa}(\hat D)$ are very close to those of the target $W_\kappa(D)$ if the estimators of the distance matrix and curvature  are consistent. Under our assumptions, we are able to bound how the estimated spectrum may deviate from the true spectrum.

\begin{prop}
\label{prop:alpha_bound}
 Suppose that $D$ is a $K \times K$ distance matrix from $K$ points on $\mathcal{M}^{p^\star}(\kappa^\star)$ satisfying Assumption \ref{ass:manifold}, with $K$ chosen such that the $\mathcal{M}^{p^\star}(\kappa^\star)$ is uniquely identified. Assume further that  there are estimators $\hat D$ and $\hat \kappa$ such that 
$\hat D_T -D \overset{p}{\rightarrow} 0$ and $\hat \kappa \overset{p}{\rightarrow} \kappa^\star$ as $T \rightarrow \infty.$ Let $\theta_\alpha$ be defined as the $\alpha$th quantile of the distribution of  $\Vert \hat W_{\hat \kappa} - W_{\kappa} \Vert_F$. Then, for every $k \in \{1,\ldots,K\}$,
	\begin{equation}
\Pr\left\{  |\lambda_{k}( \hat{W}_{\hat \kappa})- \lambda_{k}\left( W_{\kappa} \right)|  < \theta_{\alpha} \right\} \leq  \alpha \;.
\label{eq: Type1Bound}
\end{equation}
\end{prop}
Owing to this result, we can study the estimated spectrum in order to look at the metric signature and therefore estimate the manifold class, curvature, and dimension.

\subsection{Hypothesis Tests of Geometry} 
We now frame the problem of classifying the geometry of $D$ into three hypothesis tests. We will then combine the results of these three tests into a classifier of the geometry. To determine the geometry of $\mathcal{M}$, we use Lemma \ref{lem: embedding}  and develop testable statements about the spectrum of the test matrix $W_\kappa$.

For the Euclidean case, we can test positive semi-definiteness of the test matrix as
\begin{equation}
    H_{0, e}: \lambda_K(W_0) \geq 0,  \ H_{a, e}: \lambda_K(W_0) < 0.
    \label{eq: Euclidean_Hypotheses}
\end{equation}

Using the same reasoning, for the spherical case we can test the hypothesis that the embedding space is spherical for some $\kappa > 0$ as
\begin{equation}
    H_{0, s}: \lambda_K(W_\kappa) \geq 0, \ \ H_{a, s}: \lambda_K(W_\kappa)  < 0 \;.
        \label{eq: Spherical_Hypotheses}
\end{equation}
since the test matrix must be positive semi-definite.

Finally, to determine if the embedding space is hyperbolic for some $\kappa < 0$, we want to test
\begin{equation}
    H_{0, h}: \lambda_{2}(W_\kappa) =0, \ \ H_{a, h}: \lambda_{2}(W_\kappa) \neq 0 \;
        \label{eq: Hyperbolic_Hypotheses}
\end{equation}
since the signature switches sign and that the matrix does not have full rank under the assumption on dimension implies that there are zeros in the spectrum.\footnote{By Lemma \ref{lem: embedding}, failing to reject in the hyperbolic case $\lambda_{2}(W_\kappa) = 0$ is not enough to conclude that $D$ is hyperbolic, since we must test the first is positive and smallest eigenvalue is negative as well. In practice, however, we found that testing only one eigenvalue was  sufficient and, thus, use this simpler test.  Clearly, it would also be possible to test all three eigenvalues using an intersection test, which we leave to future work.}

  A natural first step in deriving a geometry classifier would be to use Proposition \ref{prop:alpha_bound}.    While this method produces a type-1 error that is below $\alpha$, the power of the method may be low. So as a classifier, that procedure can be improved upon. In Section \ref{sec: estimate_geometry} we derive more powerful tests by approximating the distribution of the eigenvalues under the assumption that the distances are computed along one of the three geometries (recall from Lemma \ref{lem: embedding} that the eigenvalues of $W_\kappa$ tell us the underlying geometry type).

To derive tests of the three geometry hypotheses, we  first need to estimate the curvature of $\mathcal{M}$, which is computed assuming the geometry of $\mathcal{M}$ is curved (i.e., not Euclidean). We use these estimated curvature values to then identify the geometry type. For this reason, we start our discussion  
by studying curvature.

\subsection{Estimating Curvature $\kappa^\star$}
\label{sec: curvature_estimate}
We begin assuming that the researcher has a consistent estimator  $\hat D_T \overset{p}{\rightarrow} D$ (which we will develop below). Equipped with $\hat D_T$, we construct a  consistent estimate of $\kappa^\star$. The core observation comes from Lemma \ref{lem: embedding}. Namely, by looking at $\mathrm{sig}(W_\kappa)$, we can use the fact that certain eigenvalues must \emph{exactly} be zero under the various geometries. For instance, both $\lambda_K(W_0(D)) = 0$ and $\lambda_K(W_{\kappa^\star}(D)) = 0$ for the Euclidean and spherical cases, respectively, as both are positive semi-definite, provided $p^\star < K$. A similar phenomenon is true for $\lambda_{2}(W_{\kappa^\star}(D)) = 0$ for the hyperbolic case.

\begin{figure}\centering
\subfloat[Spherical]{\label{spherical}\includegraphics[width=.45\linewidth]{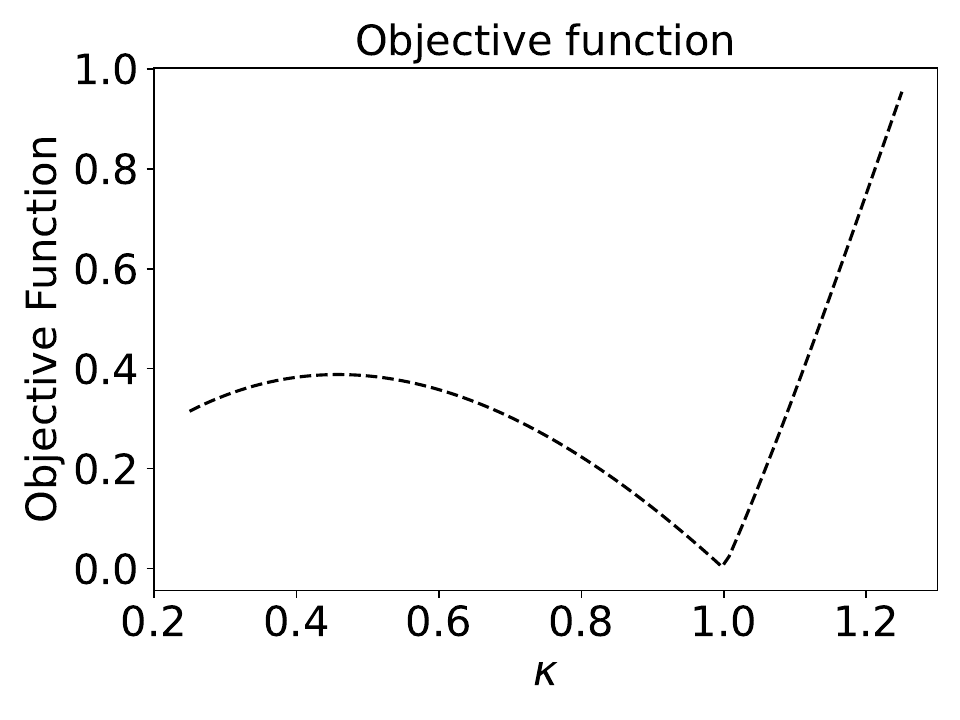}}\hfill
\subfloat[Hyperbolic]{\label{hyperbolic}\includegraphics[width=.45\linewidth]{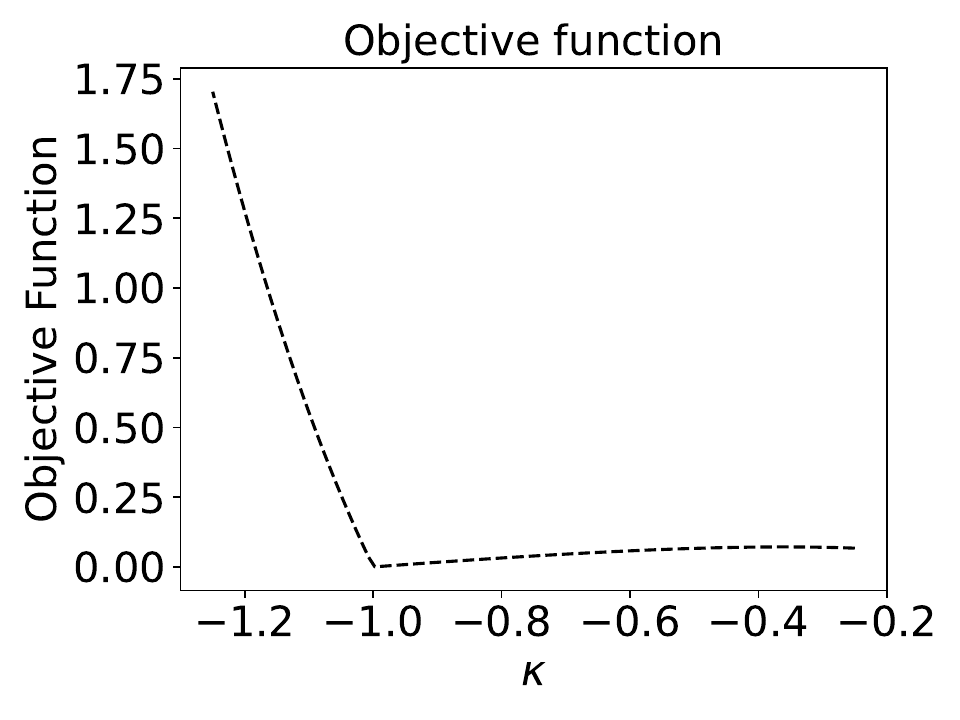}}
\caption{\footnotesize{Plot of the objective function from (\ref{eq: curvature_estimator}) when $D$ corresponds to 15 points in $\mathbf{S}^2(1)$ (left) and  $\mathbf{H}^2(-1)$ (right). We plot the curvature $\kappa$ against the value of the function $\kappa \mapsto \Big|\lambda_1\left(\cos(\sqrt{\kappa}D)\right)\Big|$ (left) and of the function $\kappa \mapsto \Big|\lambda_2\left(\cos(\sqrt{\kappa}D)\right)\Big|$ (right). We see that at the true $\kappa$, the objective function is minimized.}}
\label{fig: Objective functions}
\end{figure}

This observation leads to the following estimators of the curvature: 
\begin{equation}\label{eq: curvature_estimator}
\hat \kappa_S := \underset{\kappa \in [a, b]}{\text{arg min}} \  \Big|\lambda_1\left\{\kappa W_\kappa(\hat D_T)\right\}\Big|, \ \ \ \ \hat \kappa_H := \underset{\kappa \in [-b, -a]}{\text{arg min}} \  \Big|\lambda_{2}\left\{\kappa W_\kappa(\hat D_T)\right\}\Big|
\end{equation}
for some $0 < a < b$. In Appendix \ref{sec: PickingBounds_ab}, we discuss how to pick $a$ and $b$ in practice. The subscript $S$ indicates that this estimate is used when testing if the manifold is spherical. Similarly, the subscript $H$ indicates that this estimate is used when testing if the manifold is hyperbolic. As $T \rightarrow \infty$, the estimates approach the true curvature under the correct geometry. 
\begin{prop}[Consistency of curvature estimates]  Suppose that $D$ is a $K \times K$ matrix containing pairwise distances between points in either $\mathbf{S}^p(\kappa^\star)$ or $\mathbf{H}^p(\kappa^\star)$, where $|\kappa^\star| > 0$.  Let $Z$ denote the collection of these $K$ points. Suppose there is an estimate $\hat D_T$ such that $\hat D_T  \overset{p}{\rightarrow}D$ as $T \rightarrow \infty$. Finally, suppose that there is either 
\begin{enumerate}
    \item  a unique $\kappa^\star \in [a,b]$ such that $Z \overset{\text{isom}}{\rightarrow} \mathbf{S}^p(\kappa)$ for some $p$. Set  $\hat \kappa_T = \hat \kappa_S$.
    \item or  a unique $\kappa^\star \in [-b,-a]$ such that $Z \overset{\text{isom}}{\rightarrow} \mathbf{H}^p(\kappa)$ for some $p$.  Set  $\hat \kappa_T = \hat \kappa_H$.
\end{enumerate}
In cases (1) and (2),  $\hat \kappa_T \overset{p}{\rightarrow} \kappa^\star$ as $T \rightarrow \infty$.
\label{prop: ConsistencyKappa}
\end{prop}
 The estimators we propose for curvature are similar to those proposed in \citet{Wilson}, though \citet{Wilson} does not prove that these estimators are consistent. 

Before continuing, we want to discuss the requirements in Proposition \ref{prop: ConsistencyKappa}. First, Proposition \ref{prop: ConsistencyKappa} requires that $\kappa^\star$ is bounded away from zero (meaning that the space is not flat and hence has non-zero curvature). In other words, the manifold is either spherical or hyperbolic, and $a$ and $b$ must be chosen to include the true curvature value $\kappa^\star$. We also require that there is a unique curvature in which we can find an isometric embedding to ensure that the statement that $\hat \kappa_T \overset{p}{\rightarrow} \kappa^\star$ makes sense. We also want to emphasize that $\kappa^\star$ is fixed and again is assumed to be non-zero. The case where $\kappa^\star$ changes with $T$ is an interesting and challenging problem that we leave to future work. 

\subsection{Estimating Geometric Class $\mathcal{M}^\star$}
\label{sec: estimate_geometry}
Again, we suppose the researcher has access to a noisy distance matrix $\hat D_T$ that approximates an unknown distance matrix $D$ of interest. The matrix $D$ consists of pair-wise distances between $K$ objects along the surface of $\mathcal{M}$.

\subsubsection{Consistent Tests for Latent Geometry}
We begin by showing a consistent testing framework and then give a bootstrap method for implementation in Algorithm \ref{alg: subsample}.

To test $H_{0, e}$, first, define a rejection region $\mathcal{R}_T = (-\infty, \delta_T]$ for some real-valued sequence $\delta_T \in (-\infty, 0]$ and we define our test $\phi_T(\hat  W_{0}) \in \{0, 1\}$ as 
\begin{equation}
    \phi_T(\hat W_0) = 
    \begin{cases}
    0, \ \ \lambda_K(\hat  W_{0}) \in \mathcal{R}_T \\
    1,  \ \ \lambda_K(\hat  W_{0}) \not \in \mathcal{R}_T \;.
    \end{cases}
\label{eq: testE}
\end{equation}
If the test is 0, this indicates that we fail to reject the null hypothesis $H_{0,e}$ while if the test is 1, we reject the null hypothesis $H_{0,e}$. We use this notation throughout the paper when discussing the output of a hypothesis test.

When testing if $D$ is spherical, let $\hat \kappa$ denote an estimate of $\kappa$, defined in 
Proposition \ref{prop: ConsistencyKappa}. We define our rejection region of $H_{0, s}$ as $\mathcal{R}_T = (-\infty, \delta_T]$ and  our test as 
\begin{equation}
    \phi_T(\hat W_{\hat \kappa}) = 
    \begin{cases}
    0, \ \ \lambda_K(\hat W_{\hat \kappa}) \in \mathcal{R}_T, \\
    1,  \ \ \lambda_K(\hat W_{\hat \kappa}) \not \in \mathcal{R}_T \;,
    \end{cases}
\label{eq: testS}
\end{equation}
which is a similar test of positive semi-definiteness.

Finally, when testing if $D$ is hyperbolic, let $\hat \kappa$ denote an estimate of $\kappa < 0$, defined in
Proposition \ref{prop: ConsistencyKappa}. We define our rejection region of $H_{0, h}$ as $\mathcal{R}_T = [\delta_T, \infty)$ and define our test as 
\begin{equation}
    \phi_T(W_{\hat \kappa}) = 
    \begin{cases}
    0, \ \ \lambda_2(\hat W_{\hat \kappa}) \in \mathcal{R}_T, \\
    1,  \ \ \lambda_2(\hat W_{\hat \kappa}) \not \in \mathcal{R}_T \;,
    \end{cases}
\label{eq: testH}
\end{equation}
which looks to reject positivity of the second eigenvalue as per the  metric signature.

We now study what conditions must hold on this sequence $\delta_T$ in order for the three tests to be consistent, by which we mean that the probability the test rejects the null goes to 1 under the alternative hypothesis and that the probability it fails to reject the null goes to 1 under the null.

\begin{prop}
Let $\delta_T = o_P(1)$ be a random or deterministic sequence and let Assumption \ref{ass:manifold} hold. Let $\hat D_T \overset{p}{\rightarrow} D$ as $T \rightarrow \infty$. Then, 
\begin{enumerate}
    \item If $\delta_T \in (-\infty, 0]$, $\delta_T = o_P(1)$ and $ \mathbb{P}\left\{\lambda_K(\hat W_0) \leq \delta_T\right\} = 1 - o(1)$, then the test for $H_{0, e}$ in (\ref{eq: Euclidean_Hypotheses}) with rejection region $\mathcal{R}_T := (-\infty, \delta_T]$ is consistent.
    \item If $\delta_T \in (-\infty, 0]$, $\delta_T = o_P(1)$ and $ \mathbb{P}\left\{\lambda_K(\hat W_{\hat \kappa}) \leq \delta_T\right\} = 1 - o(1)$ with $\hat \kappa \in [a,b]$, then the test for $H_{0, s}$ in (\ref{eq: Spherical_Hypotheses}) with rejection region $\mathcal{R}_T := (-\infty, \delta_T]$ is consistent.
    \item  If $\delta_T \in [0, \infty)$, $\delta_T = o_P(1)$ and $ \mathbb{P}\left\{\lambda_2(\hat W_{\hat \kappa}) \geq \delta_T\right\} = 1 - o(1)$ with $\hat \kappa \in [-b,-a]$, then the test for $H_{0, h}$ in (\ref{eq: Hyperbolic_Hypotheses}) with rejection region $\mathcal{R}_T := [\delta_T, \infty)$ is consistent. 
\end{enumerate}
\label{prop: consistent_test}
\end{prop}

In order to combine these tests into a single estimate of the latent space geometry, we suggest using an ordered test. There are 6 possible orderings of such a test (e.g, Euclidean, then spherical, then hyperbolic). The proof is simple and uses the fact that each of the three geometry tests is consistent, under suitable assumptions on the threshold sequence $\delta_T$.

\begin{prop}[Consistent estimation of geometry type]
\label{prop: ordered_test}
Under the assumptions on the sequence $\delta$ in Proposition \ref{prop: consistent_test}, any of the 6 ordered tests return a consistent estimate of the latent space geometry.
\end{prop}

{
 We omit the proof of Proposition \ref{prop: ordered_test} which follows immediately from Proposition \ref{prop: consistent_test}.  We have shown that we can use the observed distance matrix $\hat D_T$ to test the hypotheses that the latent space is Euclidean, spherical, or hyperbolic.  From these tests we define $\hat{\mathcal{M}}^{\hat p}(\hat \kappa)$ as the intersection of the three tests.  That is, the estimated latent geometry based on $\hat{D}_T$ is defined by the result of three hypothesis tests in equations~\eqref{eq: testE},~\eqref{eq: testS}, and~\eqref{eq: testH}. More specifically, we can select any estimator that preserves consistency. As noted, for example, 
we can use an ordered test to estimate the geometry type. 
 Thanks to Proposition \ref{prop: consistent_test}, with sufficiently large $T$ the probability that more than one of these tests will fail to reject the null goes to zero, leading to a consistent test.}

\subsection{Estimating Dimension $p^\star$}
\label{sec: dim_estimate}
Given $\hat D$, $\hat \kappa$, and $\hat{\mathcal{M}}$,  we develop a consistent estimate of $p^\star$, the minimal dimension of the manifold class in which the points can be embedded. We focus on the minimum dimension since, trivially, if one can embed in $p$ one can embed in $p'>p$ for all $p'$. As we noted in Lemma \ref{lem: embedding}, we see that $p^\star$ relates to the rank. So we proceed by estimating the rank of $W_\kappa(D)$.

We present two approaches. The first continues our use of the logic of  Weyl's inequality to propose a consistent estimate as $T \rightarrow \infty$. The second uses the laddle plot method of \cite{Wei}. We did not verify the required assumptions in \cite{Wei} so we do not claim their estimator is consistent in our problem, but we find in practice (as do they) that the estimator performs well, so we suggest practitioners actually use this.

\subsubsection{Spectral estimate of dimension}
We are interested in finding the rank of $W_\kappa$ using $\hat W_{\hat \kappa}$. To do this, we let $\epsilon_T$ be a (potentially random) sequence such that $\epsilon_T$ goes to zero slower than any $\lambda_j(\hat W_{\hat \kappa})$ for which $\lambda_j(W_\kappa) = 0$. In other words, we select $\epsilon_T$ to go zero slower than the slowest zero eigenvalue of the test matrix $W_\kappa$.

Since the rank of a matrix is the number of non-zero eigenvalues, we will estimate the number of non-zero eigenvalues of $W_\kappa$. To do this, we define a rejection region $\mathcal{R}_T = (-\epsilon_T, \epsilon_T).$ For any index $T$, define the estimated rank to be
\begin{equation}
\label{eq: spectral_rank}
   \widehat{\text{rank}}(W_\kappa) = \#\{j = 1, \dotsc, K: \lambda_j(\hat W_{\hat \kappa}) \notin \mathcal{R}_T\} \;.
\end{equation}
Our estimate of the rank is then the number of observed eigenvalues that are sufficiently far away from zero, as measured by the threshold sequence $\epsilon_T$. 

Clearly, the performance of this estimator depends  on the choice of the threshold sequence $\epsilon_T$. If $\epsilon_T$ does not converge to zero (in probability), then this estimate cannot be consistent, since it will eventually start to count zero eigenvalues of $W_\kappa$ as being non-zero. Hence convergence to zero is a necessary condition. It must also converge fast enough to zero. For example, if $\lambda_j(W_\kappa) = 0$ and we have access to $\lambda_j(\hat W_{\hat \kappa}) = 1/T^2$ (i.e., a deterministic sequence), and we use $\epsilon_T = 1/T$, then we will under-count the rank of $W_\kappa$ because we will classify the eigenvalue $\lambda_j(W_\kappa)$ as non-zero at every $T$. 

We therefore must pick an $\epsilon_T$ that goes to zero slower than all estimates of eigenvalues for which their true counterpart is zero. From Weyl's inequality, we know what such sequences look like. For any index $k$ for which $\lambda_k(W_\kappa) = 0$,
\begin{equation*}
    |\lambda_w(\hat W_{\hat \kappa})| = |\lambda_k(\hat W_{\hat \kappa}) - \lambda_k(\hat W_{\hat \kappa})| \leq ||\hat W_{\hat \kappa} - W_{\kappa}||_F \;,
\end{equation*}
where the inequality is due to Weyl's. By defining $r_T := ||\hat W_{\hat \kappa} - W_{\kappa}||_F$, from the conditions in Theorem~\ref{thm: intro_thm}, we know that $r_T = o_P(1)$. We need to select $\epsilon_T \rightarrow 0$ with $r_T/\epsilon_T \rightarrow 0$, which means that $\epsilon_T$ goes to zero slower than $r_T$ does.
\begin{prop}[Consistency of minimum dimension estimate]
\label{prop: consistent_rank}
Choose $\epsilon_T \rightarrow 0$ such that $r_T/\epsilon_T \rightarrow 0$. If the assumptions in Theorem~\ref{thm: intro_thm} hold,  then $\widehat{\mathrm{rank}}(W_{\kappa}) \overset{p}{\rightarrow} \mathrm{rank}(W_\kappa)$ as $T \rightarrow \infty$. Using Lemma \ref{lem: embedding}, we can therefore consistently estimate $p^\star$ as $T \rightarrow \infty$.
\end{prop}

While the rank estimator above leads to a consistent estimate for suitable chosen $\epsilon_T$, choosing a sequence that satisfies these conditions in practice is challenging. We therefore recommend in practice to estimate the rank with a different estimator.

\subsubsection{\cite{Wei}: Laddle Plot}
Recent work, by \citet{Wei}, however, has been shown to have more appealing finite sample performance and in Appendix \ref{sec: rank_appendix}, we provide the algorithm to estimate the rank with this method. The intuition for their approach is as follows. Looking at the scree plot (related to our above approach) is consistent. And a bootstrap procedure, leveraging the fact that the eigenvectors corresponding to indices beyond the rank will be uncorrelated in  a bootstrap) also performs well. They note that under certain regularity conditions (a sufficiently fast estimate of the matrix $W_\kappa$, a self-similar bootstrap estimator), the combination of these two---a scree plot together with a bootstrap evaluation of eigenvector uncorrelatedness---performs better than either.

\citet{Wei} prove that in a number of problems, their rank estimate is consistent. We are not able to verify these conditions in practice, since they assume that their data consists of i.i.d. data. However, in our case, our data is independent but \emph{not} identically distributed. Therefore, we do not make a claim about the consistency of this approach when applied to our problem. We do, however, note that in simulations this approach has better finite sample properties.

To summarize, through Propositions \ref{prop: ConsistencyKappa}, \ref{prop: ordered_test}, and \ref{prop: consistent_rank} we have established that the procedure in Algorithm \ref{alg: statistical_geometry} is consistent for its estimands, as claimed in Theorem \ref{thm: intro_thm}.

Returning to our four examples, it is easy to see that  Examples \ref{ex: longitudinal} and \ref{ex: trait_groups} immediately satisfy the assumptions in Theorem~\ref{thm: intro_thm}, and as a consequence in each of those cases we can consistently recover the geometry. The situation is more subtle for Example \ref{example: network}, where rather than directly observing $K$ units and noisy distances among them, the researcher observes some graph $G$ is more subtle. From the graph, distances must be constructed and then geometry estimated. This is the subject of Sections \ref{sec: network_geometry_identification}.

\section{Identifying the Latent Space using only  Graph Data}
\label{sec: network_geometry_identification}
Having developed statistical tests to estimate the geometry behind a collection of points when we observe and arbitrary noisy distance matrix (the statistical geometry problem from Section \ref{sec: stat_geom_prob}), we now turn the original network problem. Specifically, our goal is to use a graph drawn from the latent space model in (\ref{eq:main_model}) and estimate the type, curvature, and dimension of the latent space. Theorem \ref{thm: main_cliques} presents the answer to this problem and demonstrate how the manifold can be consistently estimated. The proof of the theorem is provided in Appendix \ref{sec:proofs}. In this section, we lay out the ingredients: how one constructs the noisy distance matrix and adjusts for fixed effects. With these we are able to prove Theorem \ref{thm: main_cliques}, which states that Algorithm \ref{alg: cliques} returns a consistent estimate of $\mathcal{M}^{p^\star}(\kappa^\star)$.

Importantly, we also provide a number of examples of node location distributions, motivated by empirically-relevant models, to demonstrate that our core Assumption \ref{ass:z} for the clique-based method holds (Section \ref{sec: ass_1_3}). We conclude by discussing some practical choices for implementation.

  \subsection{Estimating Distances using Graph Data}
  \label{sec: LS_Application_Geometry}
  In order to apply the geometric test, we  use the graph $G$ to estimate a set of distances on $\mathcal{M}^{p^\star}(\kappa^\star).$ Since the  probabilities defined by the latent space model in (\ref{eq:main_model}), our approach is to use estimated linking frequencies in order to identify a system of implied distances between $K$ points in the manifold.

  To motivate our approach to constructing a distance matrix, we consider a simplification of the main model in (\ref{eq:main_model}). We make two assumptions for illustration, which are subsequently relaxed in the main analysis.  First, suppose there are no individual effects (so  $\nu_i^\star = 0 \ 
\forall i$). Second, suppose that nodes are assigned to one of $K$ distinct points in the latent space, which we denote by $\zeta_1^\star, \dotsc, \zeta_K^\star$.
Define $V_k = \{j \in \{1, \dotsc, n\}: z_j^\star = \zeta_k^\star\}$ to be the set of nodes at location $z_k^\star$. Under this simplification, we can write the distance between points $z_k^\star$ and $z_{k'}^\star$ using the definition of the latent space model in ~(\ref{eq:main_model}) as
\begin{equation}
    d_{k,k'} = -\log(p_{k,k'})
    \label{eq: simplified_model}
\end{equation}
where $p_{k,k'} := \mathbb{P}\left(G_{k,k'} = 1|z^\star\right)$ is the probability that nodes at locations $z_k^\star$ and $z_{k'}^\star$ connect for any $k, k' \in \{1, \dotsc, K\}.$\footnote{Clearly in this simplified model this corresponds to a stochastic block model with $K$ communities with a linking probability law $p_{kk'}$ that satisfies a geometric restriction.}  Then, we can estimate the probability $p_{k,k'}$ by
\begin{equation}
\label{eq: hatP_SBM}
    \hat p_{k,k'} := \frac{1}{|V_k| |V_{k'}|}\sum_{(i,j) \in V_k \times V_{k'}} G_{ij} \;.
\end{equation}
In words, this estimator counts the number of observed edges between $z_k^\star$ and $z_{k'}^\star$ and divides by the number of possible edges, given by $|V_k||V_{k'}|.$ Since $G_{ij} \overset{\text{i.i.d.}}{\sim} \text{Bernoulli}(p_{k,k'})$ for $(i, j) \in V_k \times V_{k'}$, this estimator is unbiased for $p_{k,k'}$. In addition, supposing that $|V_i| \rightarrow \infty$ as $n \rightarrow \infty$, the weak law of large numbers implies that $\hat p_{k,k'}-  p_{k,k'} \overset{p}{\rightarrow}0$.

By (\ref{eq: simplified_model}),  we can estimate a $K \times K$ distance matrix $\hat D = \{\hat d_{kk'}\}$ comprised of entries with distances between $\zeta_k$ and $\zeta_{k'}$ by
\[
\hat d_{kk'} = -\left[ \log(\hat p_{k,k'}) \right]_{k,k'}.
\]
We can then apply the continuous mapping theorem to show that $\hat d_{k,k'} - d_{k,k'}\overset{p}{\rightarrow} 0$. That is, the assumption in Theorem~\ref{thm: intro_thm} that we have access to a consistent estimate of distances along the latent space holds. Here, the sample size $T$ is the number of edges between the $K$ points on the latent space, given by $|V_k| |V_{k'}|$ for every pair of points $\zeta_k^\star$ and $\zeta_{k'}^\star$.

To summarize, in this example we have used the edges in $G$ to estimate a distance matrix $\hat D$ between $K$ points in the unobserved latent space $\mathcal{M}^{p^\star}(\kappa^\star)$. From this, we can apply Theorem \ref{thm: intro_thm} to consistently estimate the geometry. However, this example made two simplifying assumptions: no fixed effects and every node is located on exactly one of $K$ finite points on the manifold. Our more general result, which we now describe, relaxes these assumptions considerably.

In  (\ref{eq:main_model}),   nodes have individual fixed effects $\nu_i^\star$ as well as latent positions 
$z_i^\star$. The individual effects describe heterogeneity in the propensity for an individual to form connections and are not directly related to which connections 
will form, which is what the latent space captures.  We would prefer to estimate the distances used to test hypotheses about latent geometry without potential confounding by individual effects not specifically related to the geometry.  We accomplish this by marginalizing over the individual effects in (\ref{eq:main_model}). Recalling that the support of $\nu_i^\star$ is $(-\infty, 0]$, we integrate out the node effects to find that
\begin{align*}
    \mathbb{P}\left\{G_{ij} = 1| z^\star, \mathcal{M}^{p^\star}(\kappa^\star)\right\} &= E\{\exp(\nu_i)\}^2 \exp\{-d(z_i, z_j)\} \;.
\end{align*}

Solving for the distance $d(z_i^\star, z_j^\star)$, we have 
\begin{equation}
    d(z_i^\star, z_j^\star) = -\log(p_{i,j}) +2 \log[E\left\{\exp(\nu_i)\right\}] \;.
    \label{eq: Complex_Model_Solved_1}
\end{equation}
Note that if $\nu_i = 0$ with probability 1, which we assumed in the simplified model above, then $E\{\exp(\nu)\} = 1$, so that (\ref{eq: Complex_Model_Solved_1}) becomes (\ref{eq: simplified_model}). Here we use properties of the exponential function to isolate the node effect term $E\{\exp(\nu)^2\}$. See Appendix \ref{sec: other_graph_models} for a discussion on how to marginalize out the node effects when other link functions are used. We must now estimate (i) the term $p_{k,k}$ and then (ii) the term $\log[E\{\exp(\nu)\}].$ 

Our strategy therefore has two components. The first is to identify some $K \times K$ distance matrix $D$  among $K$ points on
the manifold that could be used to test the geometry and then develop
a consistent estimator of the pair-wise distances between these $K$ points.  By showing that the regularity conditions for the geometric result (Theorem \ref{thm: intro_thm}) are
met under the network formation model (Assumptions \ref{ass:manifold}-\ref{ass:z}), we can   identify the geometry. The second (lesser) component is to recognize and adjust for the fact that
in going from linking probabilities to distance matrix estimation, we need to adjust for the nuisance of the fixed effects parameters.

\subsection{Estimating probability of edges}
We begin by estimating the term $p_{kk'}$. The approach exploits the clique structure in the network. Cliques are useful because under regularity conditions that
are useful in applications, a clique of size $\ell$ tells us that all nodes in that clique are extremely likely to be very close in the latent space. Therefore, we can treat them as if they are all located at the same point in the manifold. Of course, this is not exactly true, but for sufficiently large clique size $\ell$, this approximation strategy becomes more and more accurate. If we find
$K$ disjoint cliques in the graph, then we have identified $K$ distinct points on the latent space. Then, by counting the number of edges between cliques, we can approximate all pairwise distances between these cliques, as we did above in (\ref{eq: hatP_SBM}). So by looking at the clique structure, we can compute an estimate $\hat D$ which approximates distances along the surface of the latent space. We now describe this approach more formally. 

Let $C_1, \dotsc, C_K$ denote $K$ disjoint cliques of size $\ell$. We can take $\ell$ to grow like $\log(n)$, and such cliques will exist with probability tending to 1 in the graph (see Appendix \ref{sec: existence_of_cliques} for a discussion). Let $\zeta_k$ be the Fréchet mean of the locations of the nodes in clique $k$:
\begin{equation*}
\zeta_{k}^\star(\ell) =\underset{\zeta\in\mathcal{M}^{p^{\star}}\left(\kappa^{\star}\right)}{\mathrm{argmin}}\sum_{i \in C_k}^{\ell}d^2_{\mathcal{M}^{p^{\star}}\left(\kappa^{\star}\right)}\left(z_{i}^\star,\zeta\right) \;.    
\end{equation*}
Informally, this point represents the average of all node locations in the latent space. We then define a $K \times K$ distance matrix $D = \{d_{kk'}\}$ with entries
\begin{equation*}
d_{kk'}:=d_{\mathcal{M}^{p^{\star}}\left(\kappa^{\star}\right)}\left(\zeta_{k}^\star(\ell),\zeta_{k'}^\star(\ell)\right) \;.
\end{equation*}
Note that $\zeta^\star$ and $D$ are indexed by the clique size $\ell$ and therefore $n$. Conditioned on seeing cliques of size $\ell$, we show that the terms $\zeta_k^\star(\ell)$ each converge to some fixed point $\zeta_k^\star$ on the latent space under general and relevant conditions (see Assumption \ref{ass:z} and Section \ref{sec: ass_1_3})
. Thus, the matrix containing the pairwise distances is the distance matrix we wish to estimate. Recall that this is the distance term on the left hand side of (\ref{eq: Complex_Model_Solved_1}).

To estimate the probability on the right hand side of (\ref{eq: Complex_Model_Solved_1}), we compute
\begin{equation}
\hat{p}_{kk'}=\frac{1}{\ell^{2}}\sum_{i = 1}^n \sum_{j = 1}^n G_{ij}{\bf 1}\left\{ i\in C_{k},j\in C_{k'}\right\} .\label{eq: cross-clique}
\end{equation}

\subsection{Adjusting for the Expected Fixed Effect}
\label{sec: fixed_effect}
For notational convenience let us denote $\tau := E\{\exp(\nu_i^\star)\}$. The main observation here is that if we  focus on nodes that are close together in the latent space, since the distance term in such cases will be (nearly) zero in~(\ref{eq:main_model}), it we will be able to estimate and account for the individual effects.

Naively, since we use cliques to estimate $p_{k,k'}$, we may consider using these nodes to also estimate $\tau.$ We could, for example, compute the number of edges in the cliques out of the number of possible edges. However, this estimate will be 1, since by definition all edges exist between nodes in the same clique. Therefore, we define a closely related idea, which we call the ``almost-clique."

Fix an $\ell$-clique $C_k$ and a number $t < \ell$. We define an ``almost-clique" $I_k(t)$ by 
\begin{equation*}
    I_k(t) := \big\{j: j \not \in C_k,  |C_k| >  \sum_{i \in C_k} G_{i,j} \geq t\big\}
\end{equation*}
to be the set of nodes not in $C_k$ that connect to at least $t$ nodes in $C_k$. The intuition behind this definition is that if $t\approx \ell$, then the distance between nodes in $I_k(t)$ should be close to zero, but since they are not in the clique not all connections will be realized.

We can estimate the probability that nodes in the sub-graph induced by $I_k(t)$ connect, 
\begin{equation*}
    \hat E(t, k) = {|I_k(t)| \choose 2}^{-1}\sum_{(i, j) \in I_k(t)\times I_k(t)}G_{i,j} \;.
\end{equation*}
To estimate $E(\exp(\nu))$, we average the above term over all cliques, leading to an estimate 
\begin{equation*}
    \hat E(\exp(\nu)^2) := \frac{1}{K}\sum_{k =1}^K \hat E_\nu(t, k) \;.
\end{equation*}
This approach suffers from selection bias when the clique size is large.  That is, by Assumption~\ref{ass:nu} all individuals have independent and identically distributed $\nu_i^\star$ terms.  Conditional on being part of a large clique, however, an individual is likely to be on the right tail of the $v_i$ distribution.  We could adjust for this bias by, for example, assuming a parametric model for $\nu_i$.  If we made such an assumption we could compute a correction for the selection bias based on the tail of the assumed distribution.  In practice, we found our non-parametric estimator worked sufficiently well without such a correction. We suggest taking $t$ to be large, for example $t = \ell - 1$ because our simulations suggest that large values of $t$ reduce the selection bias and therefore increases the accuracy of our method.

\subsection{Examples Satisfying Assumption \ref{ass:z}} \label{sec: ass_1_3}
{

Part (a) of Assumption \ref{ass:z} is innocuous. It simply requires that   $K$ points identify the manifold. Note that the existence of $K$ points out of $n \rightarrow \infty$ random points with continuous distribution over any ball on the manifold will 
identify it outside a measure zero event in the usual measure. To see this, it is easy to observe that with continuous mass on any patch of a sphere, $K$ points will exist that are not simply forming an arc or are arbitrarily local with probability tending to one.

Part (b) of Assumption \ref{ass:z} requires  
that for sufficiently large $
 \ell$, nodes in a clique are at approximately the same location on the latent surface, which allows us to conclude consistency and asymptotic normality of our distance estimates.  
{We provide a general set of assumptions that are sufficient 
 and show several   models used in applied work are covered. 
\begin{ass}\label{ass: general_support}
Let $\Omega_{n}=\left[0,A_{n}^{1/p^{\star}}\right]^{p^{\star}}\subset\mathcal{M}^{p^{\star}}\left(\kappa^{\star}\right)$
be such that $A_{n}=o\left(n\right)$ is growing. Assume either 
\begin{enumerate}
\item \emph{Bounded Support:} $\Omega_{n}$ is the support of the distribution $F_{n}\left(z\right)$
of locations, or
\item \emph{Thin Tails:} $F_{n}\left(z\right)$ satisfies, for some constant $b>0$,
\[
\Pr\left(\max_{i, j}d_{\mathcal{M}^{p^{\star}}\left(\kappa^{\star}\right)}\left(z_{i},z_{j}\right)>A_{n}^{1/p^{\star}}\right)\leq\exp\left(-bA_{n}^{1/p^{\star}}+\log n\right)\rightarrow0.
\]
\end{enumerate}
\end{ass}
Obviously (1) implies (2), but (2) allows for sub-Gaussian tails with
unbounded support. 
The next assumption is 
natural and general, holding  as long as distributions are not
taking mass only on some sub-manifolds. It says that the odds that
all $\ell$ independent draws are within $\delta$-distance of each
other is inversely related to the volume of that ball. We verify this for examples below.

For any set of $\ell$ points in the latent space, define the event 
\[\mathcal{E}_{\delta}:=\left\{ \max_{1\leq i<j\leq\ell}d_{\mathcal{M}^{p^{\star}}\left(\kappa^{\star}\right)}\left(z_{i},z_{j}\right)<\delta\right\}.\]
In words, $\mathcal{E}_\delta$ is the event that the largest distance between these $\ell$ points is less than $\delta$. We omit the dependence on $\ell$ when writing $E_\delta$ for convenience.

\begin{ass} \label{ass: E_delta}
The location distribution sequence satisfies, for any $\delta$
and $\ell$, 
\[
\Pr\left(\mathcal{E}_{\delta}\right)=\frac{a(\delta)}{A_{n}^{\ell}}\left(1+o\left(1\right)\right).
\]
for a positive constant $a(\delta)$, which can depend on $\delta$.
\end{ass}
}

Let $\mu_{d,n}:=E\left\{ d_{\mathcal{M}^{p^{\star}}\left(\kappa^{\star}\right)}\left(z_{i},z_{j}\right)\right\}$, where $z_i$ is independent of $z_j$, be the expected distance between two location draws.
\begin{ass} \label{ass: clique_rate}
For any $\delta>0$, assume a sequence of clique sizes $\ell_{n}\rightarrow\infty$
satisfying
\[
\ell_{n}\geq2\frac{\log A_{n}}{a(\delta)(\mu_{d,n}-\delta)}+1
\]
 for all $n$ sufficiently large.
\end{ass}
The above assumption is written for any $\delta > 0$, but in fact we are only concerned with $\delta < \mu_{d, n}$, the average distance between two arbitrary points drawn on the latent surface. For $\delta > \mu_{d, n}$, it is unlikely that two nodes that are at least $\mu_{d,n}$ apart would form an edge. Therefore, we are only interested in checking the condition for sufficiently small $\delta$. 

We show that Assumptions \ref{ass:z} hold under the above assumptions. 
\begin{prop} \label{prop: general_clique}
Let Assumptions \ref{ass:manifold}, \ref{ass:nu}, \ref{ass: general_support}, \ref{ass: E_delta}, and \ref{ass: clique_rate} hold. Then Assumptions \ref{ass:z} holds.
\end{prop}

We now consider three common ways of modeling node locations in latent space.   For each model, we 
 show that Assumptions \ref{ass:manifold}, \ref{ass:nu}, \ref{ass: general_support}, and  \ref{ass: E_delta} hold.

These three examples cover a wide span of location distribution models.
The first, a lattice, is a stylized example that simply corresponds to a community
block model with a geometric structure, since all nodes exactly live
at one of 
several locations. The second, a uniform distribution, is the other extreme 
where nodes have no bias towards any specific locations that help
organize cliques. This is the most adversarial case. The Gaussian Mixture Model lives in-between and
interpolates between these models and is frequently used in practical statistical modeling. It functions much like the uniform if the dispersion of nodes about their type-centers tends to infinity 
and is similar to the lattice if the dispersion tends to zero.

\begin{example}[Lattice]
\label{ex: lattice}
Every node is a member of some community $t_{i}\in\left\{ 1,\ldots,T_{n}\right\} $.
Node locations on the manifold are determined as follows. Let a lattice
$\Lambda_{n}={\{ 0,\ldots,T_{n}^{1/p^{\star}}\}} ^{p^{\star}}\subset\Omega_{n}$
be a list of coordinates which serve as the support for node placements
with a spacing of 1 between points along any axis. Node locations
$z_{i}$ are placed i.i.d. across these $T_{n}$ points on the
manifold, i.e., on the lattice $\Lambda_{n}$, so Assumption \ref{ass: general_support} holds
and $t_{i}$ corresponds to the location drawn. Assumption \ref{ass: E_delta} holds
for any $\delta<1$, with $A_{n}=T_{n}$, since every location $t\in\left\{ 0,\ldots,T_{n}\right\} $
is equally likely:
\(
\Pr\left(\mathcal{E}_{\delta}\right)=\left({1}/{T_{n}}\right)^{\ell}.
\)
\end{example}

To understand the growth-rate restrictions on $\ell(n)$, 
recall that $\ell(n)$ has to grow slowly enough so that there are cliques of size $\ell$ with probability 1. Consider the lattice above. Let us assume $A_n$ is fixed at $A > 0$. Then, at each location in the lattice, the nodes connect independently with some fixed probability determined by the fixed effects, so these nodes are in an Erdos-Renyi model. {So we need $\ell(n) < \log(n)$} in order for there to be an $\ell$-clique in the graph with probability approaching one (recall Appendix \ref{sec: existence_of_cliques}). And according to Assumption \ref{ass: clique_rate}, we need $\ell > \log(A_n) = \log(A)$, a constant. So taking $\ell(n) \rightarrow \infty$ to be growing slower than $\log(n)$ 
satisfies Assumption \ref{ass: clique_rate} for sufficiently large $n$.  

The next example is in some sense the most adversarial model for our method. By placing nodes uniformly over the support, the odds of accumulation, and therefore clique formation, at any given point are minimized. Still, the result holds.

\begin{example}[Uniform Distribution]
Node locations $z_{i}\sim U\left(\Omega_{n}\right)$ are assumed
to be drawn independently, uniformly over $\Omega_{n}$ such that Assumption \ref{ass: general_support} holds. It immediately follows by a calculation that
\(
\Pr\left(\mathcal{E}_{\delta}\right)=\left({\delta^{p^{\star}}}/{\text{volume}(\Omega_n)}\right)^{\ell},
\)
so Assumption \ref{ass: E_delta} holds as well.
\end{example}

The final example interpolates between the extremes of the lattice and uniform models.
\begin{example}[Gaussian Mixture Model]\label{ex: GMM}
 As in the lattice example, every node is a member of some community
$t_{i}\in\left\{ 1,\ldots,T_{n}\right\} $. Node locations on the
manifold are determined as follows. Let $\Lambda_{n}\subset\Omega_{n}$
be a lattice of $T_{n}$ points, which designate community centers,
randomly drawn as follows. There is a support $\Omega_{n}'\subset\Omega_{n}$,
with $\Omega_{n}'=\left[0,B_{n}^{1/p^{\star}}\right]^{p^{\star}}$
and the community centers $\zeta_{t}\sim U\left(\Omega_{n}'\right)$
are drawn independently, uniformly over this support. 

Given the community centers, every node conditional on the community
it is assigned to has a location that is dispersed about the center
\[
z_{i}\vert t_{i}=t\sim F_{n}\left(z;\zeta_{t},\sigma_{t}^{2}\right)
\]
independently. Examples include Gaussian in Euclidean space, the von Mises-Fisher
on the sphere, and the wrapped-normal distribution on hyperbolic space. Note that community centers reside within $\Omega_{n}'\subset\Omega_{n}$
but given the distribution $F_{n}$, the random variable $z_{i}$
may have full support over $\mathcal{M}^{p^{\star}}\left(\kappa^{\star}\right)$. 

The distance between the outer boundary of $\Omega_{n}'$ and $\Omega$,
denoted by $\Delta_{n}:=A_{n}^{1/p^{\star}}-B_{n}^{1/p^{\star}}$
is assumed to be growing at a sufficiently fast, possibly sub-logarithmic
rate $\Delta_{n}=\omega\left(\sqrt{\log n}\right)$. Then one can
calculate that Condition 2 of Assumption \ref{ass: general_support}. The function
of this is to ensure that even if the $z_{i}$ have full support,
they are going to essentially all live within $\Omega_{n}$.

In this setup, we can calculate that 
\[
\Pr\left(\mathcal{E}_{\delta}\right)=\left(\frac{\delta^{p^{\star}}}{A_{n}}\right)^{\ell}\times\left(1+o\left(1\right)\right).
\]
so that Assumption \ref{ass: E_delta} holds too. Complete calculations are in the Appendix \ref{sec:proofs}.
\end{example}

\subsection{Consistent Estimates of Location and Fixed Effects}
Suppose that the researcher has access to estimates of the node locations fixed effects $(\hat z_i, \hat \nu_i)$. Our result is agnostic to how these estimators are constructed and allow any consistent ones from the literature.
\begin{corr}
\label{corr: consistent_paramters_cliques}
Let $\hat{\mathcal{M}}^{\hat p}(\hat \kappa)$ denote the estimate of the geometry from Algorithm \ref{alg: cliques}.
Let $\hat z_i(\mathcal{M}^{p^\star}(\kappa^\star))$ and $\nu_i(\mathcal{M}^{p^\star}(\kappa^\star))$ be any set of consistent estimators, computed using the assumed geometry $\mathcal{M}^{p^\star}(  \kappa^\star)$.
Then, $\hat z_i(\hat{\mathcal{M}}^{\hat p}(\hat \kappa))$ and $\hat \nu_i(\hat{\mathcal{M}}^{\hat p}(\hat \kappa))$ are consistent. 
\end{corr}
The proof is straightforward consequence of Theorem \ref{thm: main_cliques}, 
so we omit it. The mode of convergence in Corollary \ref{corr: consistent_paramters_cliques} depends on how the estimates of node locations and effects converge. For example, if conditioned on the right geometry, $\max_{1 \leq i \leq n} |\hat \nu_i - \nu_i^\star| \overset{p}{\rightarrow} 0$, then this same convergence holds in Corollary \ref{corr: consistent_paramters_cliques}.
}

\subsection{Practical Implementation of Manifold Hypothesis Tests via Bootstrap} Following Proposition \ref{prop: ordered_test}, 
any arbitrary implementation of hypothesis testing for \eqref{eq: testE}, \eqref{eq: testS}, and \eqref{eq: testH} 
would suffice. For instance, following the logic of Weyl's inequality and Proposition \ref{prop:alpha_bound}, since our estimate of $W$ is asymptotically normally distributed, one can readily develop an analytic conservative test. 
Since Weyl's inequality is often a loose upper bound, we suggest take a different approach and provide a specific method which is extremely fast, easy-to-implement, and effective in simulations and data work via a sub-sample  bootstrap.

 There are two non-standard features of our problem that make classical
bootstrapping challenging. First, $W_{0}$ does not have full rank,
meaning that $\lambda_{K}\left(W_{0}\right)\leq0$; under the null
$\lambda_{K}\left(W_{0}\right)=0$ meaning the parameter is on the
boundary of the parameter space. Classical bootstrap is not valid
in such a case \citep{Andrews}. Second, $W_{\kappa}$ has repeated
eigenvalues at zero under the null for both curved spaces, which again
excludes the classical bootstrap \citep{Eaton}.

We adapt the sub-sampling method from \cite{Politis} which
is valid both with parameters on the boundary and with repeated eigenvalues. Algorithm \ref{alg: subsample} in the appendix presents the method. It uses sub-sampling to generate a distribution $\left\{ D_{b}^{\star}\right\} _{b=1}^{B}$
of $B$ bootstraps of the $K\times K$ distance matrices. Then given
this distribution, the method constructs the corresponding distribution
of the eigenvalue of interest, $\lambda_{\tilde{k}}\{W_{\kappa}(D_{b}^{\star})\}$,
which is then used to test the null hypothesis for the original data. 
In Appendix \ref{sec: CM_geometry}, we investigate another way of testing geometry via the Cayley-Menger determinant. We show numerically that this procedure has lower power than the bootstrap procedure discussed above, so in the following simulations we use the bootstrap procedure.

\section{Simulation Evaluation}\label{sec: Simulations}
In this section, we examine the performance of our proposed method on simulated data from each of the three candidate geometries.  The goal is to understand how well the methods perform in a setting where we know the (simulated) true geometry.  We first examine the Type 1 error and power of the tests to select manifold class and then show the performance of our algorithm for estimating the latent dimension.  We provide additional simulation results, including results for estimating curvature, in Appendix~\ref{app:curve}.

We discuss the Type 1 error and power of our proposed tests under various values for the clique size ($\ell$) and the number of cliques we select for our estimation ($K$).  In all cases, we simulate graphs in the following way.  First, we generate a set of groups centers randomly in the latent geometry and dimension to be tested. For the type 1 error, we generate graphs on $n = 200$ nodes and use $K = 5$ and clique sizes that are realistic in the data: $\ell \in \{4, 5, 6\}$. When generating the type 1 error figures, the graph statistics are as follows. For the Euclidean graphs, the average degree is 41 and the average clustering coefficient is 0.26.  For the spherical graphs, the average degree is 56 and the average clustering coefficient is 0.35. For the hyperbolic graphs, the average degree is 40 and the average clustering coefficient is 0.26.

For the power simulations, we generated 25 sets of latent positions and then, for each set of latent positions, constructed 100 graphs.  When comparing across values of $K$ and $\ell$, we use the same graphs for all comparisons (e.g., when comparing $K=10$ vs $K=5$, the cliques in the $K=5$ set are randomly selected from among those in $K=10$). We provide specific values we used for simulations and additional results in Appendix~\ref{sec: generatePoints}.

\begin{figure}
    \centering
    \subfloat[]{{\includegraphics[scale = 0.35]{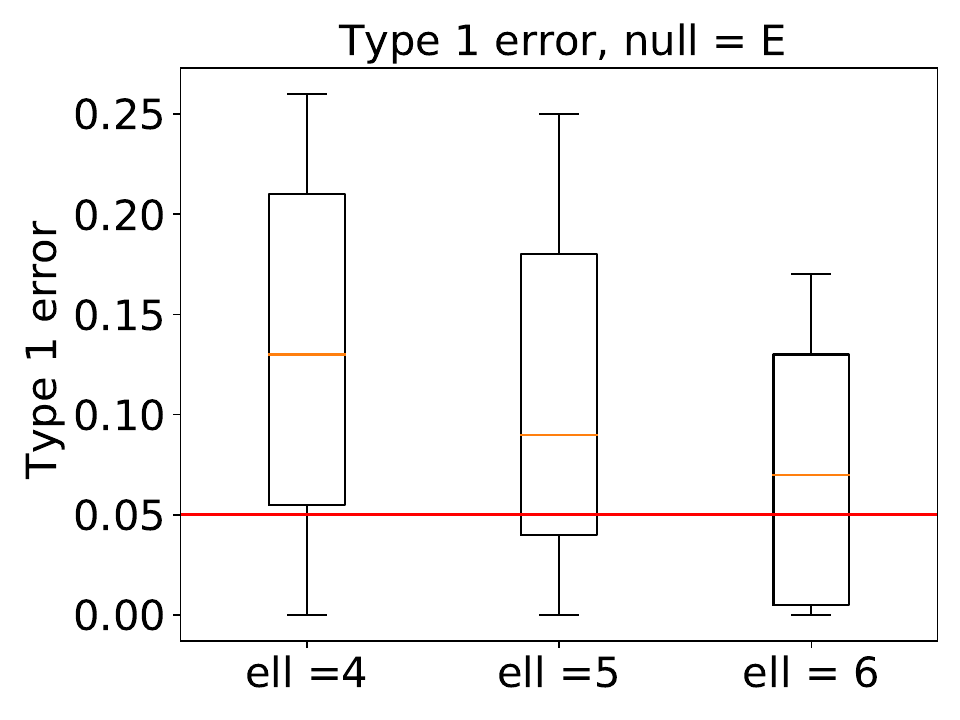} }}
    \subfloat[]{{\includegraphics[scale = 0.35]{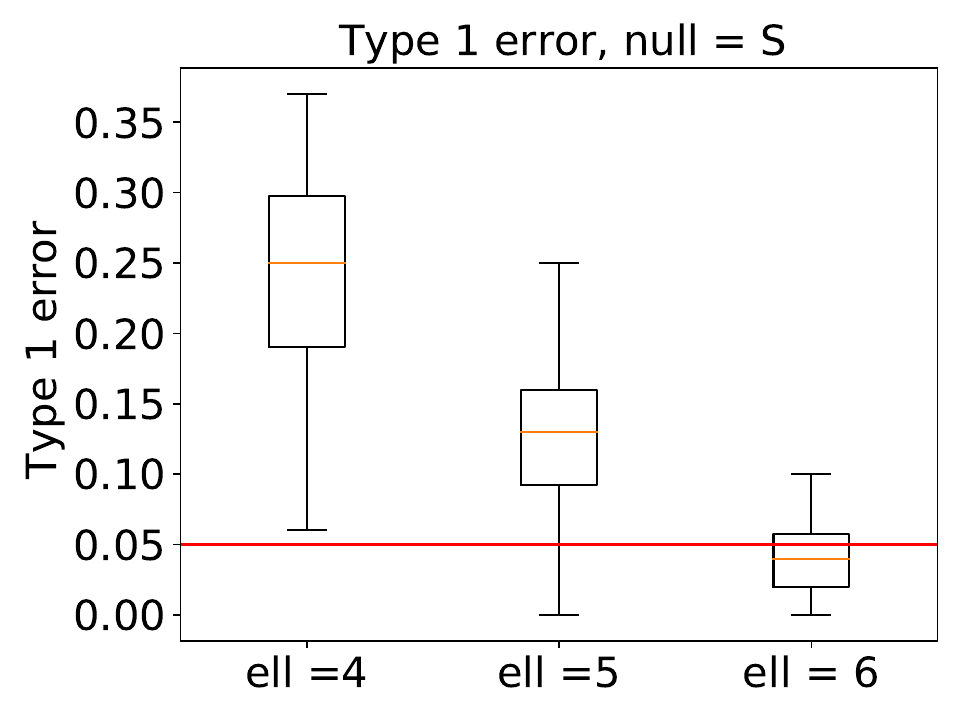} }}
    \subfloat[]{{\includegraphics[scale = 0.35]{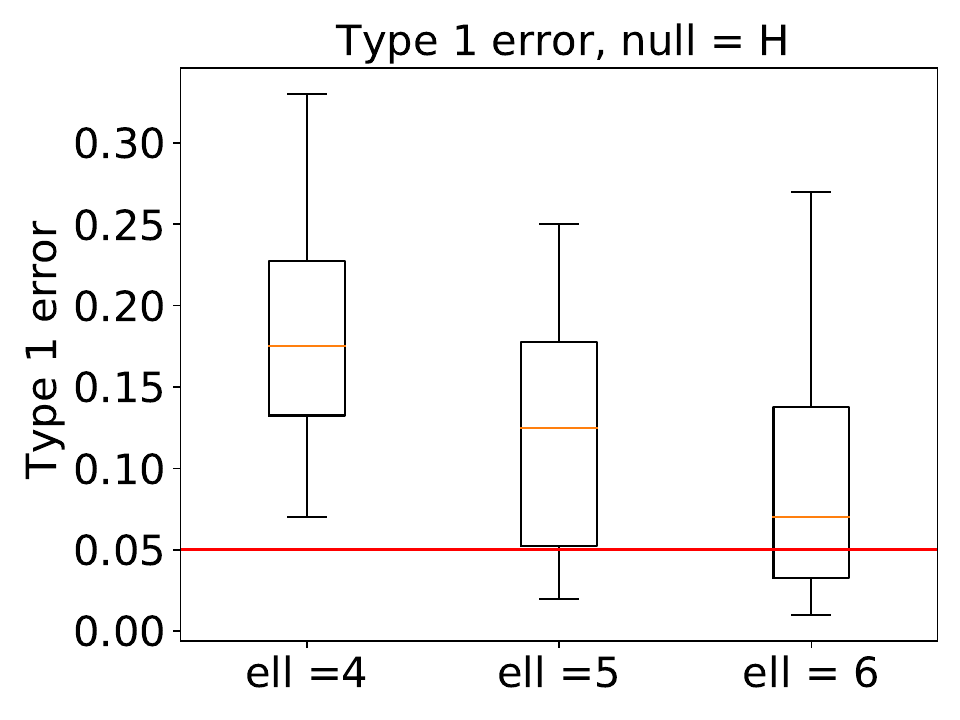} }}%
    \caption{\footnotesize{Estimated type 1 error. For each set of LS positions, we perform the test 100 times and plot the average rejection probability.}}%
    \label{fig: type1_error}%
\end{figure}

\begin{figure}\centering
\subfloat[Euclidean null]{\label{a}\includegraphics[scale = 0.35]{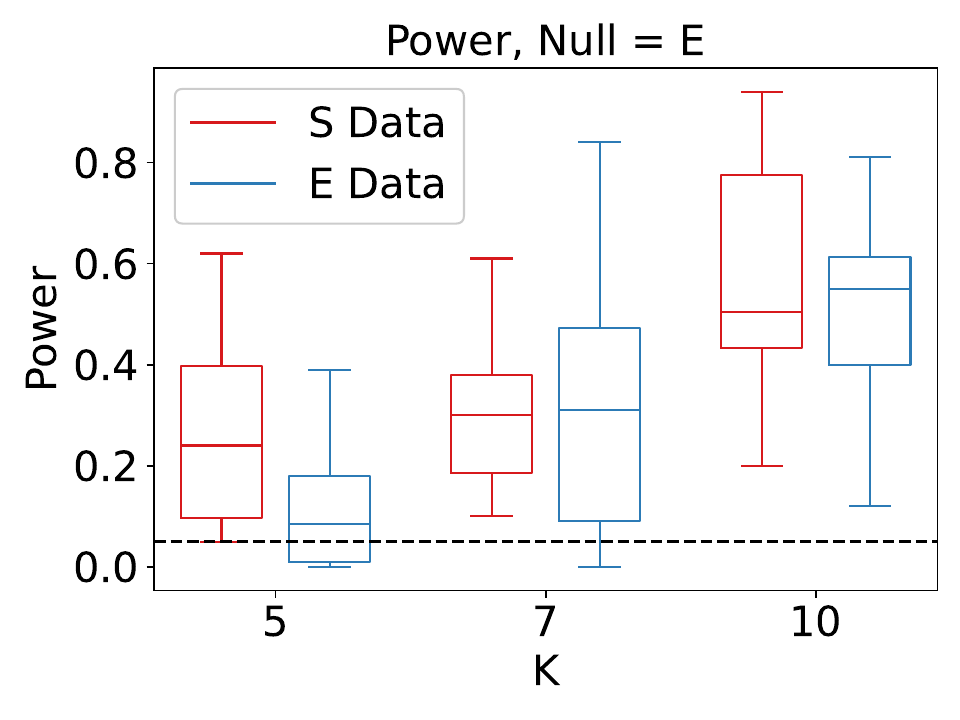}}
\subfloat[Spherical null]{\label{b}\includegraphics[scale = 0.35]{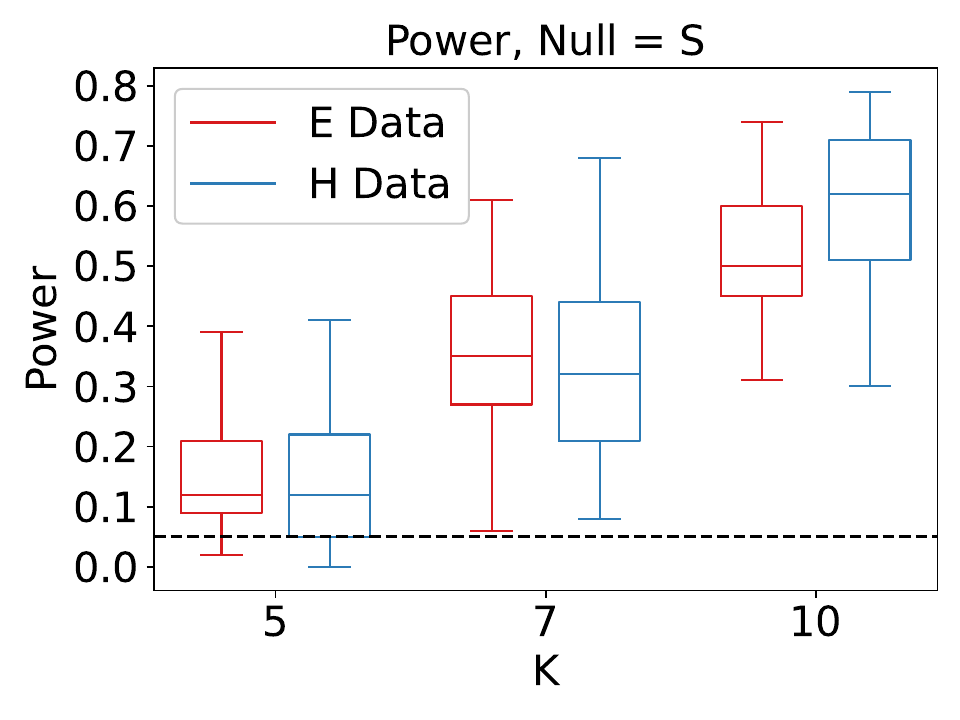}}
\subfloat[Hyperbolic null]{\label{c}\includegraphics[scale = 0.35]{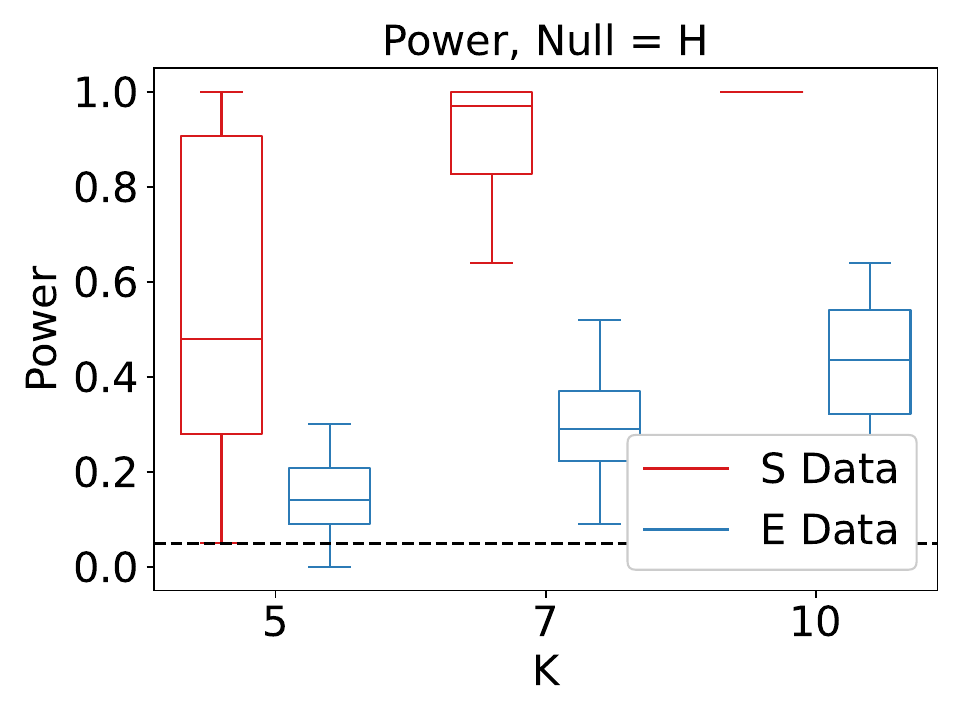}}
\caption{\footnotesize{Estimated power using simulated LS positions. For each set of LS positions, we perform the test 100 times and plot the average rejection probability. }}
\label{fig:poweracrossK}
\label{fig}
\end{figure}

Figures~\ref{fig: type1_error} and~\ref{fig:poweracrossK} show results for Type 1 error and power of the tests we propose using the simulation procedure described above.  Each point in the boxplot is the fraction of rejections out of 250 graphs for a given set of latent space positions.  The variation in the boxplot, therefore, represents heterogeneity across latent space locations that are consistent with the true underlying geometry and the simulation procedure we use.  Figure~\ref{fig: type1_error} shows boxplots of the Type 1 error for each of the three null hypotheses for three values of $\ell$.  We focus on variation in the Type 1 error across values of $\ell$ to see whether the properties of the~\citet{Politis} bootstrap procedure are preserved empirically.  We see that, in all three cases, the Type 1 error decreases as the clique size increases.  Further, for the Euclidean and hyperbolic cases the Type 1 error tends to be below the nominal level of five percent, but the spherical type 1 error is higher than five percent. In Figure \ref{fig:poweracrossK} we see that the power increases as we increase $K$ for all three geometries. In these simulations we use $\ell = 5.$ Recall that all of our manifolds are locally Euclidean---indeed that is part of their definition. So, it is unsurprising, if not expected, that power against Euclidean alternatives rises more slowly than power against alternatives of the opposite curvature. Appendix \ref{sec: lattice_sims} contains more simulations.

Moving now to the estimates of the minimal dimension, we consider $p \geq 2$ and take $\max(2, \hat p)$ as our estimate of the dimension of $\mathcal{M}^{p^\star}(\kappa^\star).$ In Table \ref{table: rank} we give our estimates of the dimension for the three geometries.

\begin{table} \centering 
  \caption{\footnotesize{Average probability of correctly predicting the dimension of the latent space, averaged across 25 different sets of $n = 200$ latent space positions. We use $K = 7$ and $\ell = 4$. For each set of latent space positions, we generate 50 networks and predict the dimension. }}
  \label{} 
\begin{tabular}{@{\extracolsep{5pt}}lccc} 

 & \multicolumn{3}{c}{\textit{True Geometry}} \\ 
\cline{2-4} 
\\[-1.8ex] & $\mathbb{R}^4$ & $\mathbf{S}^3(1) $ & $\mathbf{H}^3(-1)$ \\ 
\hline \\[-1.8ex] 
$\Prob(\hat p < p^\star)$ &  0.02 & 0.03 & 0.06 \\ 
$\Prob(\hat p \geq p^\star)$ & 0.98 & 0.97 & 0.94
\end{tabular} 
\label{table: rank}
\end{table}

\section{Examples from Economics and Biology}\label{sec: Empirics}

In this section we demonstrate the performance of our method settings with the complexity of observed data.  We demonstrate that, in  vastly distinct contexts, our approach captures features of the underlying geometry that provide contextually salient insights.  We begin by offering guidance on choices a practitioner would make when implementing the method, then provide examples from   three  
contexts. Access to data from~\citet{banerjeegossip} is restricted, however, a similar dataset is freely available here: \url{https://doi.org/10.7910/DVN/U3BIHX}. Data from our neural network example is available here: \url{https://www.dynamic-connectome.org/?page_id=25}. Replication code is available here: \url{https://zenodo.org/record/7474776#.Y6TryS-B2fU}.

\subsection{Choices for Implementation of Algorithm \ref{alg: cliques}}
A key decision for implementation is how to identify and then select cliques for a given graph.  Overall, there are several 
considerations. 

\subsubsection{Choosing Clique Size}
Algorithm \ref{alg: cliques} requires identifying $K$ cliques of size $\ell$ (again recalling that we only assume for simplicity that they are of the same size $\ell$), with $\ell$ growing slowly. $K$ is assumed to not grow with $n$, although  
more points will raise the power of our tests. This means that the researcher \emph{does not} need to identify all $\ell$-cliques in the graph, only $K$ of them which will typically be a small number (e.g., 8-12) and are easily able to be found in our empirical applications, which our simulations show leads to high power. In Appendix \ref{sec: existence_of_cliques} we show that taking $\ell < \log(n)$ is often sufficiently slow to guarantee that cliques of size $\ell$ exist in the network. Of course, one can take $\ell$ to grow slower than that, but a higher $\ell$ results in a lower Type 1 error, since larger cliques gives better estimates of the distances in the latent space. 
First, we would like to take the number of cliques $K$ and the size of the cliques $\ell$ of cliques to both be as large as possible.  In practice, we use the networkX command \texttt{enumerate\_all\_cliques} in Python, which uses a clique finding algorithm from \cite{Zhang2005}. While identifying numerous cliques can be challenging,
given the modest size of the cliques and that we only need a small,
fixed number of cliques, applied researchers can easily implement
our technique.

\subsubsection{Choosing Cliques}
As $\ell$ increases, the variance of estimates of $\hat D$ decreases, and the power of the test increases as $K$ increases, since we have more distances between points on the manifold. Figure \ref{fig:poweracrossK} from our simulations shows that  as $K$ increases, the power of our tests increase.  Second, we need cliques that are well-separated on the manifold, but connected in the graph.  Since we use cliques as ``points'' on the manifold to measure distance, the cliques should ideally not have nodes in common, since if two cliques do overlap, its possible these two ``points" on the manifold are close together or even the same point.
Third, if two cliques have no edges between them, then our estimate of the distance between the two points is $+\infty$, which contains no information about the geometry. 

Motivated by these three considerations, 
 our goal is to solve
\begin{equation}
\begin{aligned}
 \hat C_1, \dotsc, \hat C_K \in \ &\underset{C_1, \dotsc, C_K}{\text{argmin}} \sum_{i, j}^K |C_i \cap C_j|\\
\textrm{such that} \quad & |C_i| = \ell \text{ for each } i \text{ and }   \hat P(C_1, \dotsc, C_K) \text{ does not contains a 0. }
\end{aligned}
\label{eq: PickingCliques}
\end{equation}

In practice, we set $K$ and $\ell$ by first looking at the number of cliques of various sizes in the graph and choosing and $\ell$ that is close to the size of the largest cliques in the graph, but where there are still enough cliques of that size to find $K$ and are well-separated.  We then take random draws from the (very large) set of possible cliques and evaluate the objective function in (\ref{eq: PickingCliques}).  Searching over the set of possible cliques is a well-studied (NP-hard) problem in computer science and graph theory, however, we found that our relatively simple approach yielded high quality cliques after around $10^6$ draws from the clique distribution.  We evaluate the quality of the cliques we select by running the optimization independently several times.  A stable objective function value across the runs indicates high quality cliques. In the data from~\citet{banerjeegossip}, we take $K$ as either 7 or 10. The value we choose is based on how easy it is to find appropriate cliques in a given network using the problem formulation in (\ref{eq: PickingCliques}).  In the Indian village sample, the average number of cliques of size $c_i - 1$ is 80, where $c_i$ is the size of the largest clique in network $i$, known as the clique number.  It takes on average 0.005 seconds to find all cliques of size $c_i - 1$ over the 75 networks. It takes 0.004 seconds in the \emph{C. elegans} sample to find the 29 cliques of size 5 in the network.

To select $\ell$, we use the size of the largest clique found in the graph minus one. In most of the villages, choosing $\ell$ in this way resulted in dozens of possible cliques to choose from.  We present more details about cliques in the~\citet{banerjeegossip} data in Appendix~\ref{app:data}. For the \emph{C. elegans} data, we select $K = 12$ and set $\ell = 5$, which is the size of the largest clique in the graph. We reiterate that for our approach we need only $K$ cliques and do not need to enumerate all $\ell$ cliques in the graph.

\subsection{Village Risk-sharing Networks and the Introduction of Microfinance} 
\label{sec: IndianData}

We begin by studying the underlying geometries of Indian village networks. We use the Wave II village network data of \citet{banerjeegossip}, in part collected by one of the authors of the present paper. This  consists of a collection of graphs for each of 75 villages in Karnataka, India constructed by surveying 89\% of all households in each village, thereby generating a 99\% edge sample for the resulting undirected graph. There are a total of 16,451 households in the sample. In every village we have relationship data between households on each of 12 dimensions: 5 social dimensions, 4 financial dimensions, and 3 information sharing dimensions. See \citet{banerjeegossip} for more details including descriptive statistics. The links across these dimensions line up for the most part, consistent with a theory of multiplexed incentives to form links, so we study the undirected, unweighted graph following the prior literature using this data \citep{jacksonl2013,banerjeecdj2013,SavingsMonitors,banerjeegossip}.

The social networks literature has long been interested in excess closure  \citet{coleman1988}. Friends of friends tend to be friends more than one might expect and this is particularly true if network relationships substitute for formal institutions. A literature focusing on equilibrium informal financial networks, which facilities the sharing of risk between households in a village, describes why the equilibrium network shapes exhibit excess closure (e.g., \citet{ambrusms2012,jackson2013}). The 
idea is that in order to maintain cooperation, when individuals can renege on their promises to aid each other in times of need, it is useful to have friends in common to amplify punishment, thereby maintaining a good  equilibrium.

From the perspective of a latent space model, this means that we might expect excess closure in the village. There are incentives by households to ``curve'' the space, so friends of friends and so on are much more likely to themselves link, discussed in greater depth below. A natural hypothesis, therefore, is that village networks for the most part not be hyperbolic. Rather, they may be more likely to be spherical or, perhaps, Euclidean.\footnote{We briefly note that common modeling assumptions in the socio-economic literature imply constant curvature from the perspective of our model \eqref{eq:main_model}, though certainly there are perspectives that would violate constant curvature which would require future work. To see this, consider two examples. First, imagine a model in which nodes have some random locations. They can choose their efforts to link and the value of their links depends on the number of their friends who are themselves friends in expectation. There is a parameter that governs the value of closure among one's friends which can be positive, zero, or negative, which may depend on the socio-economic context. In such a model, this parameter exactly maps to curvature. Second, one can imagine a model in which agents can take an action to influence the extent to which their neighbors know each other. For instance, the action could be imagined as throwing parties (selecting positive curvature) or the opposite and ensuring ``worlds do not collide'' (selecting negative curvature) \citep{thepoolguy}. In such a model, if we study the symmetric equilibrium, then the equilibrium choice of the extent of forced socialization or barred socialization among one's friend exactly maps to constant curvature. Both of these examples also illustrate the limitations of such models. While these examples demonstrate how conventional assumptions map to constant curvature,   certainly more complex models with heterogeneity would require modeling manifolds with non-constant curvature, which we leave to future work.}

\begin{figure}
    \subfloat[Geometry Estimates]{
      \begin{tabular}{ccccc}
    {Geometry} & $\mathbb{R}^p$ & $\mathbf{S}^p$ & $\mathbf{H}^p$ & N/A\\
    \toprule
\text{$\%$ Classification} & 26\% & 36\%  & 12\% & 25\% %
\end{tabular}
}
\subfloat[Predicted Dimensions]{
    \includegraphics[scale = 0.37, trim = 0cm 0.5cm 0cm 0cm, clip = true]{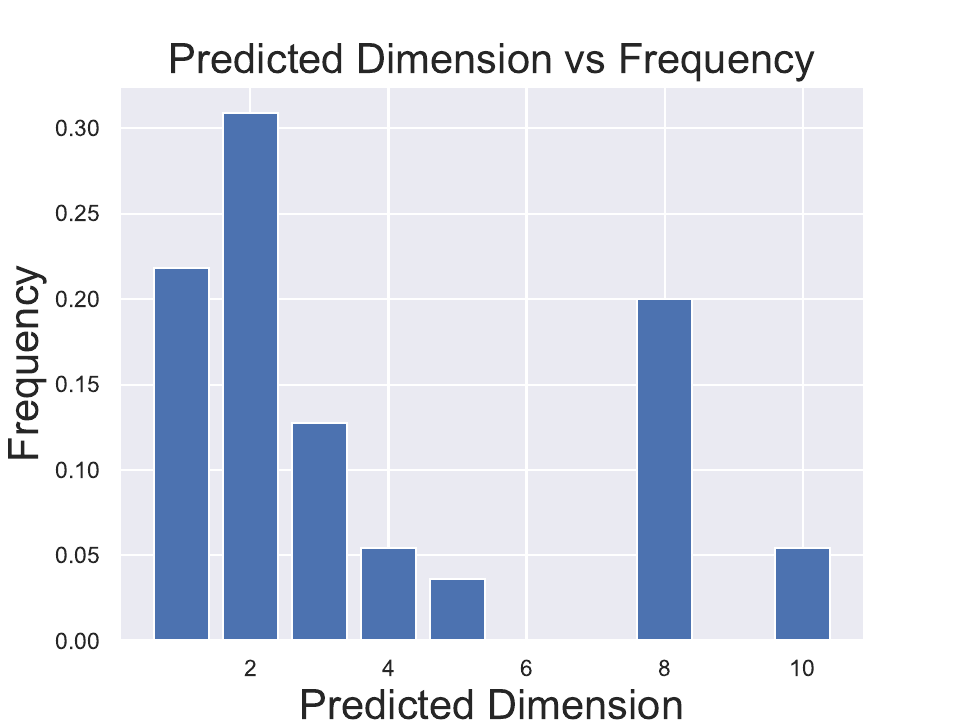}
    }
    \caption{\footnotesize{Predicted geometries and dimensions for the \cite{banerjeegossip} village networks.}    }
    \label{fig:India_dimension}
\end{figure}

Our proposed method gives hypothesis tests (and corresponding $p$-values) for each of the three candidate geometries.  As a descriptive summary, we ``classify'' each of the villages into one of the geometry types using the following procedure.  For villages where at least one village has a $p$-value over .05, we consider, for the purposes of summarizing our results, the manifold type that has the largest $p$-value.  If all three geometries reject the null at the .05 level, then we say that the village cannot be classified.  This outcome could mean a number of things, ranging from a false-rejection by chance to a village whose underlying geometry is not captured by one of the three candidates (e.g.,  curvature may be nonconstant). Figure \ref{fig:India_dimension}, Panel A 
presents classification results for the 75 villages using this descriptive approach. We see that we are able to classify 75\% of the villages, despite the fact that N/A was a possibility. Classification was not forced. Further, the results are consistent with the socio-economic hypothesis on villages needing closure. 48\% of the classified networks are spherical, 35\% are  Euclidean, and only 16\% are hyperbolic.  

We also examine the estimated dimension of the latent space.  Figure \ref{fig:India_dimension} presents the estimated dimensions, which irrespective of curvature is important to know the minimal dimension of the space required to model location decisions by agents.

We now explore the relationship between the latent geometry and socio-economic phenomena. 
This is an observational, not causal, analysis. First, we look at how the volume of informal financial transactions vary with network geometry. Specifically, we are interested in how the volume of informal loans that a household has with network neighbors (e.g., friends or members of their rotating, savings, and credit associations) varies with geometry. Both the theoretical and empirical economic literatures suggest that it is ex ante ambiguous as to the relationship between the amount of network financial flows and curvature. For example, \cite{kinnan2012kinship} study how informal financial flows efficiently allocate credit to households that experience negative shocks in the network. Theory suggests that such flows are more efficient in more expansive networks, which require negative curvature \citep{ambrusms2012}. At the same time, as discussed above, the ability to facilitate informal financial transactions may increase in the importance of closure, and therefore require positive curvature. Which force dominates is an empirical question.

To study this, we estimate the following regression:
\begin{align*}
    \text{Network Loan Amount}_i = \alpha + \beta_E {\bf 1}\{ \hat{\mathcal{M}}^{\hat p}_i = E \} + \beta_H {\bf 1}\{ \hat{\mathcal{M}}^{\hat p}_i = \mathbf{H} \} + \beta_N  {\bf 1} \{\text{N/A}_i\} + \epsilon_i
\end{align*}
where $i$ indexes the village. $\text{Network Loan Amount}_i$ is the average volume of loans from either friends or rotating savings and credit association members that a household has in the village. The loan amount is presented in INR  (USD 1 $\approx$ INR 73.5). Here, the omitted category ($\alpha$) corresponds to the loan amount for a sphere.

\begin{figure}[t]
    \centering
    \includegraphics[angle=0,origin=c,width=.24\textwidth]{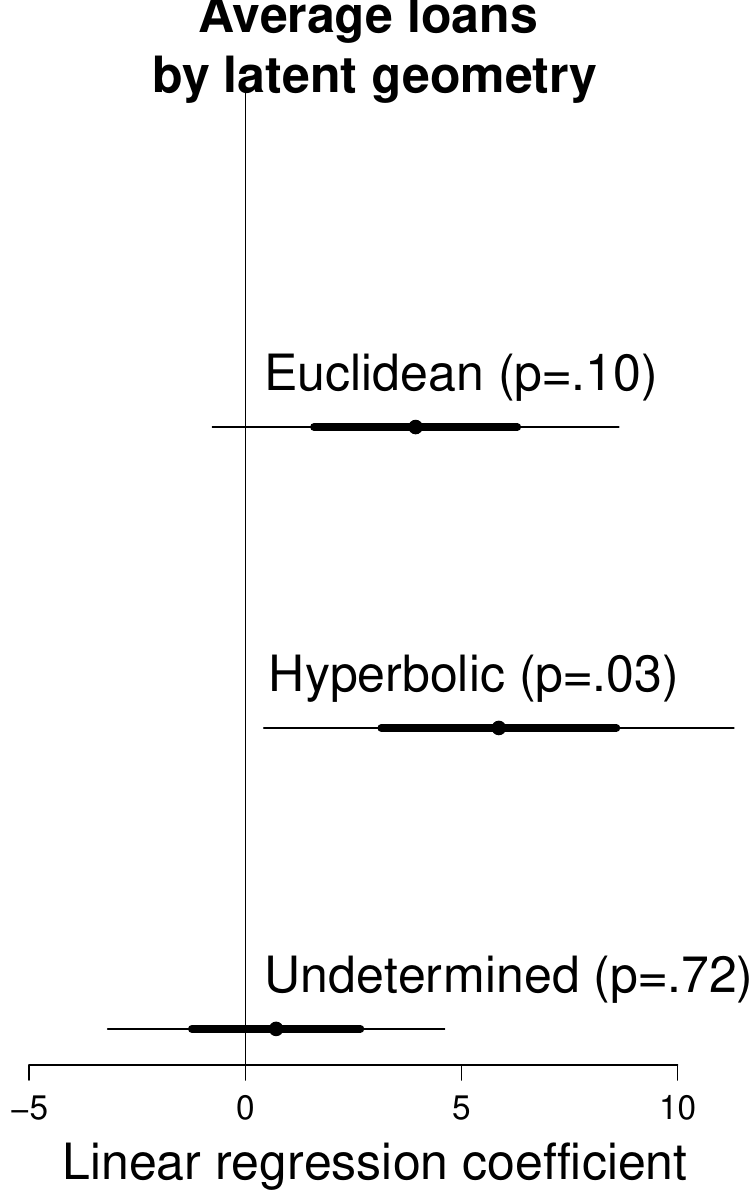}
    \includegraphics[angle=0,origin=c,width=.24\textwidth]{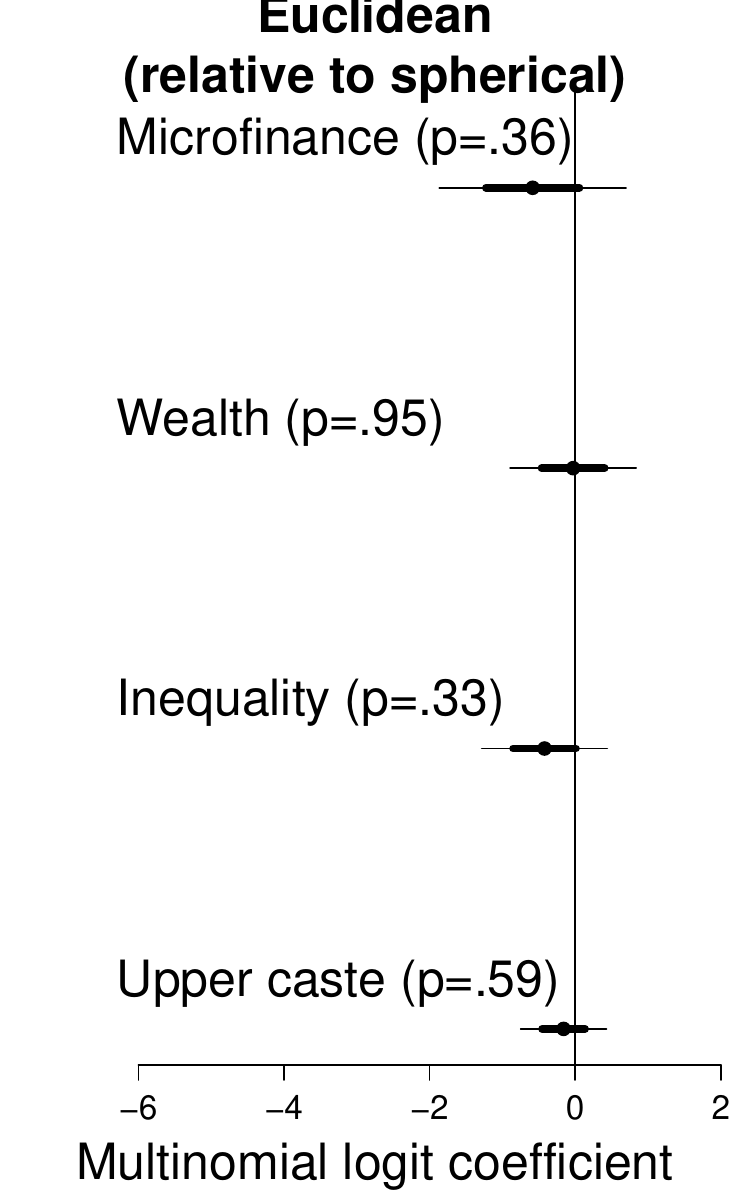}
    \includegraphics[angle=0,origin=c,width=.24\textwidth]{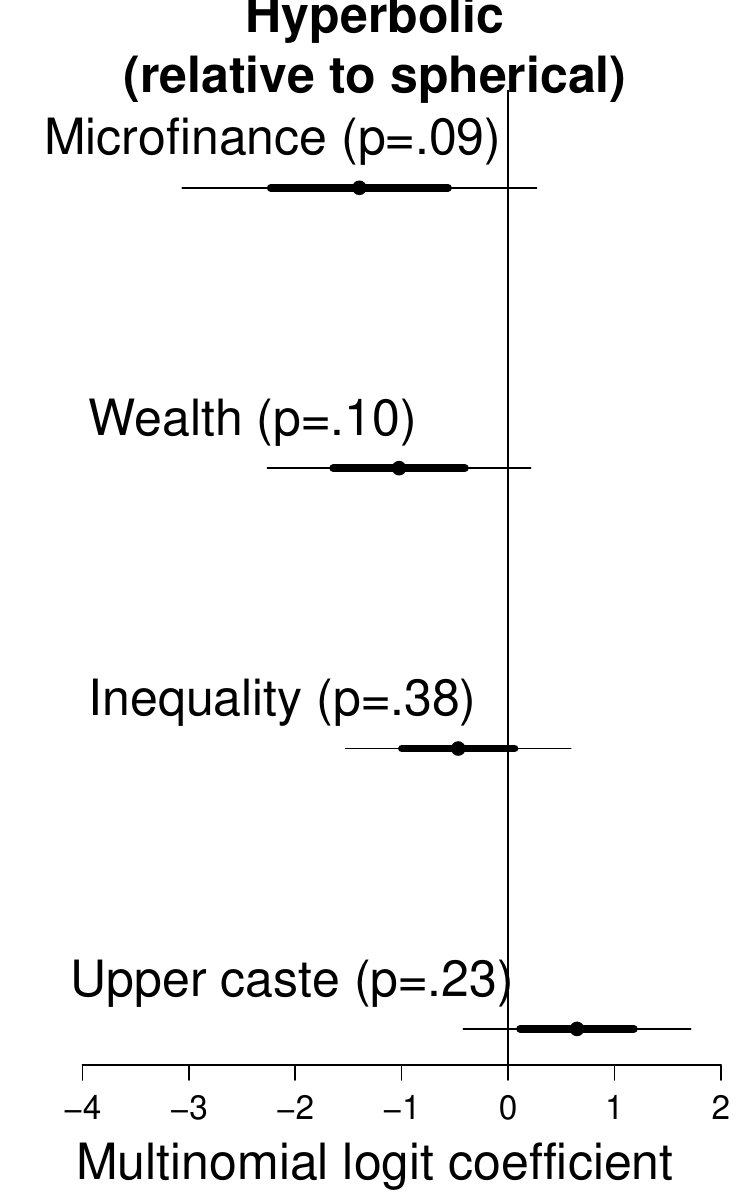}
    \includegraphics[angle=0,origin=c,width=.24\textwidth]{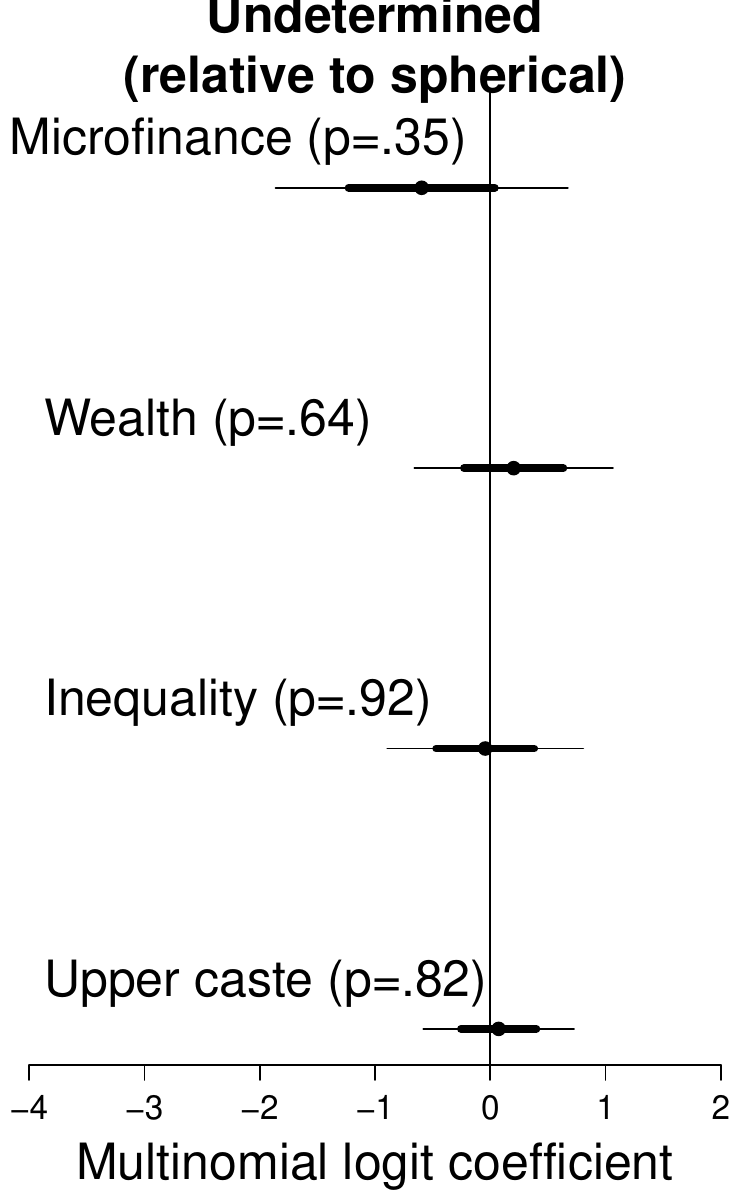}
    \caption{\footnotesize{Regression coefficients showing the determinants of geometry.  In the left figure, each line in the plot corresponds to the coefficient in a multivariate linear regression where the outcome is the average amount of loans (in thousands of INR) and the predictors are geometry types (with spherical as the reference).  The wide bars correspond to one standard error and the narrow bars represent two standard errors.  The reference value for spherical is 16.71 (again in thousands of INR). Plots 2-4 show the coefficients from a multinomial logistic regression where the outcome is the predicted geometry type for each village.  Each panel shows all coefficients for a particular geometry (with spherical as the reference).     Each line in the plot corresponds to an estimated coefficient.  The wide bars correspond to one standard error and the narrow bars represent two standard errors.  The constant values for the Euclidean, hyperbolic, and undetermined comparisons are .005, -.69, and -.01, respectively.}}
    \label{fig:mfifig}
\end{figure}

The leftmost panel of Figure~\ref{fig:mfifig} presents the results. We find that a Euclidean village relative to a spherical one has INR 3940 or  24\% ($p = 0.098$) more informal network loans. Further decreasing curvature, we compare hyperbolic villages to spherical ones and find that hyperbolic villages have INR 5865 or  35\%  ($p = 0.034)$ more in informal network loans. These increases are extremely large in real economic terms:  the difference in credit between the hyperbolic and spherical geometries  corresponds to an individual in the hyperbolic geometry receiving additional credit worth 20 days of wages. Taken together, we have seen greater financial flows precisely in geometries that permit more expansive network topologies.

Second, having studied how informal financial transaction patterns are associated with geometry, we now turn to studying  determinants of geometry. Our primary interest is in whether the introduction of a formal credit market (microfinance) to a setting otherwise dominated by only informal financial transactions changes the network structure by changing the   geometry. In our setting, as described below, microcredit was introduced to only some  villages, allowing us to compare the impact of access to microfinance on network structure.

In addition to microcredit access, we focus on three other determinants:  wealth, inequality, and caste fractionalization. It is ex ante not obvious as to how any of these might correlate with geometry and is therefore an important empirical question. For example, wealthier villages may have a reduced need to sustain informal insurance---their worst case scenario is better off than their poorer counterparts---and as a consequence may require less positive curvature. Or, in contrast, wealthier villages may be able to take on greater entrepreneurial risk as they can sustain losses, and such endeavors require group cooperation and therefore closure. Similarly, within-village wealth inequality can change incentives for triadic closure, as can ethnic fractionalization \citep{currarinijp2009}. Ultimately, the empirical correlations are of interest.
The most important relationship to study is how the introduction of formal credit to villages that otherwise used informal network transactions affects geometry.  From 2007, a microfinance institution entered 43 of the 75 villages studied here and the network data we utilize is taken after the intervention \citep{banerjeecdj2013,banerjeecdj2016change}. This allows us to study the effect of the introduction of microcredit on network geometry as a way to understand whether credit access differentially changes the need for ones' friends to maintain relationships with each other. Note that this is different from clustering or other measures of closure per se, which are also affected by the locations $z$ and fixed effects $\nu$. So we can specifically address that, all things being equal, whether the demand for one's friends to themselves be linked increases, decreases, or is unchanged when the village now has access to formal financial instruments. It is a priori not obvious. On the one hand, the new credit opportunity may encourage re-lending or joint business ventures among clients of microcredit, increasing the need for closure and generating positive curvature. On the other hand, the new credit opportunity may reduce  reliance on informal financial relationships with others in the village and push towards negative curvature. In either case, the answer as to how a large credit intervention may affect geometry is of empirical interest.

To study the determinants of geometry, we estimate a multinomial regression:
\[
\frac{\Pr(\hat{\mathcal{M}}^{\hat p}_i = m)}{\Pr(\hat{\mathcal{M}}^{\hat p}_i = \mathbf{S})} = \exp( \delta_m + \beta^m_{\text{MFI}} \text{MFI}_i + \beta_{\text{W}}^m  \text{Wealth}_i  + \beta_{\text{I}}^m   \text{Inequality}_i+ \beta_{\text{F}}^m   \text{Frac}_i) 
\]
where $m \in \{E, \mathbf{H}, \text{N/A}\}$.
$\text{MFI}_i$ denotes whether the microfinance institution entered village $i$. $\text{Wealth}_i$ denotes a wealth index measure.\footnote{The \cite{banerjeegossip} dataset does not have consumption nor expenditure measures. So we utilize the score constructed from the first principle component of a number of household features that correlate with wealth in the village. This consists of access to private electricity, home ownership, quality of roofing material, and number of rooms in the household.} $\text{Inequality}_i$ is within-village standard deviation of wealth.\footnote{Specifically, we take the score from the first principle component of the within-village standard deviation of each of the constituent wealth measures.} Finally, $\text{Frac}_i = \alpha_U(1-\alpha_U)$ where $\alpha_U$ is the share of households that are of upper caste. The score is zero if society is perfectly homogenous and 1/4 for an even split.

The right three panels of Figure~\ref{fig:mfifig} present the results. We begin by looking at microfinance.  We estimate $\hat{\beta}_{\text{MFI}}^{\mathbf{H}} = -1.40$ ($p = 0.093$). This means that a village receiving microcredit is associated with an 8.8\% decline in the probability of being hyperbolic relative to spherical. In other work, \citep{banerjeecdj2016change},  we have shown that introducing microcredit has decreased density and also the number of triads in the network. Our analysis here  demonstrates that the fundamental value of having friends in common itself \emph{increased} suggesting that the effects documented in our prior work came from shifts in node locations ($z_i$) and efforts of socializing ($\nu_i$) in the latent space, rather than changes in the relative value of closure which appears to have increased.

We also find that wealthier villages are less likely to be hyperbolic relative to spherical. We estimate $\hat{\beta}_{\text{W}}^{\mathbf{H}} = -1.02$ ($p = 0.098$). This corresponds to a 8.4\% decline in the relative probability of being hyperbolic as compared to spherical. We do not find any significant relationship between wealth inequality nor caste fractionalization and geometry.

Taken together, we have shown the empirical content of the estimation of the latent geometry. We can classify the vast majority of villages (despite allowing for N/A) and they are predominantly spherical. We find informal financial loans are higher in villages that exhibit negative curvature. Finally, and importantly,  villages where microcredit was introduced tend to have more spherical structure. This can perhaps be interpreted as showing that  access to microcredit generates, \emph{ceteris paribus}, demand for greater triadic closure.

\subsection{Network of Neurons}

Our second setting looks at a network of neurons. There is a neuroscience literature that is interested in documenting regularities in network structure as well as modeling network structure through statistical network formation models.

The first strand of the literature looks at how patterns of the graph of neurons relate to neurological mechanisms \citep{karwowski2019application}.  For instance, these networks exhibit short path lengths--disparate regions of the human brain are connected by a few steps. Further, the degree distribution reflects thick tails: certain nodes have numerous connections. Moreover, the network is dynamic: early in age the network exhibits high amounts of homophily whereas as the individual ages this declines.

The second strand 
develops low dimensional statistical representations of the neural networks since this allows for interpretability, counterfactuals, and deals with the fact that otherwise there is a litany of statistics that can be used to 
correlate with biological outcomes without any interpretable control \citep{de2014laplacian,recanatesi2019predictive}. To this end, conditional edge independence models, scale-free models, block models, and latent space models have been explored \citep{vanrullen2019reconstructing,karwowski2019application}.

Third, and particularly relevant for latent space models, is the concept of the functional graph of neurons rather than the structural graph of neurons  \citep{petersen2015brain,abdelnour2018functional}. The idea is that while a graph can be drawn of the phyiscal links between all nodes, predominantly the graph that is able to be activated--the functional network--is a network that is distinct. Much like individuals who reside in geographic space but functionally interact in a network that can be thought of as in a latent space, the functional network perspective presents an opportunity leverage latent space models.

Our specific application is to a network of neurons of \emph{Caenorhadbitis elegans},  which are soil-dwelling roundworms. There is a long history of using \emph{C. elegans} as a model organism for studying nervous systems of animals. In fact neurons of \emph{C. elegans} are extremely similar to that of humans \cite{leung2008caenorhabditis}.  For our example, we use the \emph{C. elegans} neuron data of \citet{kaiser2006nonoptimal}, which has been used a number of times in order to model neural network structure. There are several goals in modeling neural network structure. The relative location distribution, how distance affects linking rates, and the geometry all inform how signals could be passed across nodes. Moreover, though beyond the scope of our knowledge, there may be interpretations to the distribution of fixed effects---latent heterogeneity in the propensity  for certain neurons to systematically  link to others. 

  A priori it is unclear what the right latent geometry ought to be. For instance, if the network of neurons ought to have a high degree of expansiveness, it ought to be embedded in hyperbolic space. In contrast, if it ought to reflect strong, localized redundancies, or a high degree of homophily it may be better modeled as being  embedded in a spherical geometry.

The dataset contains a neural network from a single \emph{C. elegans}, consisting of a connected graph of 131 neurons, with 764 edges, and a clustering coefficient of 0.245. The clique number of this graph is 6,
but it has only one clique of size 6, but it has 29 cliques of size 5, so we use $\ell = 5$. We find $K =12$ cliques using the problem formulation in (\ref{eq: PickingCliques}) and then take a maximally disjoint clique set which is sufficient for our test. We use Algorithm \ref{alg: cliques} and compute the $p$ values for the Euclidean, spherical, and hyperbolic geometries and find these values are: $p_E = .378, \ p_S = 0.05, \text{ and } p_H = 0.267$, so we reject the spherical hypothesis (noting that we can do so despite there being a high level of clustering). The \emph{C. elegans} network of neurons, therefore, is inconsistent with a latent space with positive curvature, where neurons are excessively likely to exhibit triadic closure relative  a flat benchmark. We can only say that there is no or negative curvature, but the data are not sufficiently powered to allow us to distinguish this.

\section{Conclusion}\label{sec: Conclusion}

Latent space models are widely used in network analysis across numerous disciplines including, but not limited to sociology, economics, biology, and computer science. The predominant approach is to assume a Euclidean latent space, though there is current discussion about adopting a hyperbolic space in certain contexts. Nonetheless, the current methods employed do not provide a way to estimate the geometry itself. Unfortunately, incorrect embedding spaces can deliver misleading results and while there may be convergence to pseudo-true values, counterfactual analysis will be affected.

We develop methods which the researcher can apply in order to consistently estimate the latent space geometry from network data. Our core observation is that the observed network data encodes information on the distance between nodes in latent space. That is, a finite sample network corresponds to a noisy set of distances. So we transform our network problem to a statistical geometry problem.

In our first result we study  a more general problem: whether an observed estimate of a distance matrix among $K$ points contains enough information to consistently estimate the unobserved manifold in which the $K$ nodes can be isometrically embedded. We answer this in the affirmative: the spectrum of a distance matrix encodes the manifold's metric and therefore the manifold class, rank, and curvature. Leveraging results on eigenvalue perturbations, we prove the result.
Our second result applies to the network setting. By looking at cross-clique link frequencies, one can construct a noisy distance matrix and therefore estimate the latent manifold consistently. 

An important advantage of our approach is that, unlike other strategies, we do not need to estimate the fixed effects or the locations in a candidate manifold (nor integrate them out) in the estimation procedure. Instead, by focusing on a strategy that exploits the fundamentals of geometry, we 
 directly checks isometric embeddings, so we can estimate the geometry without ever estimating the numerous other parameters and only move to them after having estimated the geometry. 
 
We also demonstrate the  the empirical content of estimating the latent geometry which is novel in the literature. Strikingly, even though N/A is a possibility, we were able to classify (75\%) of villages, indicating the empirical relevance of our methods.  Further, consistent with theory, we show Indian risk sharing villages are often spherical. Additionally, villages that are more expansive are associated with a greater flow of informal financial loans through the network. Finally, the introduction of microcredit is associated with a shift to positive curvature: the relative value of having triadic closure increases when villages have access to formal credit. 

A number of future steps come to mind.  While our assumptions on geometry---that it is a simply connected, complete Riemannian manifold---are parsimonious and natural, they are also limited. They nest the current assumptions in the literature (we know of no empirical research that assumes a torus of genus two for instance in the networks literature) but they are still admittedly lacking. \cite{zoubouloglou2021scaled} has shown a relationship between the torus and the sphere that might allow us to apply our current spherical methodology to the torus. We speculate that there may be strategies to use local structures in the network to patch together some more global structure. That is, for instance, if it can be arranged into a pseudo-block diagonal structure, perhaps in each block there is room for a different geometry and then these can be stitched together.
See \cite{Gu2019}, among others, for related work on this topic.  Additionally, extending our results to settings where the full graph is not observed, such as Aggregated Relational Data~\citep{mccormick2015latent, breza2017using}, would allow researchers who do not have resources to collect data on all edges to leverage insights about underlying geometry.  Individual node covariates could also be leveraged to form trait groups in settings without complete network data.
Finally, an interesting area of future research involves exploring how to optimally combine 
 the results from multiple tests of the three geometries, where each test is computed using a different set of cliques.  Of course the tests based on different cliques would likely be correlated, and so an important question would be to understand how to use the correlation to increase the power of the combined test.

\bibliographystyle{ecta}
\bibliography{Networks,networks_two}

\newlength{\fig}
\settowidth{\fig}{Fig.\,99:~}
{\renewcommand*\numberline[1]{\llap{\makebox[\fig][l]{Fig.\,#1:~}}}
\makeatletter
\renewcommand*\l@figure[2]{\leftskip\fig\noindent#1\par}
\makeatother
\listoffigures}

\clearpage
\appendix

\bigskip

\appendix
\setcounter{table}{0}
\renewcommand{\thetable}{\thesection.\arabic{table}}
\setcounter{figure}{0}
\renewcommand{\thefigure}{\thesection.\arabic{figure}}

\section{Proofs}\label{sec:proofs}

\subsection{Proofs Required for Theorem \ref{thm: intro_thm} (Section \ref{sec:geometry-general})}

\begin{proof}[Proof of Proposition \ref{prop:alpha_bound}] We only prove this claim for the Euclidean case, but the same argument proves the claim for the other two geometries. We have by Weyls's inequality that $ |\lambda_1(\hat W_0) - \lambda_1(W_0)| \leq ||\hat W_0 - W_0||_F$, where $||A||_F$ is the Frobenius norm of $A$, that is $||A||_F^2 = \sum_{l,l'} a_{ll'}^2$. Then, we have that $\mathbb{P}(|\lambda_1(\hat W_0) - \lambda_1(W_0)| < \theta) \leq \mathbb{P}(||\hat W_0 - W_0||_F < \theta)$ for all $\theta$. Under $H_{0, e}$, $\lambda_1(W_0) = 0$, so we have that 
$\mathbb{P}(|\lambda_1(\hat W_0)| < \theta) \leq \mathbb{P}(||\hat W_0 - W_0||_F < \theta)$. By setting $\theta$ to be the $\alpha$ quantile of $||\hat W_0 - W_0||_F$, we conclude (\ref{eq: Type1Bound}). 
This completes the proof.
\end{proof}

\begin{proof}[Proof of Proposition \ref{prop: ConsistencyKappa}]
We prove this proposition for the spherical case. The hyperbolic case follows from a similar argument. By Theorem 2.1 in \citet{newey1994large}, we have consistency if (i) the limit objective function is uniquely maximized at the truth, (ii) the parameter space is compact, (iii) the limit objective function is continuous in the parameter, and (iv) there is uniform convergence of the empirical objective function to its limit. The latter holds if there is point-wise convergence and stochastic equicontinuity.  The parameter space is compact and since under the null $W_\kappa(D)$ is positive semi-definite, the minimum eigenvalue is 0 as long as $K > p$. Identification comes from continuity of eigenvalues in parameters of the matrix.
 Finally we check uniform convergence. First, note by hypothesis that ${\hat D}  \overset{p}{\rightarrow} {D}$ as $T \rightarrow \infty$. Since eigenvalues are continuous functions of their matrix arguments, we have by the continuous mapping theorem that $\lambda_1\left(\kappa W_\kappa({\hat D})\right) \overset{p}{\rightarrow} \lambda_1\left(\kappa  W_\kappa( {D})\right)$ for every $\kappa \in [a, b]$, and so we have pointwise convergence. To complete the proof, we will show stochastic equicontinuity to show uniform convergence. A sufficient condition is a Lipschitz condition (Lemma 2.9, \citet{newey1994large}): that for any $\kappa_1, \kappa_2$, $|\lambda_1\left(\kappa_1  W_{\kappa_1}({\hat D})\right) - \lambda_1\left(\kappa _2 W_{\kappa_2}({\hat D})\right)\Big| \leq B_T \lvert \kappa_1 - \kappa_2\rvert$ for some random variable $B_T = O_p(1).$ To do this, fix any $\kappa_1, \kappa_2 \in [a,b]$. By Weyl's inequality,
\begin{eqnarray*}
\Big|\lambda_1\left(\kappa_1 W_{\kappa_1}(\hat D )\right) - \lambda_1\left(\kappa_2 W_{\kappa_2}(\hat D)\right) \Big| \leq ||\kappa_1 W_{\kappa_1}(\hat D) - \kappa_2 W_{\kappa_2}(\hat D)||_F. \;
\end{eqnarray*}
 Since $\kappa W_\kappa(D ) = \cos(\sqrt{\kappa} D)$ and $\cos(\cdot)$  is Lipschitz continuous with Lipschitz constant 1, we have for each $l,l'$,
\begin{equation*}
\Big|\cos(\kappa^{1/2}_1 \hat d_{l,l'})- \cos(\kappa^{1/2}_2 \hat d_{l,l'}) \Big| \leq \hat d_{l,l'} \cdot  \Big| \kappa_1^{1/2} - \kappa_2^{1/2} \Big| \;.  
\end{equation*}
For $\kappa_1, \kappa_2 \in [a, b]$,
\begin{align*}
    |\sqrt{\kappa_1} - \sqrt{\kappa_2}| & = \Big|\frac{\kappa_1 - \kappa_2}{\sqrt{\kappa_1} + \sqrt{\kappa_2}}\Big| \leq \frac{1}{2\sqrt{a}} |\kappa_1 - \kappa_2| \;,
\end{align*}
so for any $\hat d_{i,j}$,
\[ \Big|\cos(\kappa^{1/2}_1 \hat d_{i,j})- \cos(\kappa^{1/2}_2 \hat d_{i,j}) \Big| \leq \frac{\hat d_{i,j}}{2a^{1/2}}  \Big| \kappa_1- \kappa_2 \Big| .\]
Putting this all together, we see that
\begin{align*}
\Big|\lambda_1\left(\kappa_1 W_{\kappa_1}(\hat D )\right) - \lambda_1\left(\kappa_2 W_{\kappa_2}(\hat D)\right) \Big| &\leq \sqrt{\sum_{i,j} \left(\kappa_1 W_{\kappa_1}(\hat D) - \kappa_2 W_{\kappa_2}(\hat D) \right)_{i,j} ^2 } \\
&\leq \sqrt{\sum_{i,j} \left(\frac{\hat d_{i,j}}{2a^{3/2}} \Big|\kappa_1-\kappa_2\Big|\right)^2} \\
&= \sqrt{\sum_{i,j} \left(\frac{\hat d_{i,j}}{2a^{3/2}}\right)^2} |\kappa_1 - \kappa_2| \;.
\end{align*}
Since $\sqrt{\sum_{i,j} \left(\frac{\hat d_{i,j}}{2a^{3/2}}\right)^2} = O_p(1)$, the desired Lipschitz condition holds, which completes the proof. The hyperbolic case is handled in a similar way.
\end{proof}

\begin{proof}[Proof of Proposition \ref{prop: consistent_test}]
We prove the Euclidean case (part a) and note that the proofs of parts b and c (spherical and hyperbolic) are nearly identical. Define $\mathcal{R}_T = (-\infty,\delta_T]$. Let $\mathbb{P}_0(A)$ denote the probability of the event $A$ under the null hypothesis that $\mathcal{M}^p(\kappa)$ is Euclidean. By (\ref{eq: testE}),
\begin{equation*}
    \mathbb{P}_0(\lambda_1(\hat W_0) \in \mathcal{R}_T) = \mathbb{P}_0(\lambda_1(\hat W_0) \leq \delta_T) = o(1) \;, 
\end{equation*}
by assumption. Under $H_1$, $\lambda_1(W_0) < 0$ by Lemma \ref{lem: embedding}. Since $\delta_T = o_P(1)$, 
\begin{equation*}
    \mathbb{P}_1(\lambda_1(\hat W_0) \in \mathcal{R}_T) = \mathbb{P}(\lambda_1(\hat W_0) \leq \delta_T) = 1 - \mathbb{P}\left(\lambda_1(\hat W_0) \geq \delta_T\right) = 1 -  o(1) \;.
\end{equation*}
This proves that the test for $(\ref{eq: Euclidean_Hypotheses})$ is consistent, as claimed.
\end{proof}

\begin{proof}[Proof of Proposition \ref{prop: consistent_rank}] For each index $j = 1, \dotsc, K$, we consider two different cases.
In case 1, $\lambda_j(W_\kappa) \neq 0$. In this case, know that since $\epsilon_T \overset{p}{\rightarrow} 0$, $\Prob(\lambda_j(\hat W_{\hat \kappa}) \in \mathcal{R}) \rightarrow 0$. Here we just use the fact that $\lambda_{j}(\hat W_{\hat \kappa}) \rightarrow c \neq 0$, so that eventually this eigenvalue is outside the rejection region. Note that this calculation does not require a particular rate on $\epsilon$; we just need $\epsilon$ to go to zero in probability. 
We now handle case 2, in which $\lambda_j(W_\kappa) = 0$. This is the more subtle case. By definition of the rejection region, \begin{equation*}
        \Prob_1(\lambda_j(\hat W_{\hat \kappa}) \in \mathcal{R}) = \Prob_1(|\lambda_j(\hat W_{\hat \kappa})| \leq \epsilon_T) \;.
    \end{equation*}
    where the subscript here indicates that the alternative hypothesis that $\lambda_j \neq 0$ is true.
    By Weyl's inequality, the above probability then becomes
    \begin{equation*}
        \Prob_1(|\lambda_j(\hat W_{\hat (\kappa})| \leq \epsilon_T) \leq  \Prob_1(||\hat W_{\hat \kappa} - W_{\kappa}||^2 \leq \epsilon_T^2) := \Prob_1(r_T \leq \epsilon_T) \;.
    \end{equation*}
    By assumption, we know that $r_T / \epsilon_T \rightarrow 0$ in probability. Therefore, for any eigenvalue that is actually zero, for large enough $T$ we will call this eigenvalue zero and it won't count in the estimated rank. This concludes the proof.
\end{proof}

\begin{proof}[Proof of Theorem \ref{thm: intro_thm}]
By assumption, we know that $\hat D \overset{p}{\rightarrow} D$, so by Proposition \ref{prop: ConsistencyKappa} we have that $\hat \kappa \overset{p}{\rightarrow} \kappa$. We will use Proposition \ref{prop: consistent_test} to argue that $\widehat{\mathcal{M}^{\hat p}}$ is consistent for $\mathcal{M}^p(\kappa)$. To do this, note that if $\mathcal{M}^p(\kappa)$ is Euclidean, then by Proposition \ref{prop: consistent_test}, $ \widehat{\mathcal{M}^{\hat p}}$ is consistent. To prove the claim for the spherical case, recall that we define $\phi(\hat W_0) = 1$ to mean that we reject the hypothesis that $\mathcal{M}^{p^\star}(\kappa^\star)$ is Euclidean. If $\phi(\hat W_0) = 0$ then we fail to reject the hypothesis that $\mathcal{M}^{p^\star}(\kappa^\star)$ is Euclidean. Similar definitions hold for the spherical and hyperbolic cases.

If $\mathcal{M}^{p^\star}(\kappa^\star)$ is spherical, then we have that
\begin{align*}
    \mathbb{P}_S(\widehat{\mathcal{M}^{\hat p}} = \mathbf{S}^p(\kappa)) &= \mathbb{P}_S(\phi(\hat W_0) = 1,\ \phi(\hat W_{\hat \kappa}) = 0)\\
   &= \mathbb{P}_S(\phi(\hat W_0) = 1) \mathbb{P}_S(\phi(\hat W_{\hat \kappa}) = 0) \\
   &\rightarrow 1 \;,
\end{align*}
where the notation $\mathbb{P}_S$ indicates that $\mathcal{M}^{p^\star}(\kappa^\star) = \mathbf{S}^p(\kappa)$ and the third line follows from Proposition \ref{prop: consistent_test}. A similar argument proves that $\widehat{\mathcal{M}^{\hat p}}$ is consistent when $\mathcal{M}^{p^\star}(\kappa^\star)$ is hyperbolic.
 Therefore, we can conclude that $\hat p$ is consistent for the true rank of $W_\kappa$. This completes the proof.
\end{proof}

\subsection{Proofs Required for Theorem \ref{thm: main_cliques} and Section \ref{sec: network_geometry_identification}}

\begin{proof}[Proof of Theorem \ref{thm: main_cliques}] %
 In order to show that estimates of distances computed using cliques are consistent, we recall the form of the estimates from (\ref{eq: Complex_Model_Solved_1}),
\begin{equation*}
    d(z_i^\star, z_j^\star) = -\log(p_{ij}) + \log(\gamma)\;,
\end{equation*}
where $\gamma := E\{\exp(\nu)\}^2$ and $p_{ij} = \Pr(G_{ij} = 1 | z_i^\star, z_j^\star)$. We estimate $d(z_i^\star, z_j^\star)$ with 
\begin{equation*}
    \hat d(z_i, z_j) = -\log(\hat p_{ij}) + \log(\hat \gamma) \;.
\end{equation*}
Under the assumptions of Theorem~\ref{thm: main_cliques}, we have a consistent estimate $\hat \gamma \rightarrow \gamma$, so we only focus on estimating the term $p_{ij}$. We estimate this term using 
\begin{equation*}
    \hat p_{kk'} = \ell^{-2} \sum_{i \in C_k(\ell)} \sum_{j \in C_{k'}(\ell)} G_{ij} \;,
\end{equation*}
where $C_k(\ell)$ is a clique of size $\ell$ and $C_{k'}(\ell)$ is another clique of size $\ell$. Suppose that each node in a clique is at the same location, say $\zeta_k$ for clique $k$. Then, $\hat p_{kk'}$ is a consistent estimate of $p_{ij}$, the probability that node $i$ at location $z_k$ connects to node $j$ at location $z_{k'}$. In practice, the locations in cliques do not fall at exactly the same location, but as $\ell \rightarrow \infty$, under Assumption \ref{ass:z},  we do know that $\max_{ij} d(z^\star_i, z^\star_j) \overset{p}{\rightarrow} 0$ for nodes $i$ and $j$ in any clique. Since the event that all nodes in a clique fall within $\delta$ of each other, for any $\delta > 0$, occurs with probability going to 1, we can condition on the event that nodes in a clique are at the same location. By the preceding argument, we can then conclude that $\hat p_{kk'} - p_{ij} \overset{p}{\rightarrow} 0 $. By the continuous mapping theorem, we can then consistently estimate $d_{ij}$. We can now apply Theorem \ref{thm: intro_thm}. This completes the proof.

\end{proof}

\begin{proof}[Proof of Proposition \ref{prop: general_clique}]
Before providing the proof, we provide a brief outline of our strategy. We will suppose that each $\nu^\star_i = 0$ to simplify notation, but the general result claimed in the Proposition holds. Our goal in this proof is to show that $\Pr(\mathcal{E}_\delta \mid \text{clique}) \rightarrow 1$ as $\ell \rightarrow \infty$. To make the proof easier, we will equivalently show that the ratio 
\begin{equation*}
   \frac{\Pr\left(\mathcal{E}_{\delta}\mid clique\right)}{\Pr\left(\mathcal{E}_{\delta}^{c}\mid clique\right)} \rightarrow \infty \;.
\end{equation*}
Showing that this ratio goes to infinity shows the numerator goes to 1 since $x/(1-x) \rightarrow \infty$ if and only if $x \rightarrow 1.$ We now turn to the proof.

We have
\begin{align*}
\frac{\Pr\left(\mathcal{E}_{\delta}\mid clique\right)}{\Pr\left(\mathcal{E}_{\delta}^{c}\mid clique\right)} & =\frac{\Pr\left(clique\mid\mathcal{E}_{\delta}\right)}{\Pr\left(clique\mid\mathcal{E}_{\delta}^{c}\right)}\cdot\frac{\Pr\left(\mathcal{E}_{\delta}\right)}{1-\Pr\left(\mathcal{E}_{\delta}\right)}\\
 & \geq\frac{\Pr\left(clique\mid\mathcal{E}_{\delta}\right)}{\Pr\left(clique\mid\mathcal{E}_{\delta}^{c}\right)}\times a(\delta)\cdot\frac{1/A_{n}^{\ell}}{1-1/A_{n}^{\ell}}\left(1+o\left(1\right)\right)
\end{align*}
for some positive constant $a(\delta)$ by Assumption \ref{ass: E_delta}.

Next, since we have $L:=\binom{\ell}{2}$ possible links and $\delta$
is the maximal distance between any two nodes,
\[
\Pr\left(clique\mid\mathcal{E}_{\delta}\right)\geq\exp\left(-L\delta\right),
\]
so we have 
\[
\frac{\Pr\left(\mathcal{E}_{\delta}\mid clique\right)}{\Pr\left(\mathcal{E}_{\delta}^{c}\mid clique\right)}\geq\frac{\exp\left(-L\delta\right)}{\Pr\left(clique\mid\mathcal{E}_{\delta}^{c}\right)}\times a(\delta) \cdot\frac{1/A_{n}^{\ell}}{1-1/A_{n}^{\ell}}\left(1+o\left(1\right)\right) \;.
\]

For bounded support, $\mu_d = \mu_{d,n} :=E\left\{ d_{\mathcal{M}^{p^{\star}}\left(\kappa^{\star}\right)}\left(z_{i},z_{j}\right)\right\}$ is finite.  Thus, by Lemma \ref{lem: reverse-jensen}, we have that
\[
\Pr\left(clique\mid\mathcal{E}_{\delta}^{c}\right)\leq\exp\left(-L\mu_{d}\right).
\]
Then, substituting this upper bound on $\Pr\left(clique\mid\mathcal{E}_{\delta}^{c}\right)$  into the ratio of interest we have
\begin{align*}
\frac{\Pr\left(\mathcal{E}_{\delta}\mid clique\right)}{\Pr\left(\mathcal{E}_{\delta}^{c}\mid clique\right)} & \geq\frac{\exp\left(-L\delta\right)}{\exp\left(-L\mu_{d}\right)}\times a(\delta)\cdot\frac{1/A_{n}^{\ell}}{1-1/A_{n}^{\ell}}\left(1+o\left(1\right)\right)\\
 & =\exp\left\{ -L\left(\delta-\mu_{d}\right)\right\} \times a(\delta)\cdot\frac{1/A_{n}^{\ell}}{1-1/A_{n}^{\ell}}\left(1+o\left(1\right)\right).
\end{align*}
We can rewrite this as
\begin{equation}
\exp\left\{ L\mu_{d}-L\delta\right\} \times\exp\left\{ \log a(\delta) +\ell\log A_{n}^{-1}-\log\left(1-A_{n}^{-\ell}\right)+o\left(1\right)\right\} 
\label{eq:lowerbound}
\end{equation}
and now need to show that this lower bound goes to infinity.  We do this by appealing to Assumption \ref{ass: clique_rate}.  Specifically, by Assumption \ref{ass: clique_rate} the following condition is true for sufficiently large $n$
\[
L\mu_{d}-L\delta\geq\ell\log A_{n}-\log\left(1-A_{n}^{\ell}\right)+\log a(\delta).
\]
Substituting the above expression into the lower bound in (\ref{eq:lowerbound}) we have 
\begin{align*}
\frac{\Pr\left(\mathcal{E}_{\delta}\mid clique\right)}{\Pr\left(\mathcal{E}_{\delta}^{c}\mid clique\right)}  \geq\exp\left\{L\mu_{d}-L\delta\right\} \times a(\delta)\cdot\frac{1/A_{n}^{\ell}}{1-1/A_{n}^{\ell}}\left(1+o\left(1\right)\right) 
   \rightarrow\infty,
\end{align*}
completing the proof in the case where each $\Omega_n$ is bounded. 

To handle the case where $\Omega_n$ is not bounded, such as when $F$ is a Gaussian distribution on $\mathbb{R}^p$, 
define an event
\[
\mathcal{F}_n:=\left\{ z_{i}\in\Omega_n,\ \text{for all }i=1,\ldots,n\right\} \;.
\]
In words, $\mathcal{F}_n$ holds whenever all $z_i$ are in the set $\Omega_n$.  Recall that under Assumption \ref{ass: general_support}(1), $\Pr(\mathcal{F}_n)=1$, and under  Assumption \ref{ass: general_support}(2) there is an exponentially thin tail. This allows us to conclude that
\begin{align*}
    \mathbb{P}({clique} | \mathcal{E}_\delta^c) &= \mathbb{P}({clique} | \mathcal{E}_\delta^c, \mathcal{F}_n) \mathbb{P}(\mathcal{F}_n) + \mathbb{P}({clique} | \mathcal{E}_\delta^c, \mathcal{F}^c_n) \mathbb{P}(\mathcal{F}^c_n) \\
    &\leq \mathbb{P}({clique} | \mathcal{E}_\delta^c, \mathcal{F}_n) \mathbb{P}(\mathcal{F}_n) +  \mathbb{P}(\mathcal{F}^c_n) 
    \end{align*}

Since the $z_i$ are i.i.d., we also know that $P(F_n^c) = P(z_i \not \in \Omega_n)^n := p(n)^n$. Combining this, we have that
\begin{equation*}
    \mathbb{P}({clique} | \mathcal{E}_\delta^c) \leq \exp(-\mu_d L) + p(n)^n
\end{equation*}
Supposing that $p(n) = \exp(-b A_n^{1/p^\star})$, where again $p^\star$ is the dimension of the latent space,  we see that
\begin{align*}
    \frac{\Pr\left(\mathcal{E}_{\delta}\mid clique\right)}{\Pr\left(\mathcal{E}_{\delta}^{c}\mid clique\right)}  &\geq\frac{\exp(-L \delta)}{\exp(-L \mu_d) +  p(n)^n}\times a(\delta)\cdot\frac{1/A_{n}^{\ell}}{1-1/A_{n}^{\ell}}\left(1+o\left(1\right)\right)  \\
    &= \frac{\exp(-L \delta)}{\exp(-L \mu_d) + \exp(-nb A_n^{1/p^\star})} \times a(\delta)\cdot\frac{1/A_{n}^{\ell}}{1-1/A_{n}^{\ell}}\left(1+o\left(1\right)\right) 
\end{align*}

Now, if we can show that $\exp(-n bA_n^{1/p^\star}) \rightarrow 0$ is negligible in the limit, then the above inequality reduces to the inequality for the bounded case, for which we proved that taking $\ell$ to grow faster than the term in this Proposition is sufficient. Since $n >> \ell$, the term $\exp(-n bA_n^{1/p^\star})$ is negligible, which allows us to re-use the proof of the bounded case. This completes the proof. 

\end{proof}

\begin{lem}\label{lem: reverse-jensen}
Consider any distribution of locations from which $z_{i}$ are drawn
i.i.d. on $\mathcal{M}^{p^{\star}}\left(\kappa^{\star}\right)$ with
finite expected distance,  $\mu_{d}:=E\left\{ d_{\mathcal{M}^{p^{\star}}\left(\kappa^{\star}\right)}\left(z_{i},z_{j}\right) \right\}<\infty$.  Then, if again $L = {\ell \choose 2}$,
\[
E\left[\prod_{i<j}\exp\left\{ -d_{\mathcal{M}^{p^{\star}}\left(\kappa^{\star}\right)}\left(z_{i},z_{j}\right)\right\} \vert\mathcal{E}_{\delta}^{c}\right]\leq\exp\left(-L\mu_{d}\right).
\]
\end{lem}

\begin{proof}
We have
\begin{align*}
E\left[\prod_{i<j}\exp\left\{ -d_{\mathcal{M}^{p^{\star}}\left(\kappa^{\star}\right)}\left(z_{i},z_{j}\right)\right\} \vert\mathcal{E}_{\delta}^{c}\right] & \leq E\left[\prod_{i<j}\exp\left\{ -d_{\mathcal{M}^{p^{\star}}\left(\kappa^{\star}\right)}\left(z_{i},z_{j}\right)\right\} \right]\\
 & \leq\prod_{i<j}\left(E\left[\exp\left\{ -L\cdot d_{\mathcal{M}^{p^{\star}}\left(\kappa^{\star}\right)}\left(z_{i},z_{j}\right)\right\} \right]\right)^{1/L}\\
 & \leq\prod_{i<j}\left(E\left[\exp\left\{ -L\cdot x_{ij}\right\} \right]\right)^{1/L}\\
 & \leq\exp\left(-L\mu_{d}\right)\times\prod_{i<j}\left(E\left[\exp\left\{ -L\cdot\eta_{ij}\right\} \right]\right)^{1/L}\\
 & \leq\exp\left(-L\mu_{d}\right)\times1
\end{align*}
where we (a) unconditioned on the event since the probabilities of
linking are higher within $\mathcal{E}$ than $\mathcal{\mathcal{E}}_{\delta}^{c}$,
(b) used Holder's generalized inequality, (c) used $\exp\left(a\right)^{L}=\exp\left(La\right)$,
(d) defined $x_{ij}:= d_{\mathcal{M}^{p^{\star}}\left(\kappa^{\star}\right)}\left(z_{i},z_{j}\right)$, 
(e) decomposed $x_{ij}=\mu_{d}+\eta_{ij}$ where $\eta_{ij} := x_{ij} - \mu_d$, and (f) used the boundedness
of linking probabilities.
\end{proof}

\begin{proof}[Proof of Example \ref{ex: GMM}]
Let $\Omega=\left[0,B^{1/p}\right]^{p}\subset\mathbb{R}^{p}$ so
${\rm vol}\left(\Omega\right)=B$ with $B=B_{n}$. Assume there are
$C=C_{n}$ communities, each with $m$ nodes, distributed uniformly at random
in $\Omega'=\left[0,b^{1/p}\right]^{p}\subset\Omega$
with $\left|B^{1/p}-b^{1/p}\right|=:t$ and $t=\omega\left(\sqrt{\log n}\right)$.
That is, the community centers---not members necessarily---reside
in a subset within the space of interest with a distance between the
boundaries of at least $\sqrt{\log n}$. The extra factor controls
for tail events.

Given these community centers $\zeta_{c}$, we have nodes distributed
\[
z_{i}\sim F_{c}\left(\zeta_{c},\sigma_{c}^{2}\right)
\]
where $F$ is a Gaussian distribution on $\mathbb{R}^{p}$ centered
at $\zeta_{c}$ with variance $\sigma_{c}^{2}$.

Note that if $\sigma_{c}^{2}=0$ then we have an example of an inhomogenous
lattice, and with $B=C$ this operates like Example \ref{ex: lattice} exactly. On the other hand, if $\sigma_{c}^{2}\rightarrow\infty$
and we restrict attention only to $\Omega$ itself, then we return
to the uniform case. In between lies the case of multimodal location
distributions with dispersion, governed by community centers. We will
identify similar rates, mildly adjusted for tail events of extreme
community or individual locations, though the bounds are not tight.

Define an event
\[
\mathcal{F}:=\left\{ z_{i}\in\Omega,\ \text{for all }i=1,\ldots,n\right\}
\]
and observe that the calculations conditional on $\mathcal{F}$ are
identical to the lattice and point process cases. We compute
\begin{align*}
\Pr\left(\mathcal{E}_{\delta}^{c}\vert clique\right) & =\Pr\left(\mathcal{E}_{\delta}^{c}\vert clique,\mathcal{F}\right)\Pr\left(\mathcal{F}\right)+\Pr\left(\mathcal{E}_{\delta}^{c}\vert clique,\mathcal{F}^{c}\right)\Pr\left(\mathcal{F}^{c}\right)\\
 & \leq\Pr\left(\mathcal{E}_{\delta}^{c}\vert clique,\mathcal{F}\right)\Pr\left(\mathcal{F}\right)+\Pr\left(\mathcal{F}^{c}\right)\\
 & \leq\exp\left(-L\mu_{d}\right)\left(1+o\left(1\right)\right)
\end{align*}
where $\mu_{d}$ is the expected distance between two points distributed
in $\Omega$ from the mixture model.

To bound $\Pr\left(\mathcal{F}^{c}\right)$, observe that
\[
\Pr\left(\max_{i}\text{dist}\left(z_{i},\zeta_{c}\right)>t\right)\leq\exp\left(-\text{const.}\times t^{2}+\log\left(n\right)\right)\rightarrow0
\]
 by the sub-Gaussian distribution of the distance function for normals (the folded normal is sub-Gaussian), the growth assumption on $t$, that it holds for all communities simultaneously (which is not tight since most nodes will not be within the $t$-shell of the boundary due to the slow expansion of Euclidean balls). Below we calculate $t>\text{const.}\times\sqrt{LB^{1/p}\sqrt{p}}$,
from which the result follows.

By the application of Lemma \ref{lem: reverse-jensen}, and a calculation of
the expected distance which is $c_{2}B^{1/p}\sqrt{p}$,
\[
\Pr\left(E_{\delta}^{c}\vert clique,\mathcal{F}\right)\leq\exp\left(-c_{2}LB^{1/p}\sqrt{p}\right),
\]
and so
\[
\frac{\Pr\left(\mathcal{E}_{\delta}\vert clique\right)}{\Pr\left(\mathcal{E}_{\delta}^{c}\vert clique\right)}=\exp\left(c_{2}LB^{1/p}\sqrt{p}-L\delta\right)\cdot\frac{\Pr\left(E_{\delta}\right)}{1-\Pr\left(E_{\delta}\right)}.
\]
We therefore have
\[
\ell\geq a(\delta)\frac{\log\left(\frac{B}{c_{3}\delta^{p}}-1\right)}{B^{1/p}\sqrt{p}-\delta}\left(1+o\left(1\right)\right)+1
\]
which gives the growth-rate bound on the clique size. Now recall the
restriction on the growth rate of $B$ relative to $b$. If $b=\alpha B$
for some $\alpha<1$, then
\[
t=B^{1/p}-b^{1/p}=B^{1/p}\left(1-\alpha^{1/p}\right)=\Theta\left(B_{n}^{1/p}\right)
\]
 and so the restriction is immediate if for instance $B_{n}=\omega\left(\left[\log n\right]^{p/2}\right)$.
This admits many simple rates such as if $C=\log n$, then $m=\frac{n}{C_{n}}=\frac{n}{\log n}$
or $B_{n}=\omega\left(\left[\log\left(\frac{n}{\log n}\right)\right]^{p/2}\right)$
which is a slow poly-logarithmic growth rate. In this case the number of communities is smaller than the domain. But if $C=\sqrt{n}$,
then $m=\sqrt{n}$ and therefore $B_{n}=\omega\left(2^{-p/2}\left(\log n\right)^{p/2}\right)$, the number of communities can grow faster than the domain.
\end{proof}

\setcounter{figure}{0}
\setcounter{table}{0}
\section{Generating latent space points}
\label{sec: generatePoints}

We now describe how we generate our points in the three latent spaces. The basic idea is to generate $K$ group centers. We then call the first $n/K$ nodes to be in group 1, the second $n/K$ nodes to be in group 2, and so on. Let $c_i \in \{1, \dotsc, K\}$ denote the group membership of node $i$. Finally, we distribute the node latent space positions centered at their group locations according to some procedure that is unique for each of the three geometries. To generate the latent space positions in the Euclidean case, we do the following:
\begin{enumerate}
    \item Generate $K$ group centers $\mu \in \mathbb{R}^p$ distributed according to $\mu \overset{\text{i.i.d.}}{\sim} \mathrm{N}(\mathbf{0}_p, \sigma^2 I_p)$. 
    \item Then simulate the positions of the nodes as $z_i|c_i \overset{\text{i.i.d.}}{\sim} \mathrm{N}\left(\mu_{c_i}, 
    \frac{\sigma^2}{K} I_p\right).$
\end{enumerate}

To generate the latent space positions in the spherical case, we do the following:

\begin{enumerate}
    \item Generate $K$ group centers $\mu \in \mathbf{S}^2(\kappa)$. To do this, we generate two angles: $\theta \overset{\text{i.i.d.}}{\sim} \text{Unif}(0, \pi)$ and $\phi \overset{\text{i.i.d.}}{\sim} \text{Unif}(0, 2\pi).$ Then compute \begin{equation*}\mu_i = \kappa^{-1/2} \left(\sin(\theta_i)\cos(\phi_i), \sin(\theta_i) \sin(\phi_i), \cos(\phi_i)\right) \in \mathbb{R}^3 \;.
    \end{equation*}
    \item Then simulate the positions of the nodes. To do this, generate 
    two angles $\theta_i \sim \text{Unif}(\theta_{c_i} -\delta, \theta_{c_i} + \delta)$ and $\phi_i \sim \text{Unif}(\phi_{c_i} -\delta, \phi_{c_i} + \delta)$ and compute 
    \begin{equation*}\mu_i = \kappa^{-1/2} \left(\sin(\theta_i)\cos(\phi_i), \sin(\theta_i) \sin(\phi_i), \cos(\phi_i)\right) \in \mathbb{R}^3 \;.
    \end{equation*}
\end{enumerate}

To generate the latent space positions in the Hyperbolic case, we do the following:

\begin{enumerate}
    \item Generate $K$ group centers $\mu \in \mathbf{H}^2(\kappa)$. To do this, we generate two locations $x_i$ and $y_i$ distributed uniformly on $[-s, s] \times [-s, s]$ and select the third coordinate $z = \sqrt{1/\kappa + x_i^2 + y_i^2}$ so by construction $(x, y, z) \in \mathbf{H}^2(\kappa).$
    \item
    Then simulate the positions of the nodes. To do this, generate 
    two coordinates $x_i$ and $y_i$ distributed uniformly on $[x_{c_i} - \delta, x_{c_i}+ \delta] \times [y_{c_i} - \delta, y_{c_i}+ \delta]$ then set $z_i = \sqrt{1/\kappa + x_i^2 + y_i^2}.$
\end{enumerate}

We next present the parameters used for the simulations in Section~\ref{sec: Simulations}.  In the table below, $\kappa$ is the curvature used for the Spherical geometry.  The $\sigma$ parameter determines the spread of the points in the Euclidean geometry.  For the Hyperbolic geometry the scale refers to the scale of the first two coordinates of the space.  In all of these results, we use rate = 1/3.  We draw the node effects $\nu_i \overset{\text{i.i.d.}}{\sim} \text{Unif}(\beta, 0)$, where we set $\beta = -0.01.$ 

\begin{table}[h]
\caption{The parameter values used to make the results in Section~\ref{sec: Simulations}. The rows correspond to the true data generating process and the columns correspond to the null hypothesis being tested.} 
\begin{tabular}{l|l|l|l}
  & \bf{E} & \bf{S} & \bf{H}   \\ \hline
\bf{E} & $\sigma = 0.5$ & $\sigma = 0.8$ & $\sigma = 0.8$    \\ \hline
\bf{S} & $\kappa = 0.75$  & $\kappa =1 $   & $\kappa = 0.75$   \\ \hline
\bf{H} & scale = 2.5, $\kappa = 0.75$  & scale = 2.5, \ $\kappa = 0.75$ & scale = 2.5, $\kappa = 1$ \\
\end{tabular}
\end{table}

\setcounter{figure}{0}
\setcounter{table}{0}
\section{Choosing bounds for curvature estimate}
\label{sec: PickingBounds_ab}
We now discuss a way to pick $a$, the lower bound in the spherical method to pick $\kappa$. Note that the maximum distance between any two points is $r \pi = \pi/\sqrt{\kappa}$, which occurs when the points are antipodal. This shows that for a distance matrix $D = \{d_{ij}\}$, which contains distances between $K$ points on $\mathbf{S}^p(\kappa)$, it must be that 
\begin{equation*}
    \max_{1\leq i, j \leq K} d_{i,j} \leq \frac{\pi}{\sqrt{\kappa}} \;.
\end{equation*}
By solving for $\kappa$, we see that $\kappa$ satisfies
\begin{equation*}
    \kappa \leq \left(\frac{\pi}{\underset{{1\leq i, j \leq K}}{\max} d_{ij}}\right)^2 := b \;.
\end{equation*}
Based on the discussion in \citet{Wilson}, we set 
\begin{equation*}
    a := \left(\frac{1}{3 \min_{i,j} d_{i,j}}\right)^2 \;.
\end{equation*}
The suggestion for $a$ comes from \citet{Wilson}, which says that for curvature values less than $a$, the space is essentially Euclidean. We use the same bounds for the hyperbolic case, but we flip the signs so that $[a, b] \subseteq (-\infty, 0]$. Future work could more thoroughly investigate how to pick the bounds for the hyperbolic case.

Figure \ref{spherical}  plots the function $\kappa \mapsto \Big|\lambda_1\left(
\kappa W_\kappa \right)\Big|
$ when $D$ corresponds to $K = 15$ points drawn randomly on $\mathbf{S}^2(1)$. Figure \ref{hyperbolic} plots the function $\kappa \mapsto \Big|\lambda_{K - 1}\left(
\kappa W_\kappa \right)\Big|
$ when $D$ corresponds to $K = 15$ points drawn randomly on $\mathbf{H}^2(-1)$. The functions are both minimized at the true curvature $(\kappa_0 = 1$ for the spherical case and $\kappa_0 = -1$ for the hyperbolic case).

\setcounter{figure}{0}
\setcounter{table}{0}
\section{Additional details on the bootstrap procedure}
\label{sec:apboot}

\begin{algorithm}
\SetAlgoLined
Input: adjacency matrix $G$, sub-sample size $m$, and number of bootstrap samples $B$, and rate $r \geq 0.$
\begin{enumerate}
\item Compute the observed eigenvalue $\lambda_{\tilde{k}}(\hat{W}_{\hat{\kappa}}(\hat{D}))$ for $\tilde k = K$ for Euclidean/spherical \\ and $\tilde k = 2$ for hyperbolic.
\item Construct a bootstrap distribution of eigenvalues. 
For $b=1,\ldots,B$:
\begin{enumerate}
\item Sample $D_{b}^{\star}$
\begin{enumerate}
\item Let ${\bf I}:=\left\{ I_{1},\ldots,I_{m}\right\} $ and ${\bf J}:=\left\{ J_{1},\ldots,J_{m}\right\} $
be two sets of integers of length $m$ drawn independently and uniformly
from $\left\{ 1,\ldots,\ell\right\} $ with replacement.
\item Calculate $P_{b}^{\star}$ with entries
\[
p_{b,kk'}^{\star}=\max\left\{ \frac{1}{m^2}\sum_{i,j}G_{ij}\cdot{\bf 1}\left\{ i\in C_{k}\cap{\bf I},\ j\in C_{k'}\cap{\bf J}\right\} ,\frac{1}{\ell^{2}}\right\} .
\]
\item Calculate $D_{b}^{\star}=-\log\left(P^{\star}/\hat{\tau}^{2}\right)$
where division is component-wise.
\end{enumerate}
\item Calculate $W_{b}^{\star}=W_{\hat{\kappa}}\left(D_{b}^{\star}\right)$
and eigenvalue $\lambda_{\tilde{k}}(W_{\kappa}(D_{b}^{\star}))$.
\end{enumerate}
\item Compute the CDF of the deviation in the bootstrapped and empirical
eigenvalue
\[
\hat{L}_{n}\left(x\right):=\frac{1}{B}\sum_{b=1}^{B}{\bf 1}\left\{ m^{2 r}\cdot\left(\lambda_{\tilde{k}}(W_{\hat{\kappa}}(D_{b}^{\star}))-\lambda_{\tilde{k}}(\hat{W}_{\hat{\kappa}}(\hat{D}))\right)\leq x\right\} ,\ \text{for any }x\in\mathbb{R}.
\]
\item Compute critical values
\(
c_{n,1-\alpha}=\inf\left\{ x:\ \hat{L}_n\left(x\right)\geq1-\alpha\right\} .
\)
\item Test hypotheses:
\begin{enumerate}
\item Reject $H_{0,e}$ and $H_{0,s}$ when, for each of their respective
test matrices,
\[
\ell^{2r}\cdot\lambda_{K}(\hat{W}_{\hat{\kappa}}(\hat{D}))<c_{n,\alpha}.
\]
\item Reject $H_{0,h}$ when
\[
\ell^{2r}\cdot\lambda_{2}(\hat{W}_{\hat{\kappa}}(\hat{D}))>c_{n,1-\alpha}.
\]
\end{enumerate}
\end{enumerate}
 \caption{Hypothesis Testing via Sub-sample Bootstrap}
 \label{alg: subsample}
\end{algorithm}

Given $n$ independent and identically distributed data points $X_1, \dotsc, X_n$ drawn from a distribution we want to estimate a parameter $\theta$ with an estimator $\hat \theta_n$. We make the following assumption about $\hat \theta_n$, which appears in \cite{Politis}.
\begin{ass}
    There exists a deterministic sequence $\tau_n$ such that $\tau_n(\hat \theta_n - \theta)$ converges in distribution to some random variable $L.$
    \label{ass: ThetaDistribution}
\end{ass}
Suppose that the goal is to construct confidence intervals for $\theta$ using $X_1, \dotsc, X_n$. To do this, we select a sub-sample rate $m = m(n)$, where $m \leq n$. Then, let $Y_1, \dotsc, Y_{{n\choose b}}$ be all the subsets of ${X}$ of size $b$, and let $\hat \theta_{n, i}$ be the estimate of $\theta$ using the $i$th subset $Y_i$.  Using the rate $\tau_n$ from Assumption \ref{ass: ThetaDistribution}, with $n$ replaced by the ``sub-sample" size $b$, we can form the empirical CDF of $\tau_b(\hat \theta_{n, i} - \hat \theta_n),$ 
\begin{equation*}
    L_n(x) := \frac{1}{{n\choose b}}\sum_{i = 1}^{n\choose b} \mathbf{1}\Big\{\tau_b(\hat \theta_{n, i} - \hat \theta_n) \leq x\Big\} \;.
\end{equation*}
Intuitively, as $n$ and $b \rightarrow \infty$, we expect that $L_n$ converges to the CDF of $\tau_n(\hat \theta_n - \theta)$, denoted by $L$. If this were true, then we could use the quantiles of $\tau_b(\hat \theta_{n, i} - \hat \theta_n)$ as estimates of the quantiles of $\tau_n(\hat \theta_{n, i} - \hat \theta_n)$, which would allow us to compute confidence intervals for $\theta$.  The following result shows when we can use $L_n$ to construct asymptotically correct confidence intervals for $\theta.$
\begin{prop}[Theorem 2, (iii) of \citet{Politis}]
Let $c_n(1-\alpha) := \inf\{x: \hat L_n(x)\geq 1-\alpha\}$. Similarly, let $c(1-\alpha) = \inf\{x: L(x) \geq 1-\alpha\}$ where $L$ is the CDF of $X_1$. If the CDF of $X_1$ is continuous at $c(1-\alpha)$ and $\tau_b/\tau_n \rightarrow 0$ and $b/n \rightarrow 0$ then 
\begin{equation*}
    \mathbb{P}\left(\tau_n(\hat \theta_n - \theta)\leq c_n(1-\alpha)\right) \rightarrow 1 - \alpha \;.
\end{equation*}
\end{prop}
This proposition allows us to construction asymptotically correct confidence intervals for $\theta$ from the sub-sampled data. Note that when $n$ is large, computing  all ${n \choose b}$ subsets of ${X}$ is computationally infeasible, so we instead select a collection $\{Y_1, \dotsc, Y_s\}$ for some integer $s \leq {n\choose b}$, and compute
\begin{equation*}
    \hat L_n(x) :=\frac{1}{s} \sum_{i = 1}^{s} \mathbf{1}\Big\{\tau_b(\hat \theta_{n, i} - \hat \theta_n) \leq x\Big\} \;.
\end{equation*}
According to \citet{Politis}, we have the following result: 
\begin{prop}
[Theorem 2, (iii) of \citet{Politis}]
Let $c_n(1-\alpha) := \inf\{x: \hat L_n(x)\geq 1-\alpha\}$. Similarly, let $c(1-\alpha) = \inf\{x: L(x) \geq 1-\alpha\}$ where $L$ is the CDF of $X_1$. If the CDF of $X_1$ is continuous at $c(1-\alpha)$ and $\tau_b/\tau_n \rightarrow 0$ and $b/n \rightarrow 0$,  then
\begin{equation*}
    \mathbb{P}\left(\tau_n(\hat \theta_n - \theta)\leq \hat c_n(1-\alpha)\right) \rightarrow 1 - \alpha \;.
\end{equation*}
\end{prop}
This result allows us to construct confidence intervals for $\theta$. Having described the sub-sampling method from \cite{Politis}, we now return to our original problem and show how to apply this method to our problem. The parameter interest $\theta$ is the eigenvalue $\lambda_{k^\star}(W)$. To study this, we will show how to use the \cite{Politis} method to sub-sample the distance matrix $D$. Using this sub-sampled distance matrix, we can then compute sub-sampled matrices $W_\kappa$ and compute their eigenvalues, since $W_\kappa$ is just a simple transformation of $D$.

The data in our problem is the adjacency matrix $G$. More concretely, it is the adjacency matrix for the subgraph with nodes $\bigcup_{k =1}^K C_i(\ell)$, the union of all $K$ cliques. We fix some sub-sample rate $m$. With the sub-sample rate, we then want to re-sample the entries of $D$. To do this, we will focus on how to do this for the $(k,k')$ entry of $D$. This process is repeated for all the entries of $D$. Let $\tilde G_{k,k'}$ denote the adjacency matrix corresponding to the sub-graph induced by the nodes in $C_k(\ell) \cup C_{k'}(\ell).$ For example,  if $\ell = 3$, then a potential $\tilde Y_{k,k'}$ might take the form
\begin{equation*}
    \tilde G_{k,k'} = 
    \begin{pmatrix}
    1 & 0 & 0 \\
    1 & 1 & 0 \\
    0 & 1 & 0
    \end{pmatrix} \;.
\end{equation*}
This indicates that the first node in $C_k$ connects to the first node in $C_{k,k'}$ but not to the second or third nodes in $C_{k'}.$ We then sample two sets of integers of length $m$, denoted by $I_k$ and $I_{k'}$, independently and uniformly from $\{1, \dotsc, \ell\}$, without replacement. These indices will be the re-sampled nodes. We then compute 
\begin{equation*}
    P^\star_{k,k'} = \frac{1}{m^2}\sum_{}[\tilde G_{k,k'}]_{ij} \mathbf{1}\{(i,j) \in I_k \times I_{k'}\} \;.
\end{equation*}
Since it is possible that $P^\star_{k,k'}$ is zero (meaning that the re-sampled pairs of nodes do not connect), we use $P^\star_{k,k'} = \max(1/\ell^2, P^\star_{k,k'})$, since we observe at least one edge in $\tilde G_{k,k'}$.  We repeat this procedure for all pairs of edges $(k,k').$ We then compute $D^\star$ using (\ref{eq:main_model}). We provide a step-by-step implementation of the sub-sampling method in Algorithm \ref{alg: subsample}.

Recalling that our parameter of interest is the eigenvalue $\lambda_{k^\star}(W)$, we use the above procedure to compute $\lambda_{k^\star}(W^\star_b)$ for $b =1, \dotsc, B.$ We then define
\begin{equation*}
    \hat L_n(x) = \frac{1}{B}\sum_{i = 1}^B \mathbf{1}\{m^{2r}\left(\lambda_{k^\star}(W^\star_i) - \lambda_{k^\star}(\hat W) \right)\leq x\}, \ \ \ \text{ for any } x \in \mathbb{R} \;.
\end{equation*}

We then perform hypothesis testing. To do this, we let $c_n(1-\alpha) = \inf\{x: \hat L_n(x) \geq 1 - \alpha\}$ be the $(1-\alpha)100\%$ percentile of $m^{2r}\left(\lambda_{k^\star}(W^\star_i) - \lambda_{k^\star}(\hat W) \right)$. Then, from Proposition~\ref{prop:alpha_bound}, we know that $P(m^{2r}(\lambda_{k^\star}(\hat W) - \lambda_{k^\star}(W)) \leq c_n(1-\alpha)) \approx 1 - \alpha + o(1)$ for large $\ell.$ This motivates the bootstrapping method we summarize in Algorithm \ref{alg: subsample}.

\subsection{Sensitivity of bootstrapping algorithm}
\begin{figure}[]
\minipage{0.32\textwidth}
  \includegraphics[width=\linewidth]{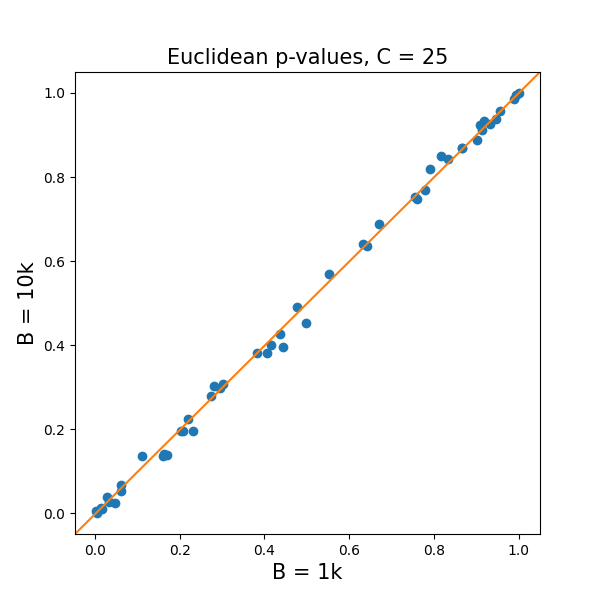}
\endminipage\hfill
\minipage{0.32\textwidth}
  \includegraphics[width=\linewidth]{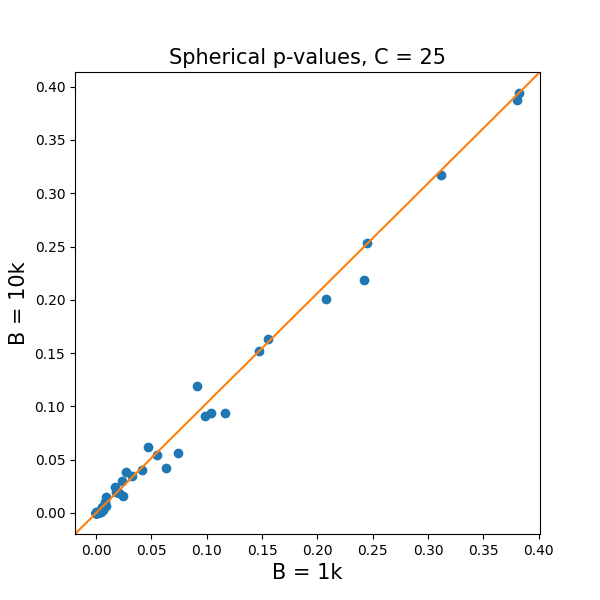}
\endminipage\hfill
\minipage{0.32\textwidth}%
  \includegraphics[width=\linewidth]{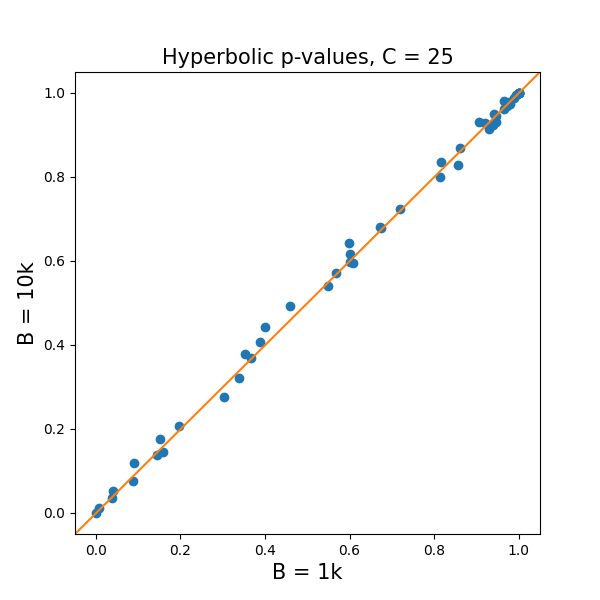}
 
\endminipage
\caption{Plot of the $p$-values for the three geometries (Euclidean, spherical, hyperbolic from left to right). Each point $(x, y)$ corresponds to two $p$-values. The $x$ coordinate is the $p$-value computed using $B = 1000$ and the $y$ coordinate is the $p$-value computed using $B = 10000$. The graph is computed using the graph model in (\ref{eq:main_model}) and we use 9 latent space positions drawn randomly from a $5 \times 5$ lattice in $\mathbb{R}^2$. Since most points fall on or near the diagonal, this is evidence that in this scenario, the sub-sampling algorithm in Algorithm \ref{alg: subsample} is not very sensitive to the choice of $B$, the number of distance matrices we sub-sample.}
\label{fig: bootstrap_stability}
\end{figure}

\begin{figure}[!htb]
\minipage{0.32\textwidth}
  \includegraphics[width=\linewidth]{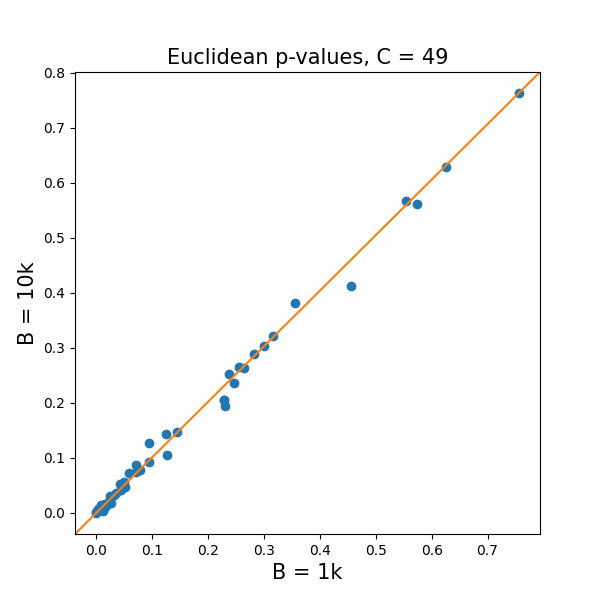}
\endminipage\hfill
\minipage{0.32\textwidth}
  \includegraphics[width=\linewidth]{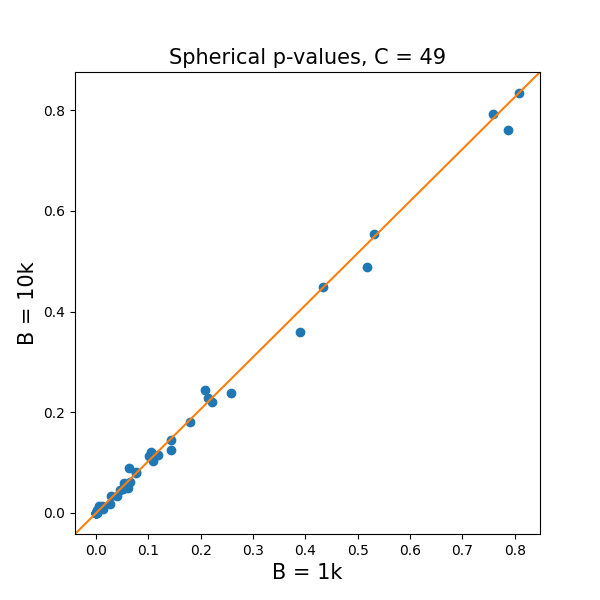}
\endminipage\hfill
\minipage{0.32\textwidth}%
  \includegraphics[width=\linewidth]{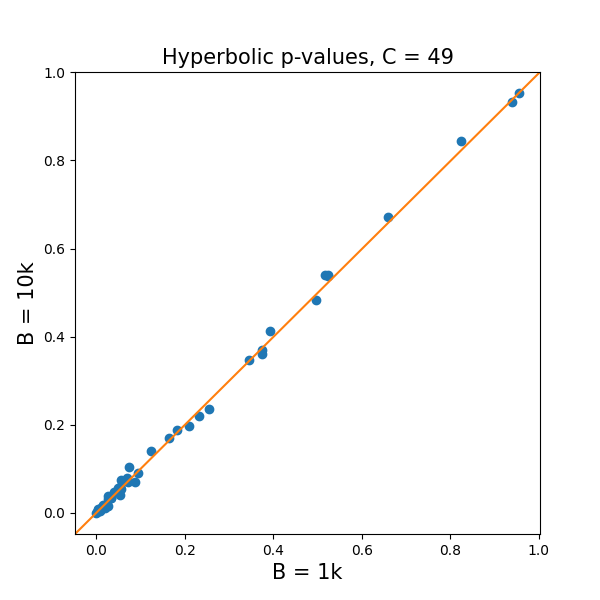}
 
\endminipage
\caption{Plot of the $p$-values for the three geometries (Euclidean, spherical, hyperbolic from left to right). Each point $(x, y)$ corresponds to two $p$-values. The $x$ coordinate is the $p$-value computed using $B = 1000$ and the $y$ coordinate is the $p$-value computed using $B = 10000$. The graph is computed using the graph model in (\ref{eq:main_model}) and we use 9 latent space positions drawn randomly from a $7 \times 7$ lattice in $\mathbb{R}^2$. Since most points fall on or near the diagonal, this is evidence that in this scenario, the sub-sampling algorithm in Algorithm \ref{alg: subsample} is not very sensitive to the choice of $B$, the number of distance matrices we sub-sample.}
\label{fig: bootstrap_stability2}
\end{figure}

We now analyze the sensitivity of the sub-sampling algorithm in \ref{alg: subsample} to the parameter $B$, the number of matrices $D^\star$ we generate. To do this, we generate a 2-dimensional lattice in $\mathbb{R}^2$ of length $5$ and 7 and randomly select $9$ points from these $C^2$ points. Using these $9$ points, we generate 50 networks and compute the $p$-values for the Euclidean, spherical, and hyperbolic geometries using $B = 1000$ and $B = 10,000$.  In Figures \ref{fig: bootstrap_stability} and \ref{fig: bootstrap_stability2}, we plot the 50 $p$-values for this simulation, and a diagonal line from $(0,0)$ to $(1,1)$. We see that most $p$-values lie on or very near to the diagonal line, which indicates that the sub-sampling algorithm is not very sensitive to the choice of the parameter $B$. 

\setcounter{figure}{0}
\setcounter{table}{0}
\section{Rank Estimator}
\label{sec: rank_appendix}

In Algorithm \ref{alg: Rank_Estimate} we formally describe the algorithm and the estimate of the rank of $W_\kappa.$  

\begin{algorithm}[b]
\SetAlgoLined
\begin{enumerate} \item Compute the scree function $\phi_T(j) := \frac{\hat \lambda_{K - j - 1}}{\sum_{i = 1}^K \hat \lambda_i}$ for $j \in \{0, 1, \dotsc, K-1\}$.
\item Sample $B$ bootstrapped $D^\star_1, \dotsc, D^\star_B$ matrices from Algorithm \ref{alg: subsample} and use them to compute $W^\star_1, \dotsc, W_B^\star.$ 
\item For $j \in \{0, 1, \dotsc, K - 1\}$,
\begin{enumerate}
    \item  Define $\hat A_j \in \mathbb{R}^{K \times j}$ with $\hat A_j = (\hat v_{K -j + 1}, \dotsc, \hat v_{K})$.
    \item Let $v^\star_1, \dotsc v^\star_K$ denote the eigenvectors of $W^\star_i$ corresponding to its eigenvalues $\lambda^\star_1\leq \dotsc \leq \lambda_K^\star.$ 
    \item Set $A^\star_{j,i} \in \mathbb{R}^{K \times j}$ with $A^\star_{j,i} = (v^\star_{K -j + 1}, \dotsc,  v^\star_{K})$.
\end{enumerate}
\item Compute 
\begin{equation*}
    f_n^0(j) = 1 - \frac{1}{B}\sum_{i = 1}^B |\det(\hat A_j^T A^\star_{j,i})|
\end{equation*}
\item Compute
\begin{equation*}
    f_n(j) = \frac{f_n^0(j)}{\sum_{i = 0}^{K-1} f_n^0(i)}  \;.
\end{equation*}

\item The estimate $\hat r$ of the rank of $W_\kappa$ is
\begin{equation}
    \hat r = \underset{j \in \{0, 1, \dotsc, K-2\}}{\arg\min} \left(\phi_T(j) + f_n(j)\right) \;.
    \label{eq: rankEstimate}
\end{equation} 
\end{enumerate}
 \caption{Estimating Rank of $W_\kappa$}
 \label{alg: Rank_Estimate}
\end{algorithm}

\begin{figure}
    \centering
    \includegraphics[scale =  0.5]{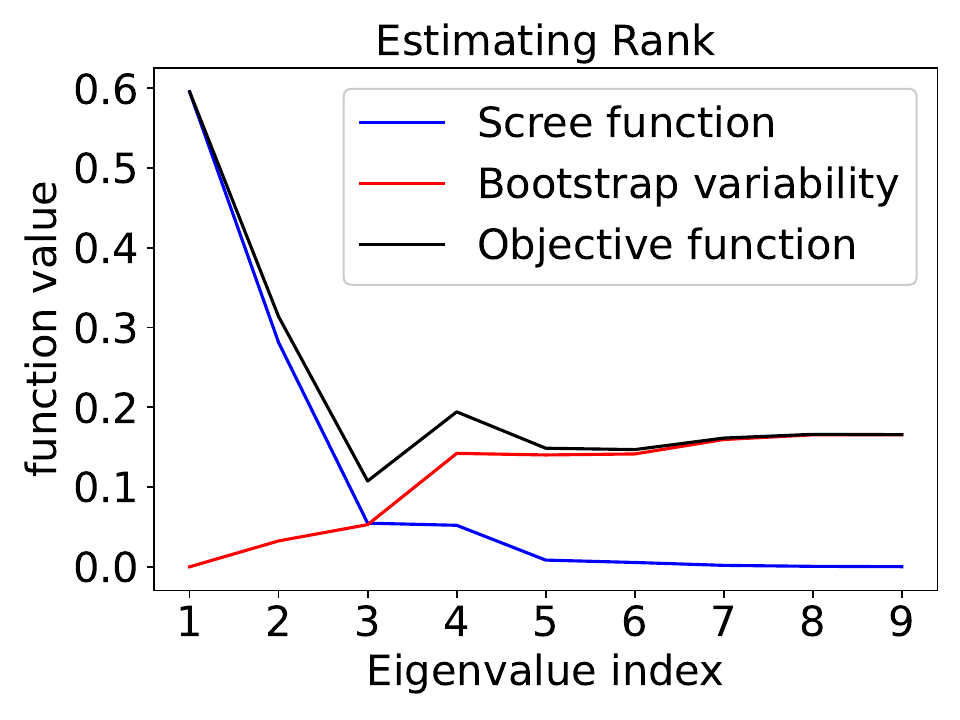}
    \caption{\footnotesize{We generate a graph using a 3-dimensional Euclidean latent space with $K = 10$ cliques. 
 We plot the scree function $\phi$ and the bootstrap variability function $f_n$ defined in Algorithm \ref{alg: Rank_Estimate}. We also plot their sum, defined as the objective function. The horizontal axis represents the possible ranks of the matrix. We see the objective function has a minimum at 3, so we estimate the rank of the matrix to be 3, which is the true dimension of the latent space.}}
    \label{fig: Rank_Plot}
\end{figure}

The \citet{Wei} estimator uses two pieces of information. The first is the scree function, which plots the sample eigenvalues in order from larges to smallest. In Figure \ref{fig: Rank_Plot} we plot the scree function for a distance matrix computed between $K = 15$ points on a 3-dimensional Euclidean latent space. We see that the scree plot is large but decreasing for the first three eigenvalues but becomes flat after that point. The second piece of information this estimator uses is the variability of the bootstrapped eigenvectors of the matrix $W_\kappa$, given in step (4) of Algorithm \ref{alg: Rank_Estimate}.  \citet{Wei} argues that the for $j < r$, the true rank of $W_\kappa$, there is little variation in the term $f_n(j)$ in step (4) but for $j \geq r$, this function increases. We see this behavior in Figure \ref{fig: Rank_Plot}: For $j < 3$, the bootstrap variability is lower than when $j \geq 3$. See \citet{Wei} for a more thorough explanation of why this phenomenon occurs. Based on these two pieces of information, \citet{Wei} suggests adding the two functions together to produce a final objective function. They claim that this new function has a ``ladle" shape. The minimum of  this new function is our estimate of the rank of $W_\kappa$.

\setcounter{figure}{0}
\setcounter{table}{0}
\section{Additional details for the ~\citet{banerjeegossip} data}
\label{app:data}

\begin{figure}%
     \centering
     \subfloat[]{{\includegraphics[width=7cm]{{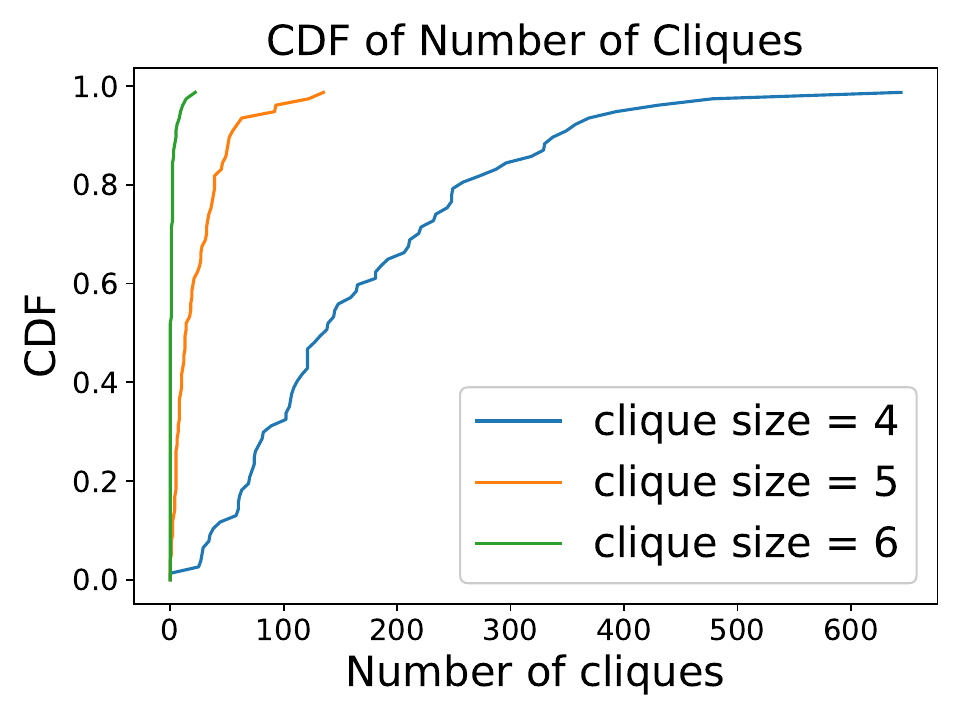}} }}%
     \qquad
     \subfloat[]
     {{\includegraphics[width=7cm]{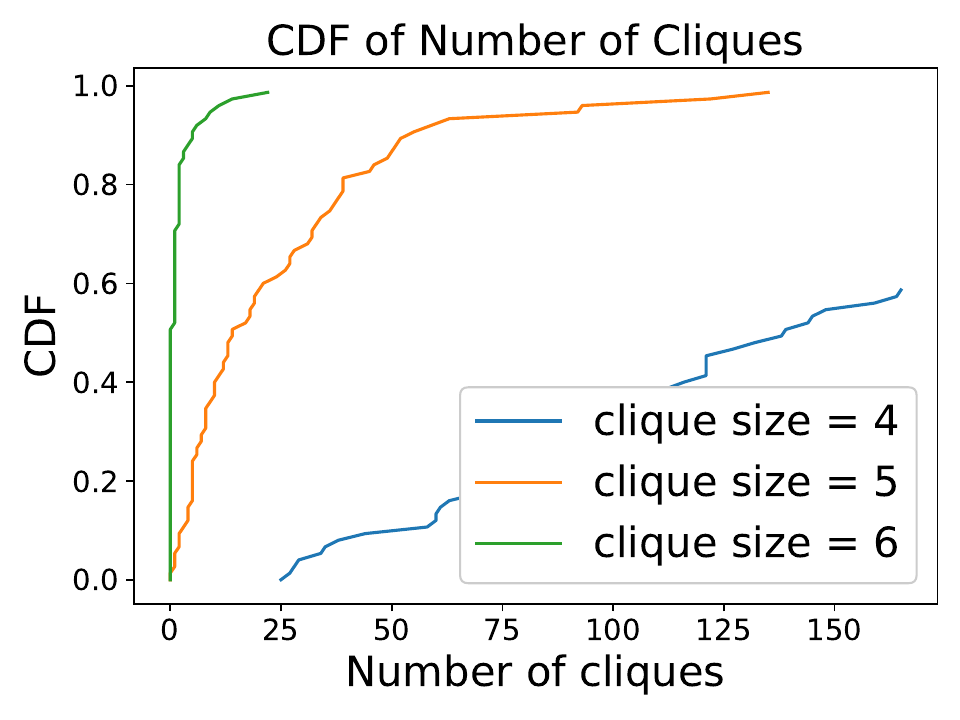} }}%
     \caption{\footnotesize{CDF of number of cliques for clique sizes $\ell \in \{4, 5, 6\}$ for the 75 Indian villages.} }%
\label{fig: cliquecdf}
 \end{figure}

We show cumulative distribution plots of the number of cliques (sizes 4, 5, and 6)  across the 75 villages in Figure~\ref{fig: cliquecdf}. 

\setcounter{figure}{0}
\setcounter{table}{0}
\section{Additional simulation results}
\label{app:curve}
\begin{figure}
\subfloat[Spherical]{\label{a}\includegraphics[width=.45\linewidth]{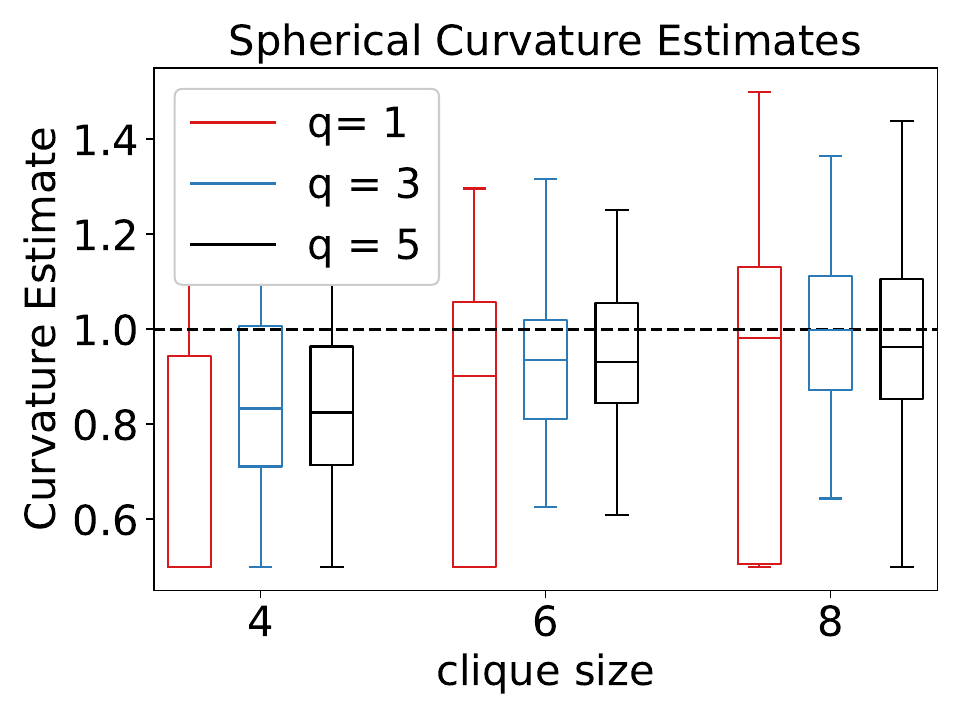}}\hfill
\subfloat[Hyperbolic]{\label{b}\includegraphics[width=.45\linewidth]{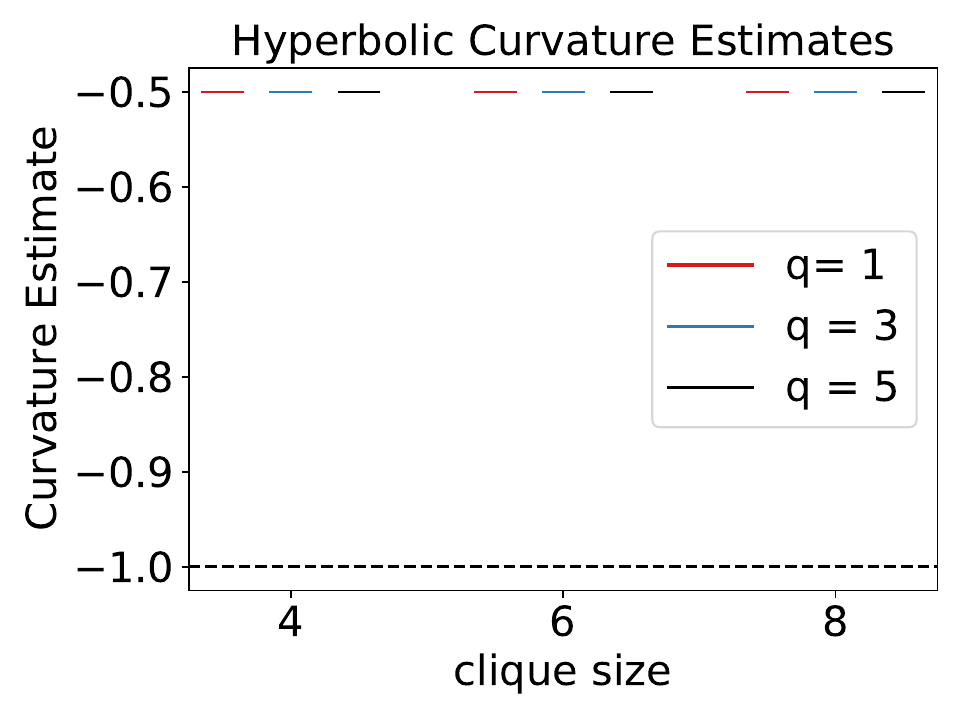}}
\caption{\footnotesize{Left: Curvature estimates for $\mathbf{S}^2(1)$ using $K = 10$ cliques, with clique size $\ell = 4, 6, 8$ on the horizontal axis. We use $q = 1, 3, 5$ where $q$ is defined in (\ref{eq: Q_Curvature_Sphere}). We plot the true curvature $\kappa = 1$ in the black dashed line. Right: Curvature estimates for $\mathbf{H}^2(-1)$ using $K = 10$ cliques, with clique size $\ell = 4, 6, 8$ on the horizontal axis. In Figure \ref{fig: H_curvature_vs_required_clique}, we analyze how large the clique size must be for the hyperbolic curvature estimator to perform better.}}
   \label{fig: Curvature}%
\end{figure}

We plot curvature estimates for 100 simulated graphs using cliques of size $\ell \in \{5, 7, 9\}.$ We see that as $\ell$ increases, the variance and bias of $\hat \kappa_S$ decreases in the spherical case (Figure \ref{fig: Curvature}).

\begin{figure}
    \centering
    \includegraphics[width= .45\linewidth]{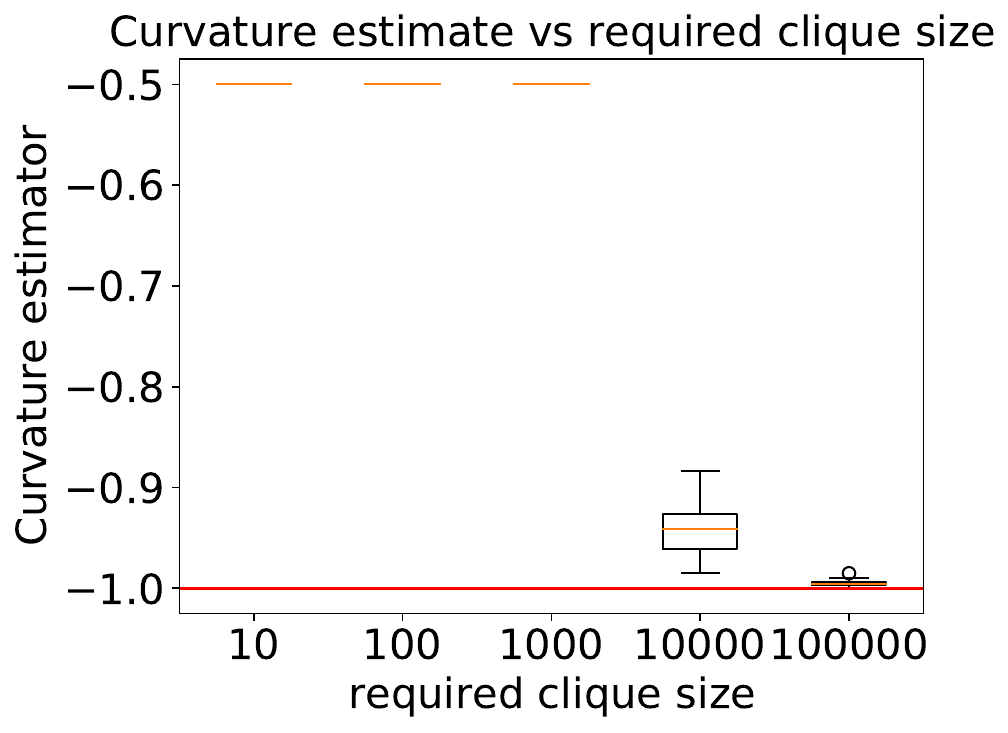}
    \caption{\footnotesize We plot the estimated curvature using distances computed from $K = 10$ points in $\mathbb{H}^2(-1)$ for various clique sizes on the $x$-axis. To simuluate this figure, we fix a set of $K = 10$ locations on $H^2(-1)$ and compute pairwise distances $d_{ij}$. We then simulate 50 independent noisy realizations $\hat d_{ij} \sim N(d_{ij}, \sigma^2)$ for $\sigma \in \{0.1, 0.01, 0.001, 0.0001, 0.00001\}$. We then compute the estimate of the curvature from (\ref{eq: curvature_estimator}) using $\hat D = \{\hat d_{ij}\}$. Given a certain noise level, we use that fact that when using cliques to estimate distances, the variance of $\hat d_{ij}$ is on the order of $1/\ell^2$. So we equate $\sigma^2 = 1/\ell^2$ to compute an approximate required clique sized required. For example, we require clique sizes of approximately $10^4$ or higher to obtain an estimator that does not always select the lower bound $a$ in (\ref{eq: curvature_estimator}). }
    \label{fig: H_curvature_vs_required_clique}
\end{figure}

We now analyze the accuracy of the curvature methods for the spherical and hyperbolic latent space models. 
The estimator in Proposition \ref{prop: ConsistencyKappa} minimizes $\kappa \mapsto \lambda_1(\kappa \hat W_\kappa )$. But from Lemma \ref{lem: embedding}, we in fact know that the first few eigenvalues of $\cos(\sqrt{\kappa} \hat D)$ are zero, which suggests that we can use the estimator 
\begin{equation}
\hat \kappa(q) = \frac{1}{q} \sum_{i= 1}^q \hat \kappa_i, \ \ \hat \kappa_i = \underset{\kappa \in [a, b]}{\text{arg min}} \Big|\lambda_i\left(\kappa W(\hat D)_\kappa \right)\Big|
\label{eq: Q_Curvature_Sphere}
\end{equation}
Assuming that $q << K$, we can reasonably believe that the first through $q$th eigenvalues of $\cos(\sqrt{\kappa} D)$ are zero. In fact, it is easy to modify the proof of Proposition \ref{prop: ConsistencyKappa} to show that $\hat \kappa(q) \overset{p}{\rightarrow} \kappa$, provided that $q << K.$ Taking $q > 1$ does not always reduce the variance of $\hat \kappa(t)$, which could be because the $\hat \kappa_1, \dotsc, \hat \kappa_q$ are not necessarily independent. In Figure \ref{fig: Curvature} we plot 250 estimates of $\kappa$ when $\mathcal{M}^p(\kappa) = \mathbf{S}^2(1)$ and when $\mathcal{M}^p(\kappa) = \mathbf{H}^2(-1)$ using $K = 10$. Specifically, we generate a network and find $K$ cliques of size 4, 5, 6. We estimate distances between these $K$ groups using the number of cross-clique edges as described in Algorithm \ref{alg: cliques}. Although Proposition \ref{prop: ConsistencyKappa} says that the estimate is consistent as the sample size grows, we see that the hyperbolic curvature estimate has not reached its asymptotic behavior for cliques of size 4, 5, and 6. In Figure \ref{fig: H_curvature_vs_required_clique} we determine how big the cliques must be in order for the Hyperbolic curvature estimator to be close to the true curvature. However, as we see in Figures \ref{fig:poweracrossK}, we see that the classification of geometry is still accurate, which is our ultimate goal. 

\setcounter{figure}{0}
\setcounter{table}{0}
\section{Sectional Curvature Definitions}
\label{sec: sec_curvature}

To discuss sectional curvature, some preliminary definitions are required. We review these concepts in a self-contained way. The reader may look to \citet{o1983semi} for a more in-depth explanation of these concepts. The \emph{tangent space} at $m \in \mathcal{M}^p$  is denoted $T_m(\mathcal{M}^p)$, defined as the set of all tangent vectors to the manifold at $m$: that is, all real-valued functions $v$ that map any smooth function $f:\mathcal{M}^p \rightarrow \mathbb{R}$   to $ v(f(m)) \in \mathbb{R}$ that is $\mathbb{R}$-linear and Leibnizian.\footnote{An obvious tangent vector is the directional derivative at a point on the manifold: it maps a smooth function to its derivative in that direction evaluated at that point on the manifold.}

A Riemannian manifold  $(\mathcal{M}^p,g)$ comes equipped with a \emph{metric tensor} $g$  which at every point $m \in \mathcal{M}^p$ takes two vectors in the tangent space of the manifold at $m$, $u,v\in T_m(\mathcal{M}^p)$, and maps it to a non-negative number: $g_m(u,v)\mapsto \mathbb{R}_{\geq 0}$ and the map is symmetric, non-degenerate, and bilinear. That is, $g$ defines a scalar product over the manifold; on a smooth manifold the metric tensor smoothly varies over the manifold itself.

To define curvature, we first need to define the \emph{Riemann curvature tensor}, $R$ evaluated at point $m \in \mathcal{M}^p$, which takes three tangent vectors in the tangent space at $m$---$u,v,w\in T_m(\mathcal{M}^p)$---and returns $R_m(u,v)w \in T_m(\mathcal{M}^p)$\footnote{Here $[.,.]$ is the Lie bracket.}
\[
R_m(u,v)w := \nabla_{[u,v]}w - [\nabla_u,\nabla_v]w.
\]
Here is some intuition. Consider the vector $w$ which is tangent to the manifold at $m$. Consider the plane defined by $u$ and $v$ which are tangent at $m$ as well. Now take $w$ and parallel transport it, meaning take it along the  parallelogram in the $u$ direction and then $v$ direction and compare that to taking the same $w$ along the $v$ direction and then $u$ direction to the same point. The returned vector has entries that describe how much $w$ changes relatively across the two paths. If this is identically zero, this means of course that there was no change in this parallel transportation. Intuitively, if one does this on a flat manifold, for instance $\mathbb{R}^2$ with the usual Euclidean metric, it is clear that the vector $w$ does not change whatsoever. But on a sphere, for instance, the  reader can intuit  that things change.

Then the \emph{sectional curvature} at $m$, which we refer to simply as curvature for the remainder of this paper, is given by
\[
\kappa_m(u,v):= \frac{g_m(R_m(u,v)v,u)}{g_m(u,u) \cdot g_m(v,v) - g_m(u,v)^2}.
\]
It turns out that this is independent of basis $u,v$ whatsoever (see Lemma 39 in \citet{o1983semi} for instance) so we can simply write $\kappa_m$.  That the manifold has constant sectional curvature means that for all $m\in \mathcal{M}^p$, $\kappa_m = \kappa$ and so we simply write $\mathcal{M}^p(\kappa)$.

\setcounter{figure}{0}
\setcounter{table}{0}
\section{Lattice simulations}
\label{sec: lattice_sims}
\begin{figure}[H]
\minipage{0.32\textwidth}
  \includegraphics[width=\linewidth]{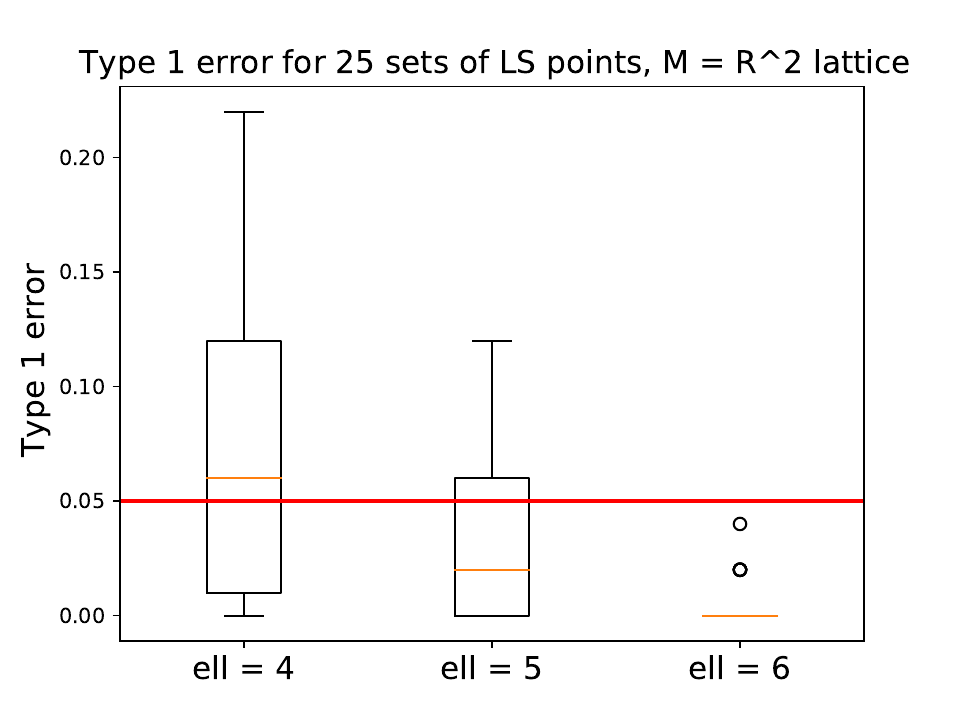}   
\endminipage\hfill
\minipage{0.32\textwidth}
  \includegraphics[width=\linewidth]{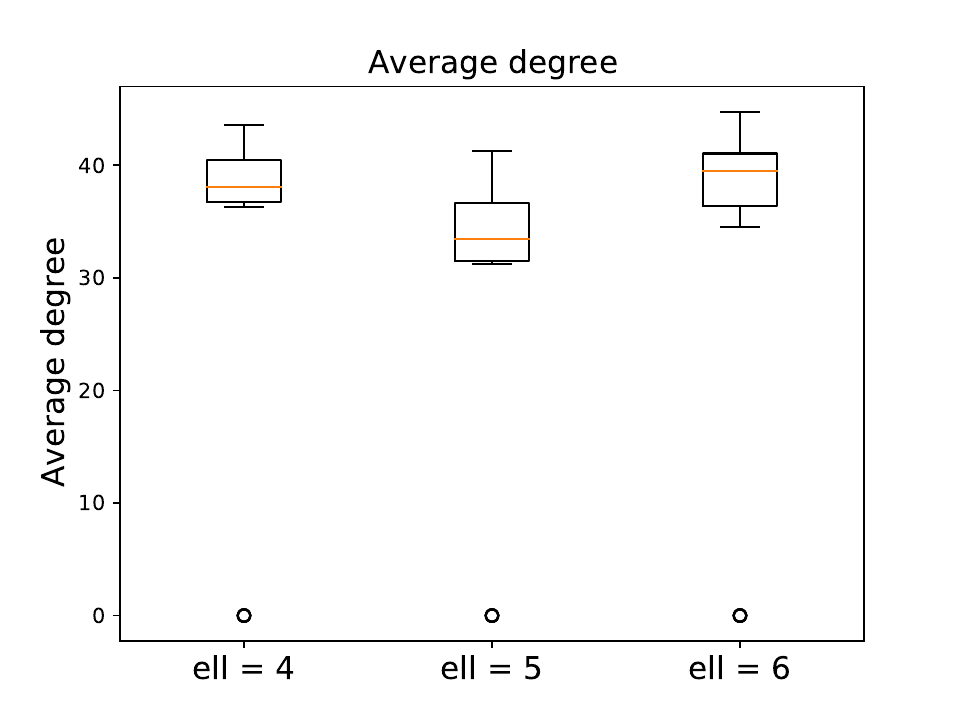}
\endminipage\hfill
\minipage{0.32\textwidth}%
  \includegraphics[width=\linewidth]{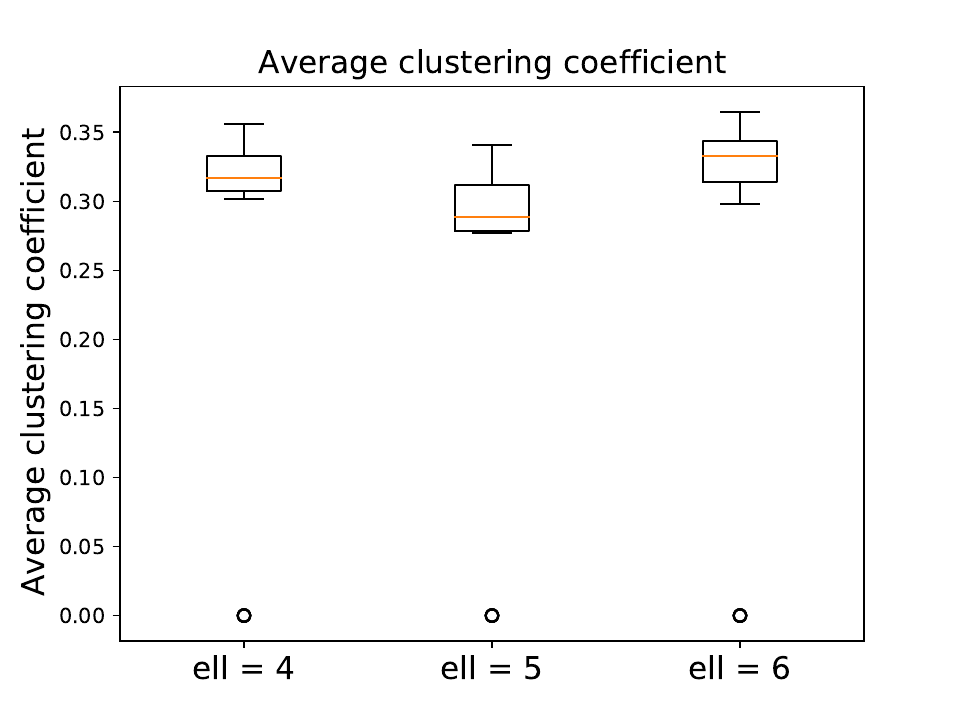}
 
\endminipage
\caption{Simulation results for the two-dimensional lattice. Using a $4 \times 4$ lattice in $\mathbb{R}^2$, we randomly select 25 sets of 9 latent space positions. For each set of positions, we generate 50 networks from using the graph model in (\ref{eq:main_model}) and calculate how many of these 50 networks we reject the null hypothesis that $\mathcal{M}$ is Euclidean. We repeat this for all 25 sets of latent space positions and plot the resulting probability of type 1 error for $\ell = 4, 5, 6$. We see that the type 1 error is at $\alpha = 0.05$ or below and decreases as $\ell$ increases. We also report the average degree (middle figure) and average clustering coefficient for the simulated networks. We use $\tau = 0.4$ and $\beta= -0.6$.\label{fig:lattice1}}
\end{figure}

\begin{figure}[H]
\minipage{0.32\textwidth}
  \includegraphics[width=\linewidth]{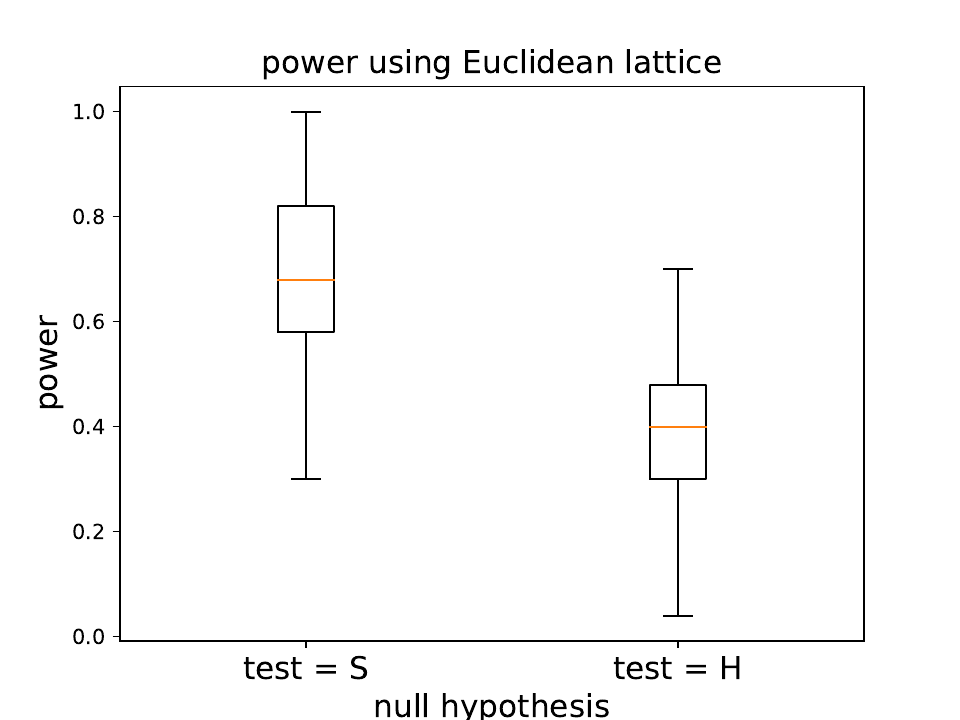}   
\endminipage\hfill
\minipage{0.32\textwidth}
  \includegraphics[width=\linewidth]{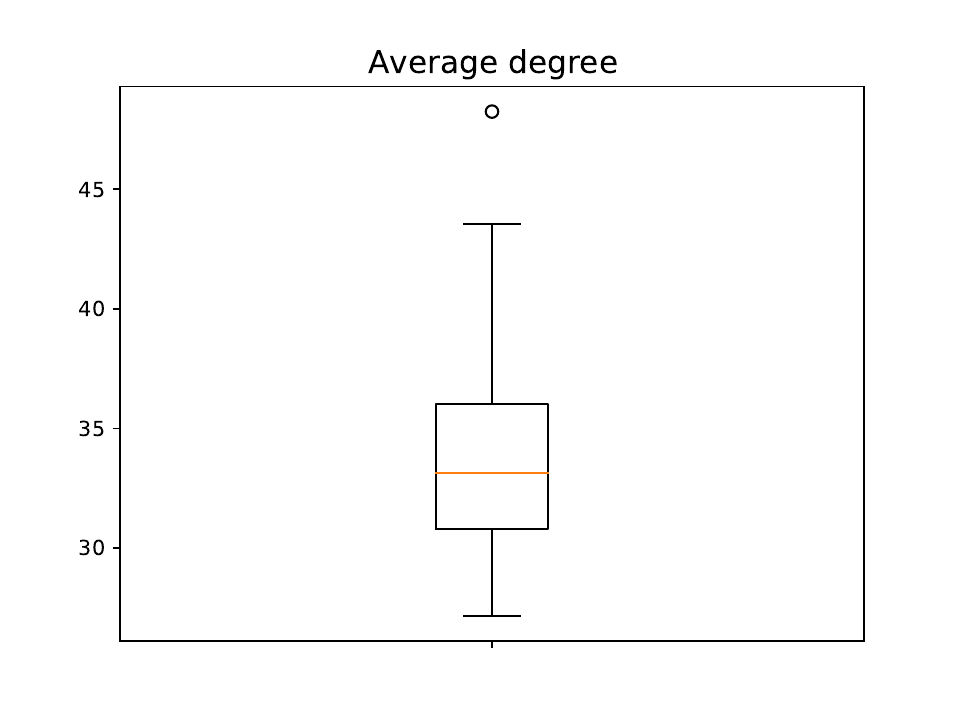}
\endminipage\hfill
\minipage{0.32\textwidth}%
  \includegraphics[width=\linewidth]{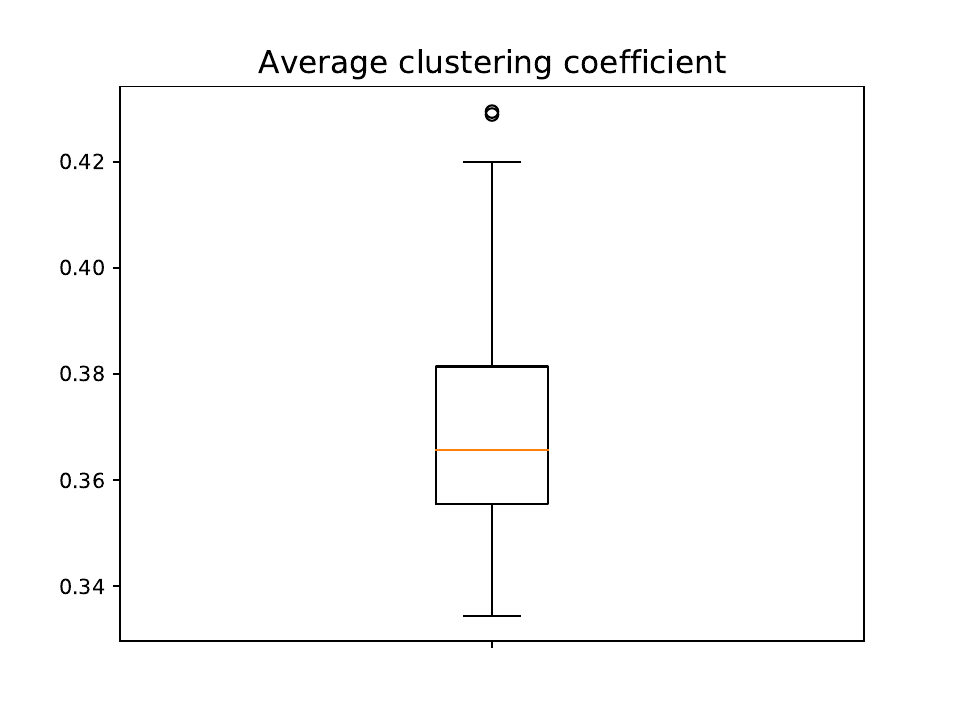}
 
\endminipage
\caption{Simulation results for the two-dimensional lattice. Using a $4 \times 4$ lattice in $\mathbb{R}^2$, we randomly select 25 sets of 9 latent space positions. For each set of positions, we generate 50 networks from using the graph model in (\ref{eq:main_model}) and calculate how many of these 50 networks we reject the null hypothesis that $\mathcal{M}$ is spherical or hyperbolic. We repeat this for all 25 sets of latent space positions and plot the resulting power.  We also report the average degree (middle figure) and average clustering coefficient for the simulated networks. We use $\tau = 0.4$.}
\label{sec: Euclidean_lattice_Type1}
\end{figure}

\begin{figure}[H]
\minipage{0.32\textwidth}
  \includegraphics[width=\linewidth]{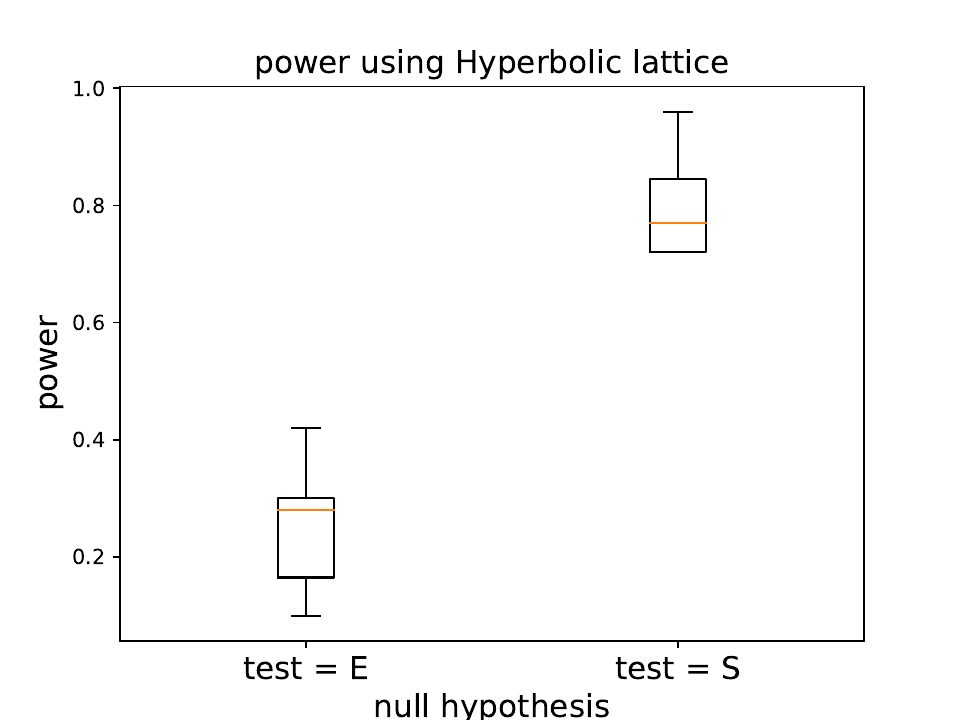}   
\endminipage\hfill
\minipage{0.32\textwidth}
  \includegraphics[width=\linewidth]{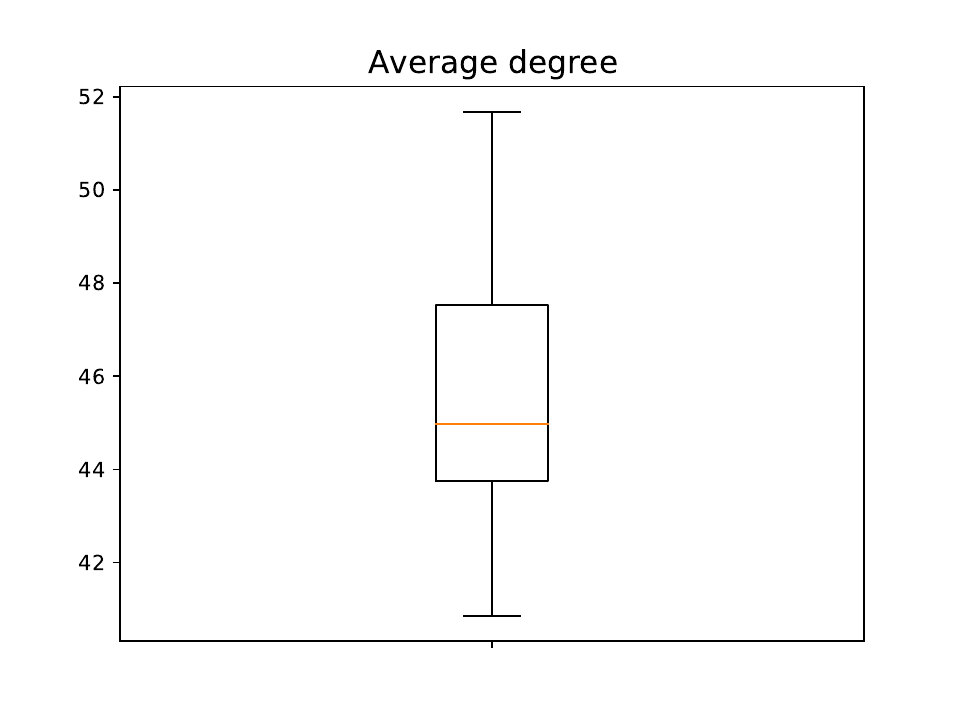}
\endminipage\hfill
\minipage{0.32\textwidth}%
  \includegraphics[width=\linewidth]{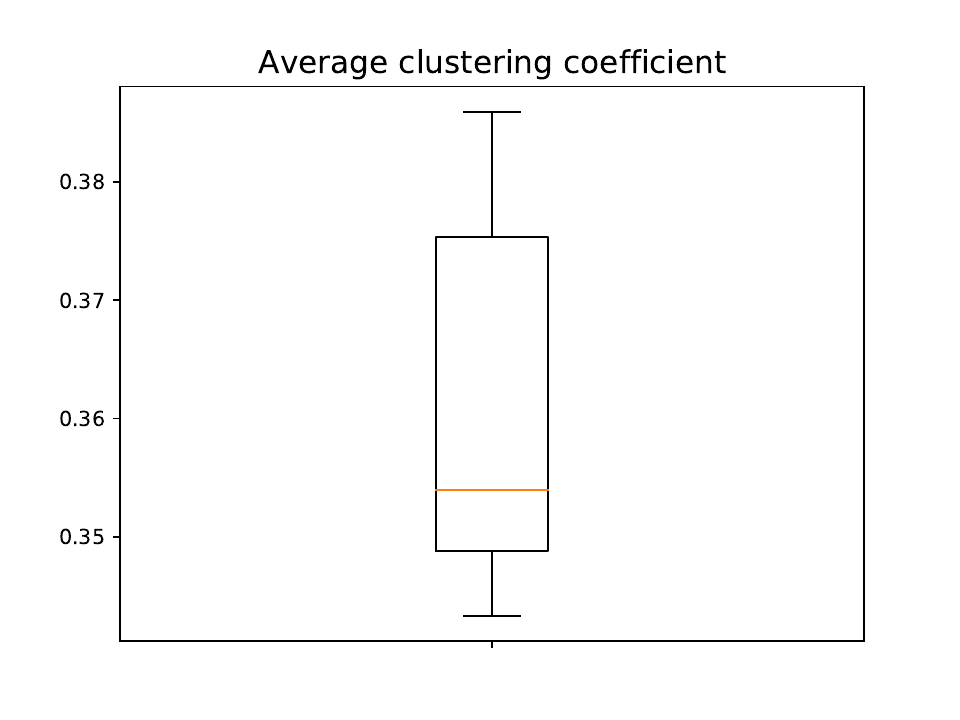}
 
\endminipage
\caption{Simulation results for the two-dimensional lattice. Using a $5 \times 5$ lattice in $\mathbb{H}^2(-1)$, we randomly select 25 sets of 9 latent space positions. For each set of positions, we generate 50 networks from using the graph model in (\ref{eq:main_model}) and calculate how many of these 50 networks we reject the null hypothesis that $\mathcal{M}$ is Euclidean. We repeat this for all 25 sets of latent space positions and plot the resulting power. We also report the average degree (middle figure) and average clustering coefficient for the simulated networks. We use $\beta= -0.2$.\label{fig:latticelast}}
\end{figure}

We now demonstrate the type 1 error and power simulations by drawing points on a lattice in Euclidean, spherical, and hyperbolic space. We present the results in Figures~\ref{fig:lattice1}-\ref{fig:latticelast}. 

\begin{figure}[H]
\minipage{0.32\textwidth}
  \includegraphics[width=\linewidth]{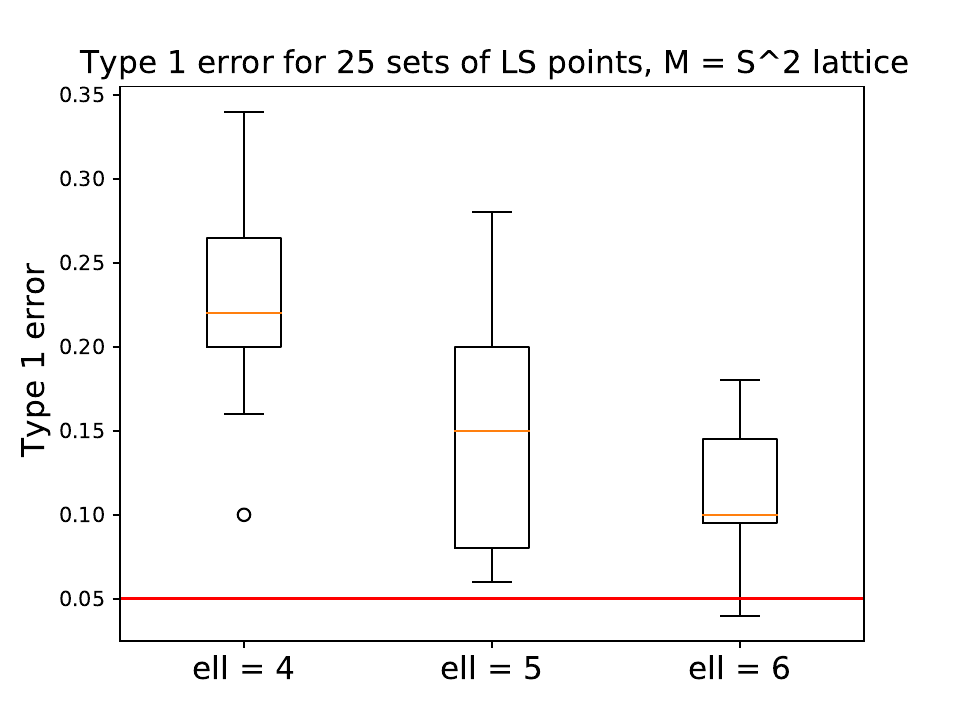}   
\endminipage\hfill
\minipage{0.32\textwidth}
  \includegraphics[width=\linewidth]{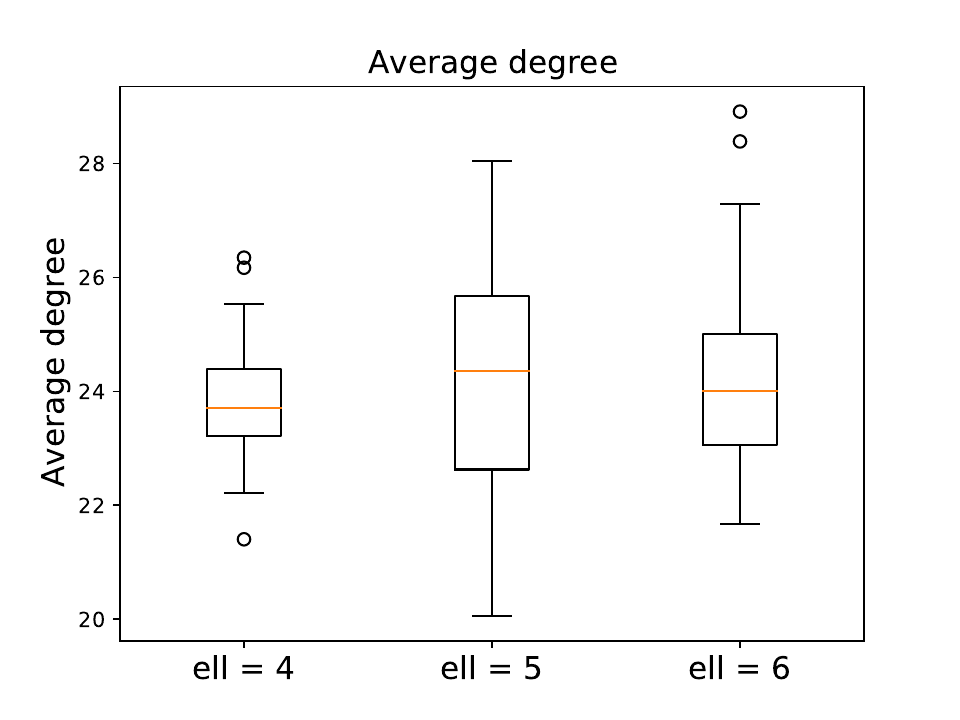}
\endminipage\hfill
\minipage{0.32\textwidth}%
  \includegraphics[width=\linewidth]{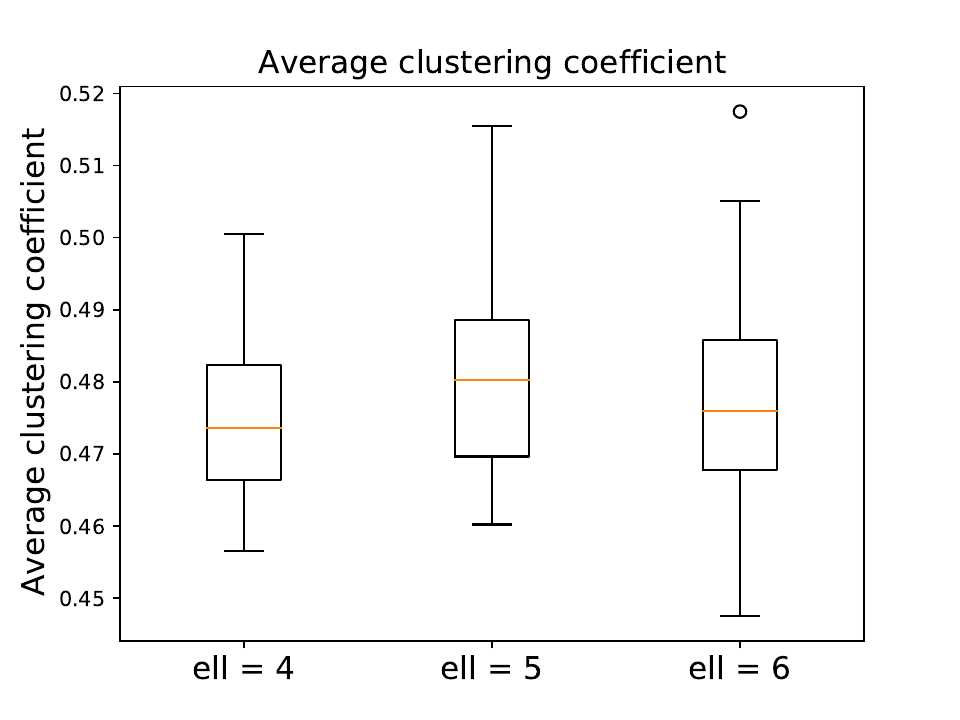}
 
\endminipage
\caption{Simulation results for the two-dimensional lattice. Using a $3 \times 3$ lattice in $\mathbb{S}^2(1)$, we randomly select 25 sets of 4 latent space positions. For each set of positions, we generate 50 networks from using the graph model in (\ref{eq:main_model}) and calculate how many of these 50 networks we reject the null hypothesis that $\mathcal{M}$ is Euclidean. We repeat this for all 25 sets of latent space positions and plot the resulting probability of type 1 error for $\ell = 4, 5, 6$. We see that the type 1 error is above $\alpha = 0.05$ but  decreases to about 0.1 as $\ell$ increases. We also report the average degree (middle figure) and average clustering coefficient for the simulated networks. We use $\beta= -0.2$.}
\end{figure}

\begin{figure}[H]
\minipage{0.32\textwidth}
  \includegraphics[width=\linewidth]{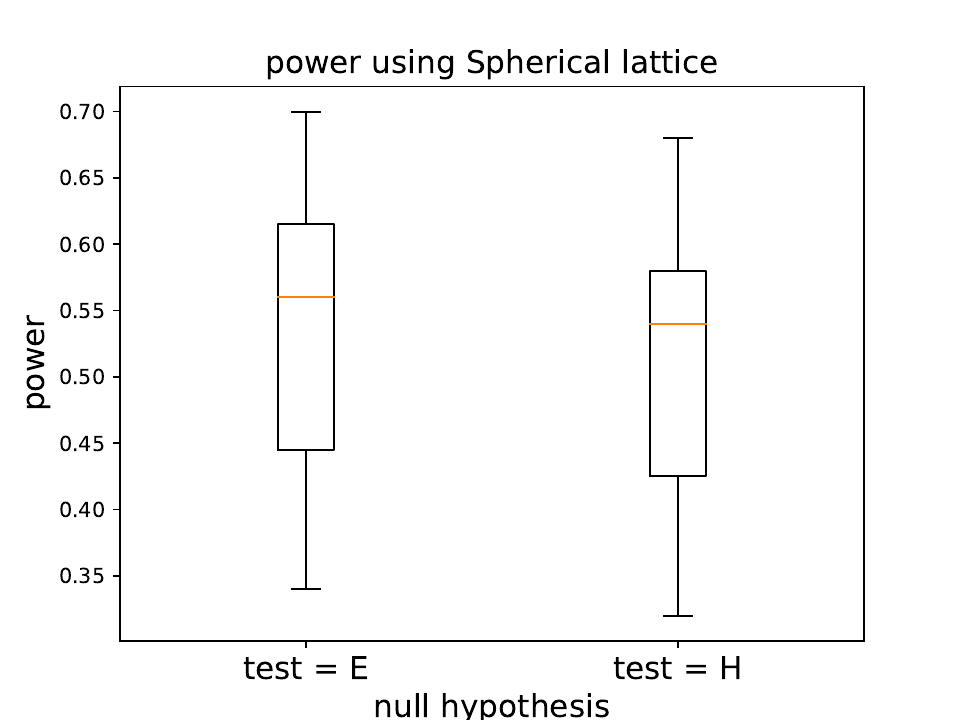}   
\endminipage\hfill
\minipage{0.32\textwidth}
  \includegraphics[width=\linewidth]{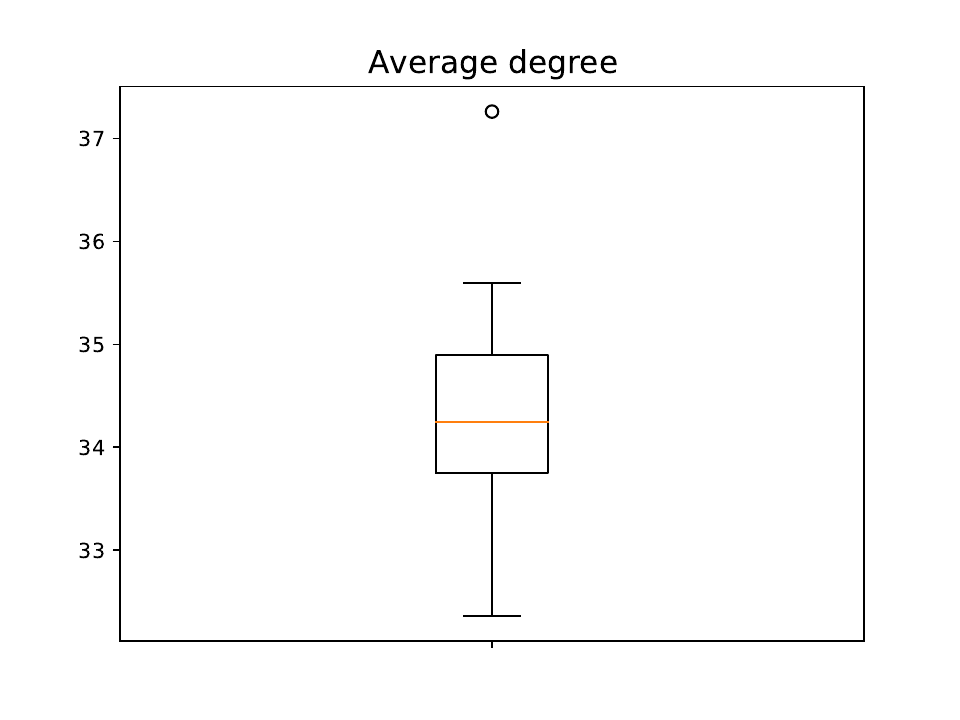}
\endminipage\hfill
\minipage{0.32\textwidth}%
  \includegraphics[width=\linewidth]{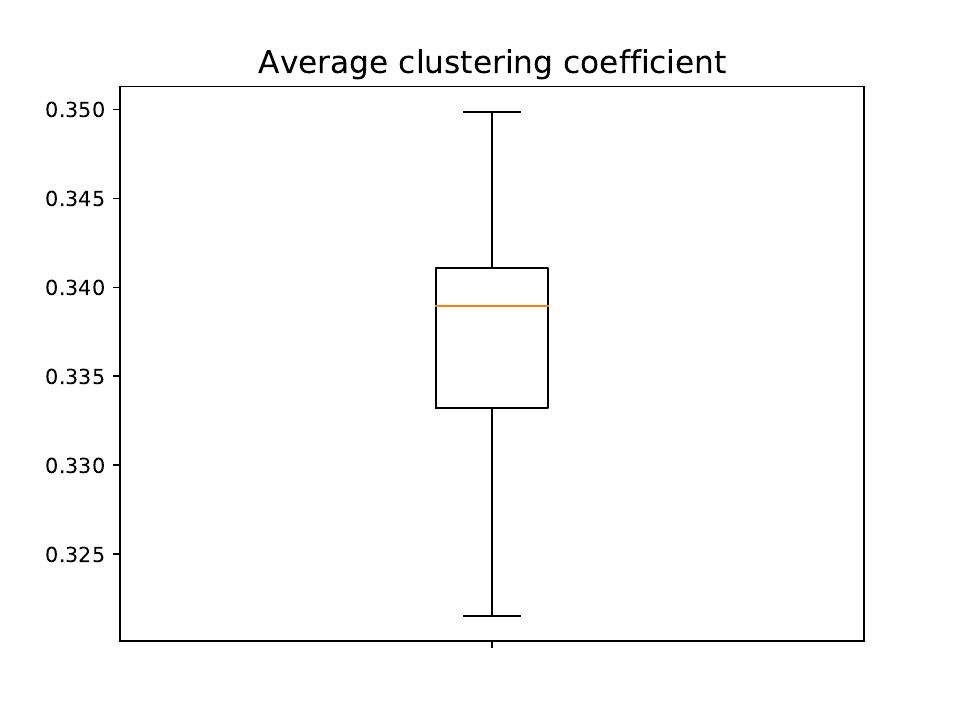}
 
\endminipage
\caption{Simulation results for the two-dimensional lattice. Using a $5 \times 5$ lattice in $\mathbb{S}^2$, we randomly select 25 sets of 9 latent space positions. For each set of positions, we generate 50 networks from using the graph model in (\ref{eq:main_model}) and calculate how many of these 50 networks we reject the null hypothesis that $\mathcal{M}$ is Euclidean. We repeat this for all 25 sets of latent space positions and plot the resulting power. We also report the average degree (middle figure) and average clustering coefficient for the simulated networks. We use $\beta= -0.2$.}
\end{figure}

\begin{figure}[H]
\minipage{0.32\textwidth}
  \includegraphics[width=\linewidth]{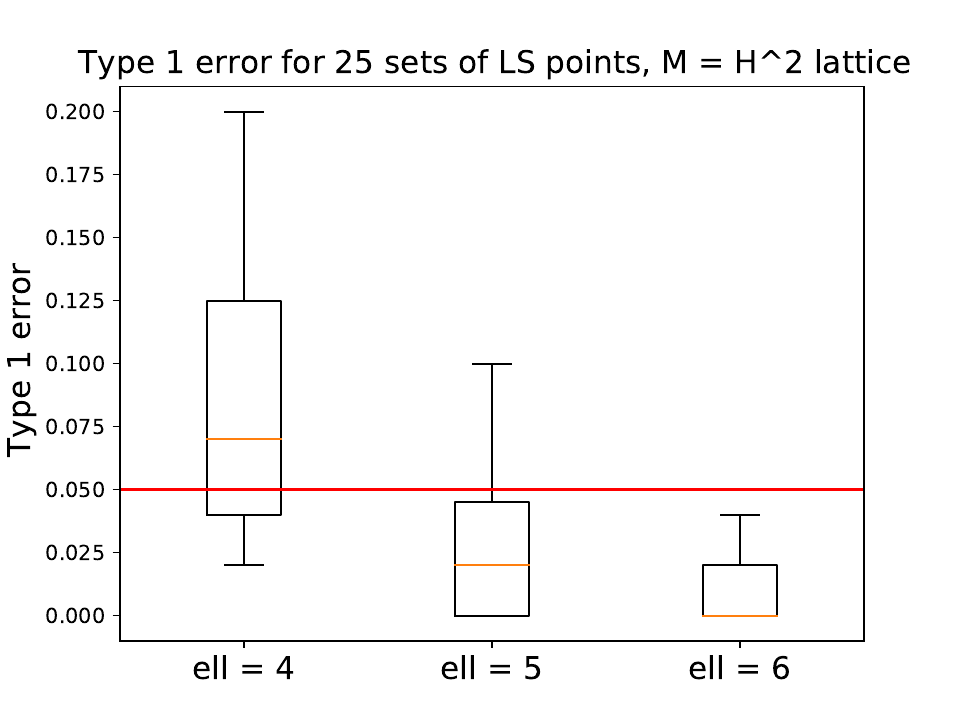}   
\endminipage\hfill
\minipage{0.32\textwidth}
  \includegraphics[width=\linewidth]{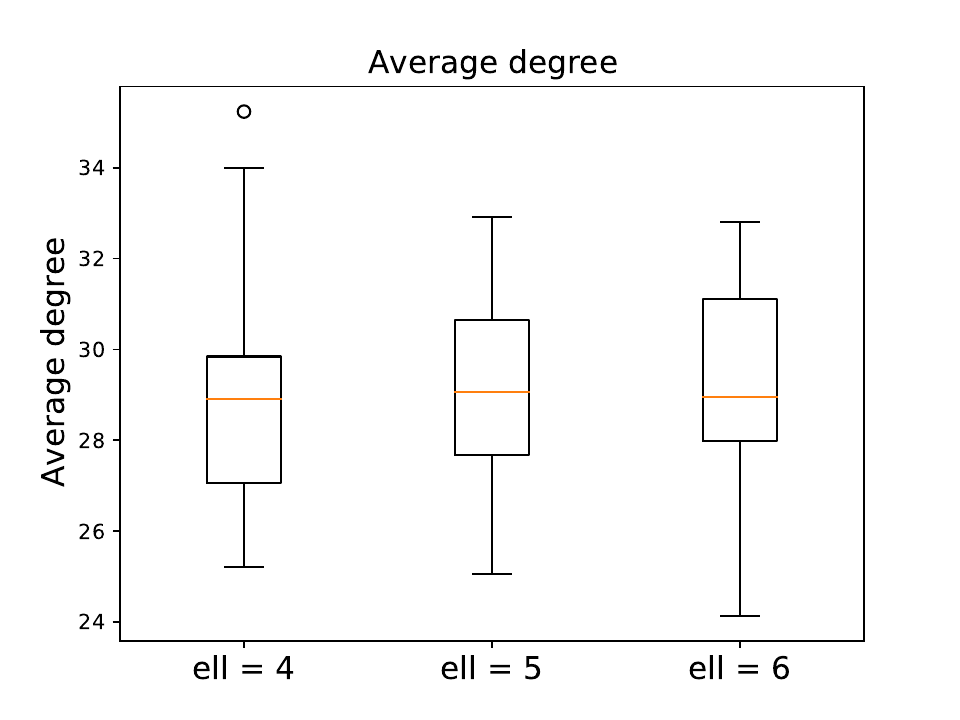}
\endminipage\hfill
\minipage{0.32\textwidth}%
  \includegraphics[width=\linewidth]{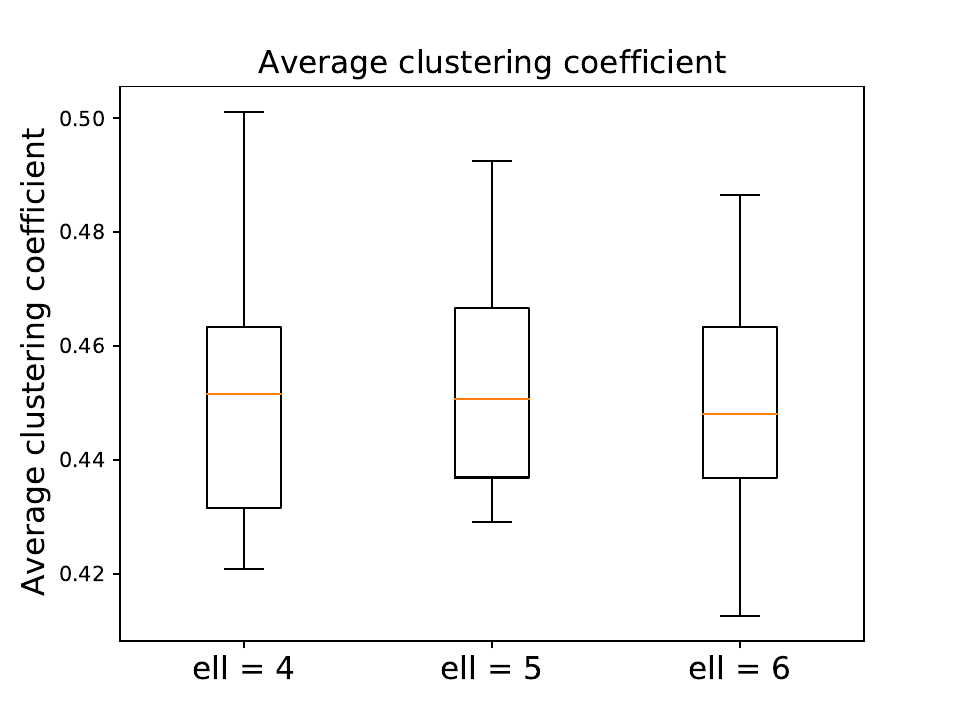}
 
\endminipage
\caption{Simulation results for the two-dimensional lattice. Using a $5 \times 5$ lattice in $\mathbb{H}^2(-1)$, we randomly select 25 sets of 9 latent space positions. For each set of positions, we generate 50 networks from using the graph model in (\ref{eq:main_model}) and calculate how many of these 50 networks we reject the null hypothesis that $\mathcal{M}$ is Euclidean. We repeat this for all 25 sets of latent space positions and plot the resulting power. We also report the average degree (middle figure) and average clustering coefficient for the simulated networks. We use $\beta= -0.2$.}
\end{figure}

\setcounter{table}{0}
\setcounter{figure}{0}

 \section{Other graph models}
 \label{sec: other_graph_models}
 In (\ref{eq:main_model}), we consider an exponential link function which connects distances in the latent space to the probability that nodes form an edge. The exponential function has the desirable property that $\exp(a+b) = \exp(a)\exp(b)$ which allows us to isolate the node effects and distances:
 \begin{equation*}
     \Pr(g_{ij} = 1 | \nu_i^\star, \nu_j^\star, z_i^\star, z_j^\star) = \exp(\nu_i^\star + \nu_j^\star) \exp\{-d(z_i^\star, z_j^\star)\} \;.
 \end{equation*}
So by integrating out the node effects, we see that
\begin{equation*}
     \Pr(g_{ij} = 1 |z_i^\star, z_j^\star) = E\{\exp(\nu)\}^2 \exp\{-d(z_i^\star, z_j^\star)\} \;.
\end{equation*}
We now illustrate how our approach can be applied to other graph models. 
\begin{example}[Expit link function]
Suppose instead of the exponential link function, we use the expit link function, which was used in \cite{hoffrh2002}, among many others:
\begin{equation*}
    \Pr(g_{ij} = 1 |  \nu_i^\star, \nu_j^\star, z_i^\star, z_j^\star) = \text{expit}\{\nu_i^\star + \nu_j^\star - d(z_i^\star, z_j^\star)\} \;,
\end{equation*}
where $\text{expit}(x) = \exp(x)/\{1 + \exp(x)\}$ is the expit function.
\end{example} 
The choice of which link function to choose is an important one, which network goodness-of-fit tests can help address \citep{Ouadah, Gesine, lubold2021spectral}. 

 Now, after integrating out the node effects, we have 
\begin{equation*}
     \Pr(g_{ij} = 1 |  \nu_i^\star, \nu_j^\star, z_i^\star, z_j^\star) = \int \int \text{expit}\{\nu_i^\star + \nu_j^\star - d(z_i^\star, z_j^\star)\} dF(\nu_i^\star) dF(\nu_j^\star) \;,
\end{equation*}
which again assumes that the node effects are drawn independently of each other. Let us define a function $H : [0, \infty) \rightarrow [0, \infty)$ with 
\begin{equation*}
    H(x) := \int \int \text{expit}\{\nu_i^\star + \nu_j^\star -x\} dF(\nu_i^\star) dF(\nu_j^\star) \;.
\end{equation*}
Using Monte Carlo integration, we can approximate $H$ to within any desired certainty. Now, suppose that $C_k$ and $C_{k'}$ are two cliques in the graph $G$ drawn from this graph model. Then, 
\begin{equation*}
    \hat p_{kk'} := \frac{1}{\ell^2} \sum_{i \in C_k} \sum_{j \in C_{k'}} G_{ij} \approx H\{d(z_k, z_{k'})\} \;.
\end{equation*}
We can then solve for the argument of $H$ that solve the above expression. That is, assuming that $H$ is invertible, we can write
$\hat d_{kk'} = H^{-1}(\hat p_{kk'}).$ Assuming $H^{-1}$ is continuous, we can estimate distances consistently.

\begin{example}[Latent space model with covariates]
Consider the model given in \cite{hoffrh2002}:
\begin{equation*}
    \Prob(g_{ij} = 1 | \alpha^\star, \beta^\star, z_i^\star, z_j^\star) = \text{expit}\{\alpha^\star + \beta^\star X_{ij} - d(z_i^\star, z_j^\star)\} \;.
\end{equation*}
where $\alpha^\star$ measures the baseline probability of connecting, $X_{ij}$ is a dyad-level covariate, $\beta^\star$ measures the effect of this covariate on the probability of edges, and $d$ measures distances in the latent space.
\end{example}
To illustrate how estimate parameters in this model, suppose that $X_{ij}$ is measuring homophily, so that $X_{ij}$ is 1 if nodes $i$ and $j$ share some common trait (like ethnicity, education level, political beliefs, etc) and is 0 otherwise. Suppose also that covariates are observable. 

Suppose we observe a clique of nodes with the same trait. Then, 
\begin{equation}
\label{eq: edge_prob_clique_same_trait}
    \frac{1}{{\ell \choose 2}} \sum_{i < j} G_{ij} \approx \text{expit}(\alpha^\star + \beta^\star) \;,
\end{equation}
since nodes in the same clique are likely to be close to each other and therefore $d(z_i^\star, z_j^\star) = 0$ for $i$ and $j$ in the same clique (Assumption \ref{ass:z}).

Suppose we also observe an $\ell$-clique $C(\ell)$ with nodes that have different traits. Partition these nodes into two groups, $C_{0}(\ell)$ and $C_{1}(\ell)$ where the subscript 0 and 1 indicates whether the covariate is 0 or 1, so that $C(\ell) = C_0(\ell) \cup C_1(\ell)$. 
Then, 
\begin{equation*}
   (|C_{0}(\ell)| |C_{1}(\ell)|)^{-1} \sum_{i \in C_{0}(\ell)} \sum_{j \in C_{1}(\ell)} G_{ij} \approx \text{expit}(\alpha^\star) \;.
\end{equation*}
Thus, we can estimate $\alpha^\star$ by solving the above equation. Using this estimate, we can then plug this value into (\ref{eq: edge_prob_clique_same_trait}) and solve for $\hat \beta$. 

Given estimates of $\alpha^\star$ and $\beta^\star$, we can now estimate distances between cliques. To do this, we define for any two cliques $C_k$ and $C_{k'}$, where all nodes have the same traits, 
\begin{equation*}
    \hat P_{kk'} :=  (|C_{k}(\ell)| |C_{k'}(\ell)|)^{-1} \sum_{i \in C_{0}(\ell)} \sum_{j \in C_{1}(\ell)} G_{ij} \;.
\end{equation*}
where $C_k(\ell)$ and $C_{k'}(\ell)$ are cliques of size $\ell$. We can then solve for $\hat D_{kk'}$, 
\begin{equation*}
    \hat D_{kk'} = \text{logit}(\hat P_{kk'}) - \hat \alpha - \hat \beta \;.
\end{equation*}

\setcounter{table}{0}
\setcounter{figure}{0}
\section{Existence of cliques and locations of nodes in a clique}
\label{sec: existence_of_cliques}
\begin{figure}
    \centering
       \subfloat[Estimated probability of $\log(n)$ clique]{{\includegraphics[scale = 0.3]{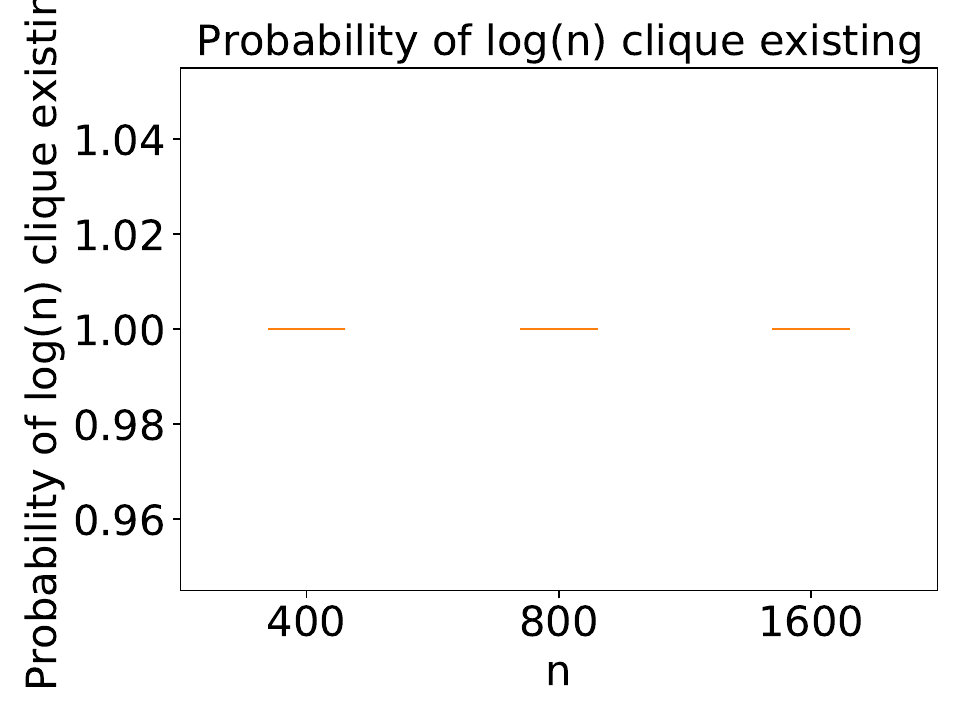} }}%
     \subfloat[Average degree]{{\includegraphics[scale = 0.3]{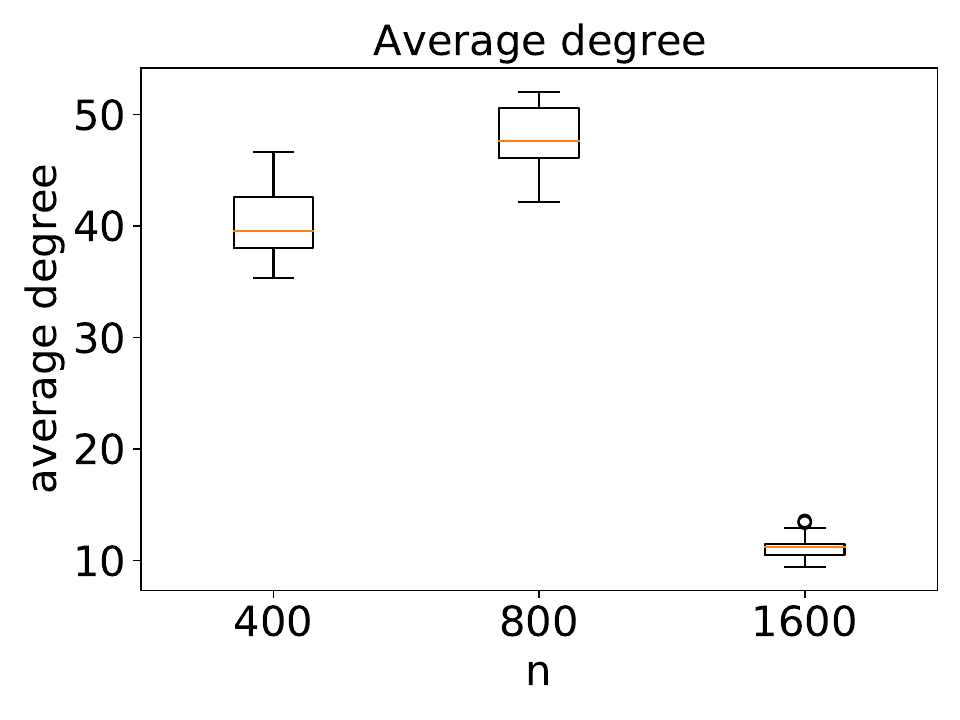} }}
    \subfloat[Average clustering coefficient]{{\includegraphics[scale = 0.3]{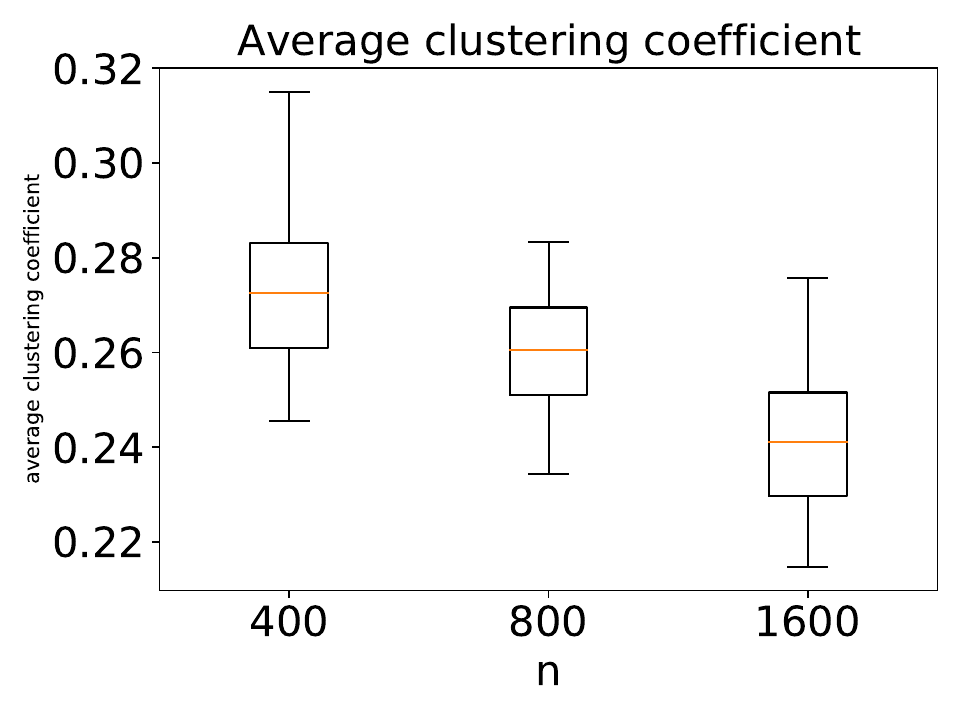} }}%
    \caption{\footnotesize{ We generate 25 sets of $n$ latent space positions using the lattice model. For each set of LS positions, we generate 50 graphs and count the number of times a clique of size $\log(n)$ exists in the graph. For each set of LS positions, we record the average degree for the 50 graphs and plot the 25 average values in (B). Similarly, in (C) we plot the average clustering coefficient.  }}%
    \label{fig: p_clique_Lattice}%
\end{figure}

\begin{figure}
    \centering
       \subfloat[Estimated probability of $\log(n)$ clique]{{\includegraphics[scale = 0.3]{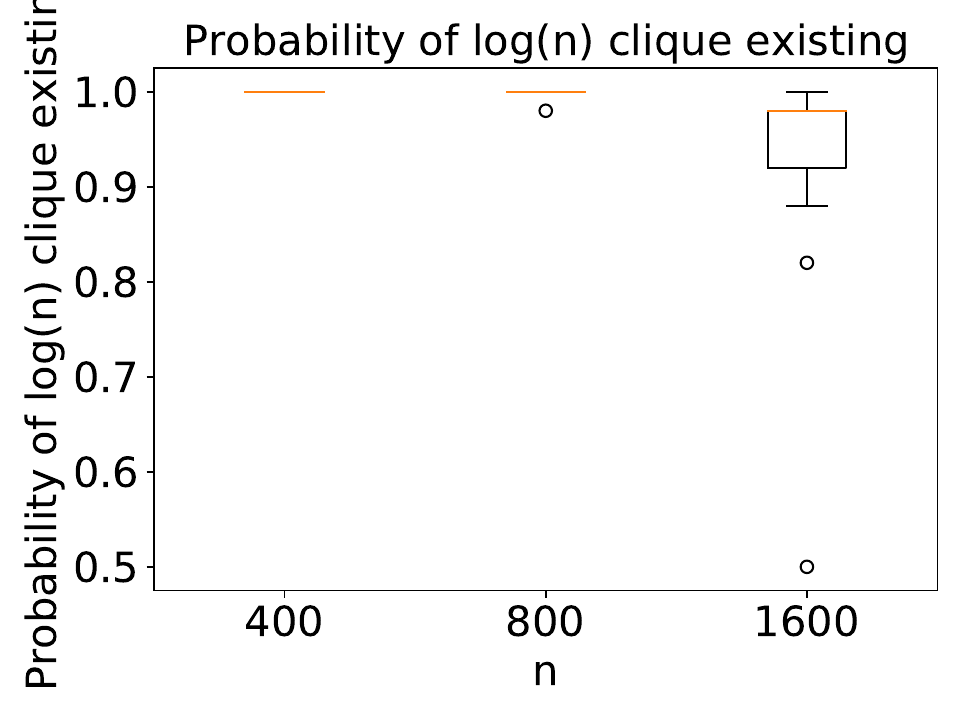} }}%
     \subfloat[Average degree]{{\includegraphics[scale = 0.3]{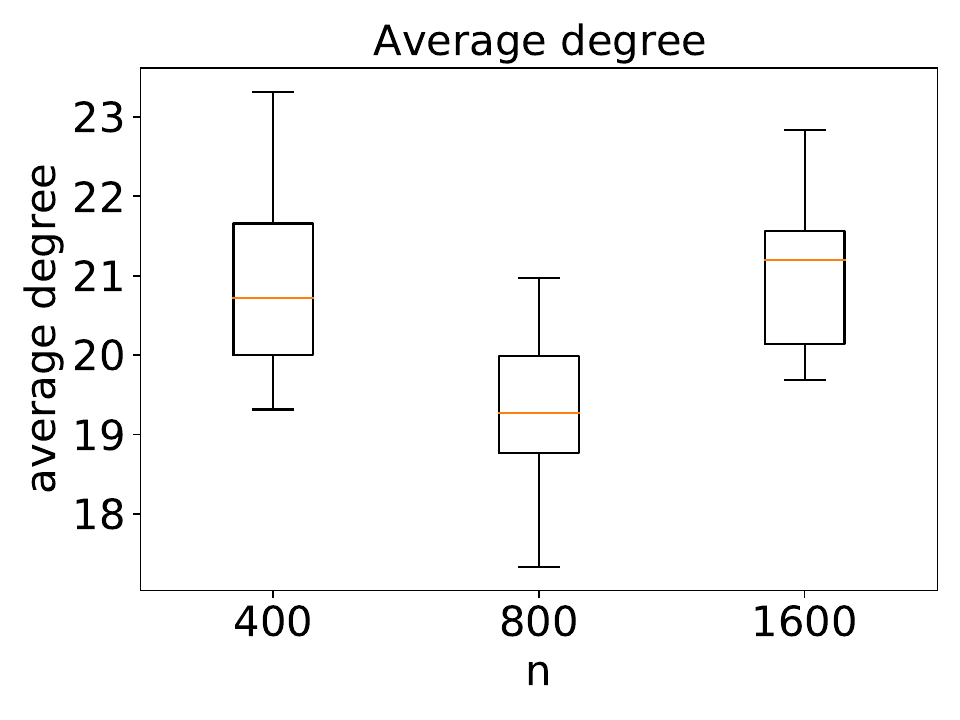} }}
    \subfloat[Average clustering coefficient]{{\includegraphics[scale = 0.3]{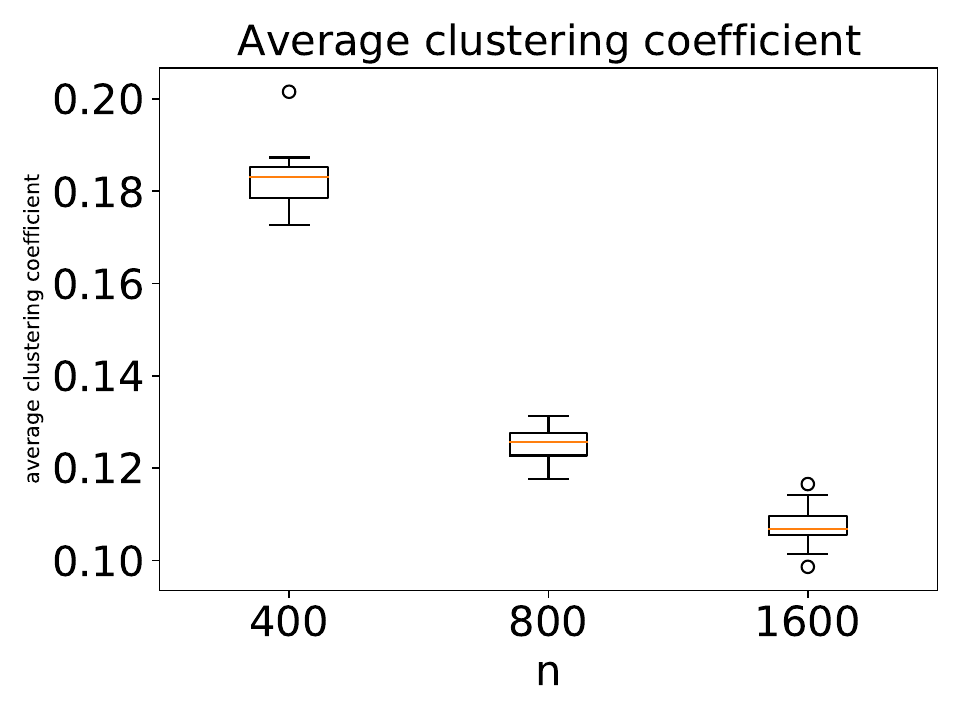} }}%
    \caption{\footnotesize{ We generate 25 sets of $n$ latent space positions using the Gaussian mixture model. For each set of LS positions, we generate 50 graphs and count the number of times a clique of size $\log(n)$ exists in the graph. For each set of LS positions, we record the average degree for the 50 graphs and plot the 25 average values in (B). Similarly, in (C) we plot the average clustering coefficient.  }}%
    \label{fig: p_clique_GMM}%
\end{figure}

\begin{figure}
    \centering
       \subfloat[Estimated probability of $\log(n)$ clique]{{\includegraphics[scale = 0.3]{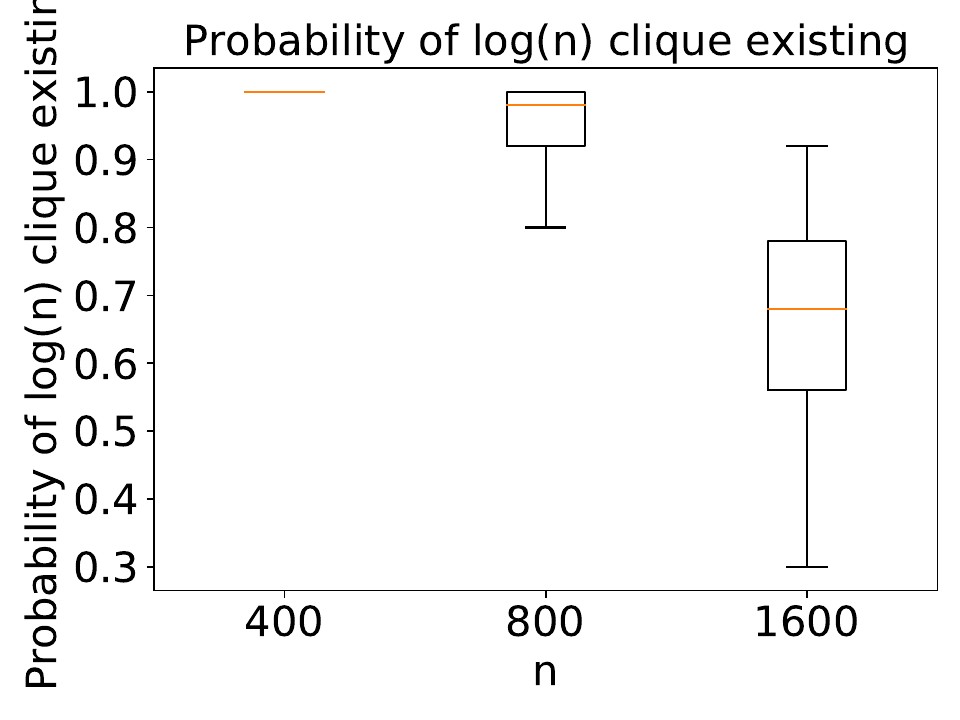} }}%
     \subfloat[Average degree]{{\includegraphics[scale = 0.3]{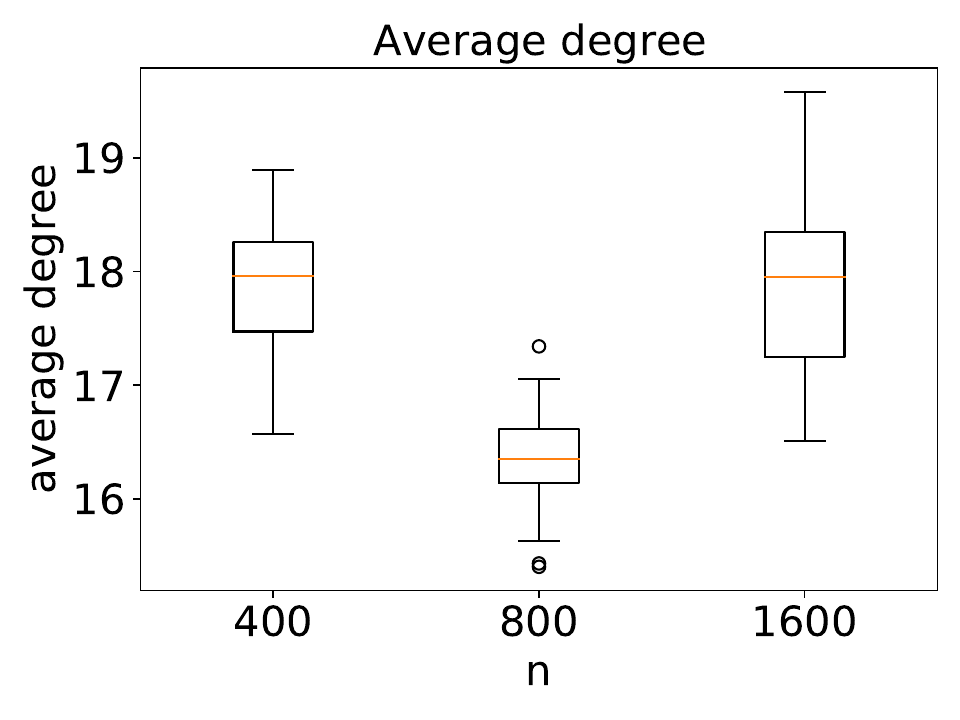} }}
    \subfloat[Average clustering coefficient]{{\includegraphics[scale = 0.3]{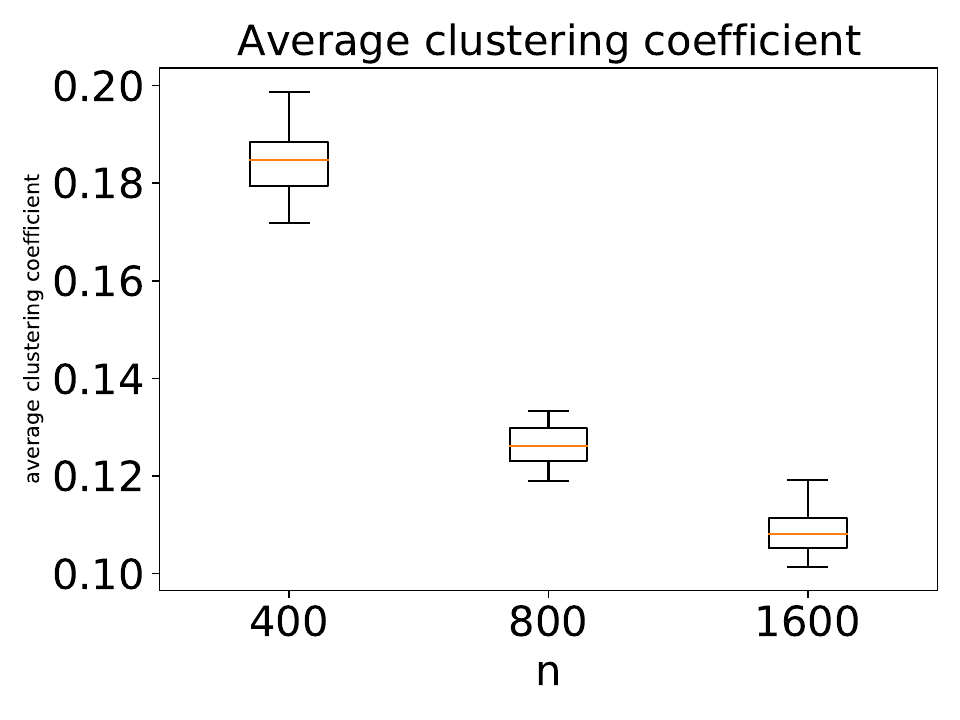} }}%
    \caption{\footnotesize{ We generate 25 sets of $n$ latent space positions from a uniform distribution. For each set of LS positions, we generate 50 graphs and count the number of times a clique of size $\log(n)$ exists in the graph. For each set of LS positions, we record the average degree for the 50 graphs and plot the 25 average values in (B). Similarly, in (C) we plot the average clustering coefficient.  }}%
    \label{fig: p_clique_Uniform}%
\end{figure}

In Theorem \ref{thm: main_cliques}, we used edges between cliques to estimate distances between points on the unknown latent surface $\mathcal{M}$. Theorem \ref{thm: main_cliques} then states that as the graph size $n$ and the clique size $\ell$ both go to infinity, we can consistently estimate all parameters of the latent space graph model. In particular, we have assumed that as $\ell, n$ both go to infinity, that we observe cliques of size $\ell$. To understand the rate at which $\ell, n$ can grow, we consider a simplifying example. Let $G$ be an undirected random graph drawn from the Erdos-Renyi model with parameters $n$ and $p$. That is, edges form independently with probability $p$. The following result is a well-known result which says that in an ER graph, the clique number grows like $\log(n)$ for large $n$. It can be found in many places, such as in \cite{grimmett_mcdiarmid_1975}.
\begin{prop}[\cite{grimmett_mcdiarmid_1975}]
\label{prop: clique_prob}
The clique number of an ER model $Z_{n,p}$ satisfies 
\begin{equation*}
    \frac{Z_{n, p}}{\log(n)} \rightarrow \frac{2}{\log(1/p)} \;.
\end{equation*}
almost surely. 
\end{prop}
In other words, the clique number $Z_{n, p}$ grows like $C \log(n)$ for the constant $C = 2 \{\log(1/p)\}^{-1}$. So by taking $\ell = C\log(n)$, we will almost surely see an $\ell$ clique in the graph as $n, \ell \rightarrow \infty$. 
Now let us return to our problem. We observe a graph drawn from (\ref{eq:main_model}), where the node locations and effects are drawn iid from two distributions. Clearly, the probability of observing an $\ell = \ell(n)$ clique depends on the distributions of the points and node effects. To our knowledge, there are no results like Proposition \ref{prop: clique_prob} that hold for arbitrary graph models. However, we would like to investigate the behavior of the clique number for three common ways of assigning points in the latent space. In particular, we want to determine if there are cliques of size $\log(n)$ in graphs where the node locations are drawn according to a lattice mode, Gaussian mixture model, and uniform model. These three models are the models we study in Section \ref{sec: ass_1_3}. We give these results in Figures \ref{fig: p_clique_Lattice}, \ref{fig: p_clique_GMM}, and \ref{fig: p_clique_Uniform}. We see cliques of size $\log(n)$ with probability going to 1 as $n \rightarrow \infty$. The models are listed in increasing ``difficulty," meaning that it should be less likely for there to be clique when points are uniformly drawn (C) than when they are drawn from a GMM (B). And, when nodes are at the same location (A), it should be even more likely that a clique exists. We see this pattern in our simulations; for any $n$, the GMM in (B) has a higher probability of containing an $\log(n)$ clique for every $n$. But, for sufficiently large $n$, all models contain cliques of size at least $\log(n)$ with relatively high probability. 

\begin{figure}
    \centering
       \subfloat[]{{\includegraphics[width=.45\textwidth]{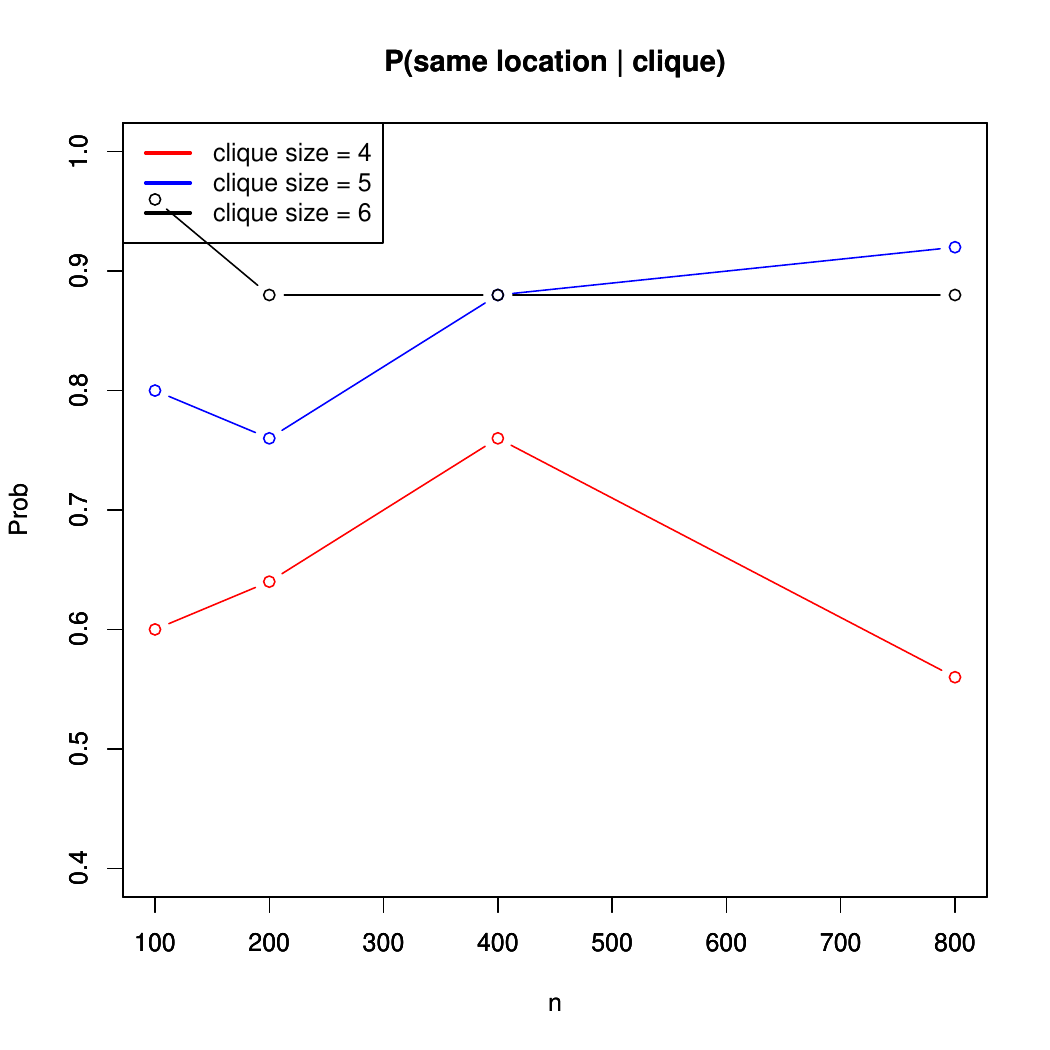} }}%
    \qquad
     \subfloat[]{{\includegraphics[width=.45\textwidth]{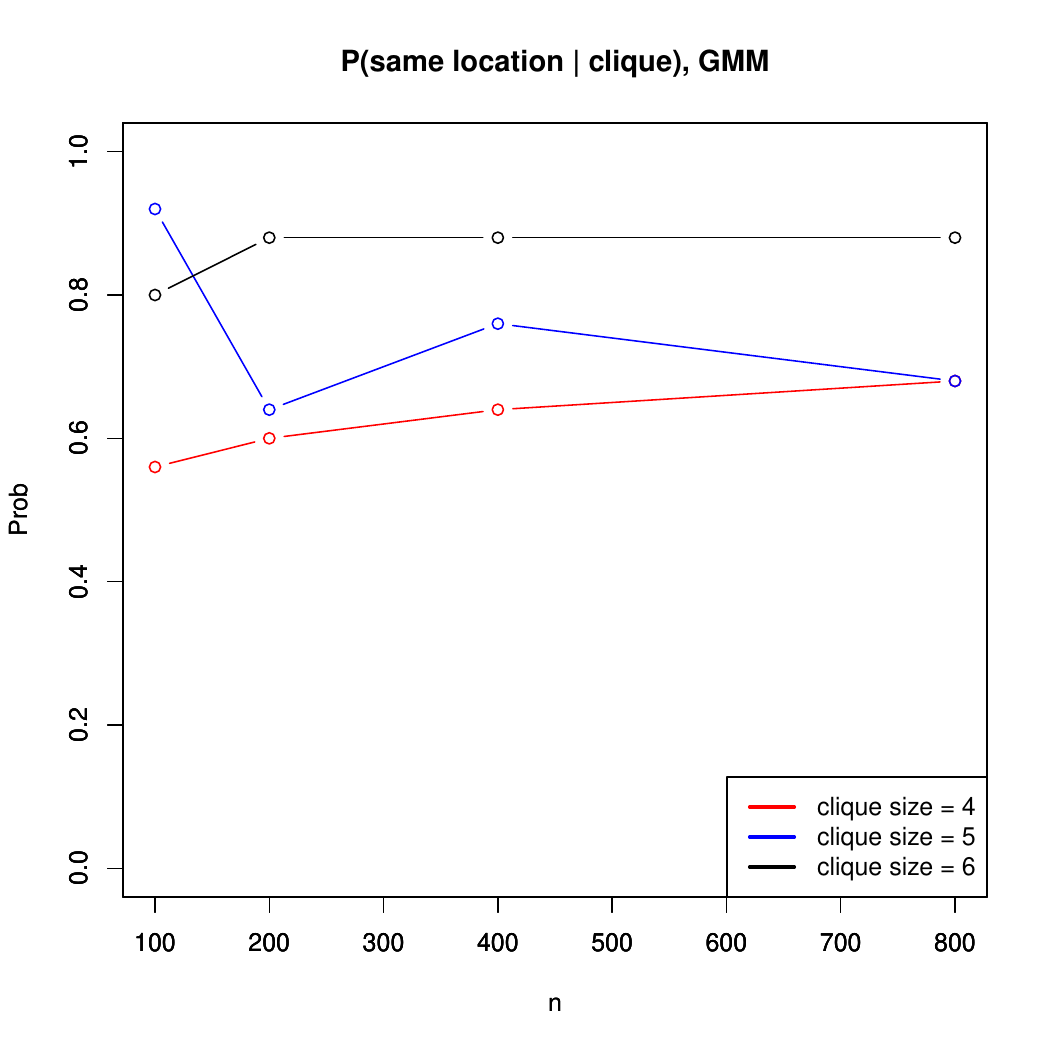} }}
        \qquad
    \subfloat[]{{\includegraphics[width=.45\textwidth]{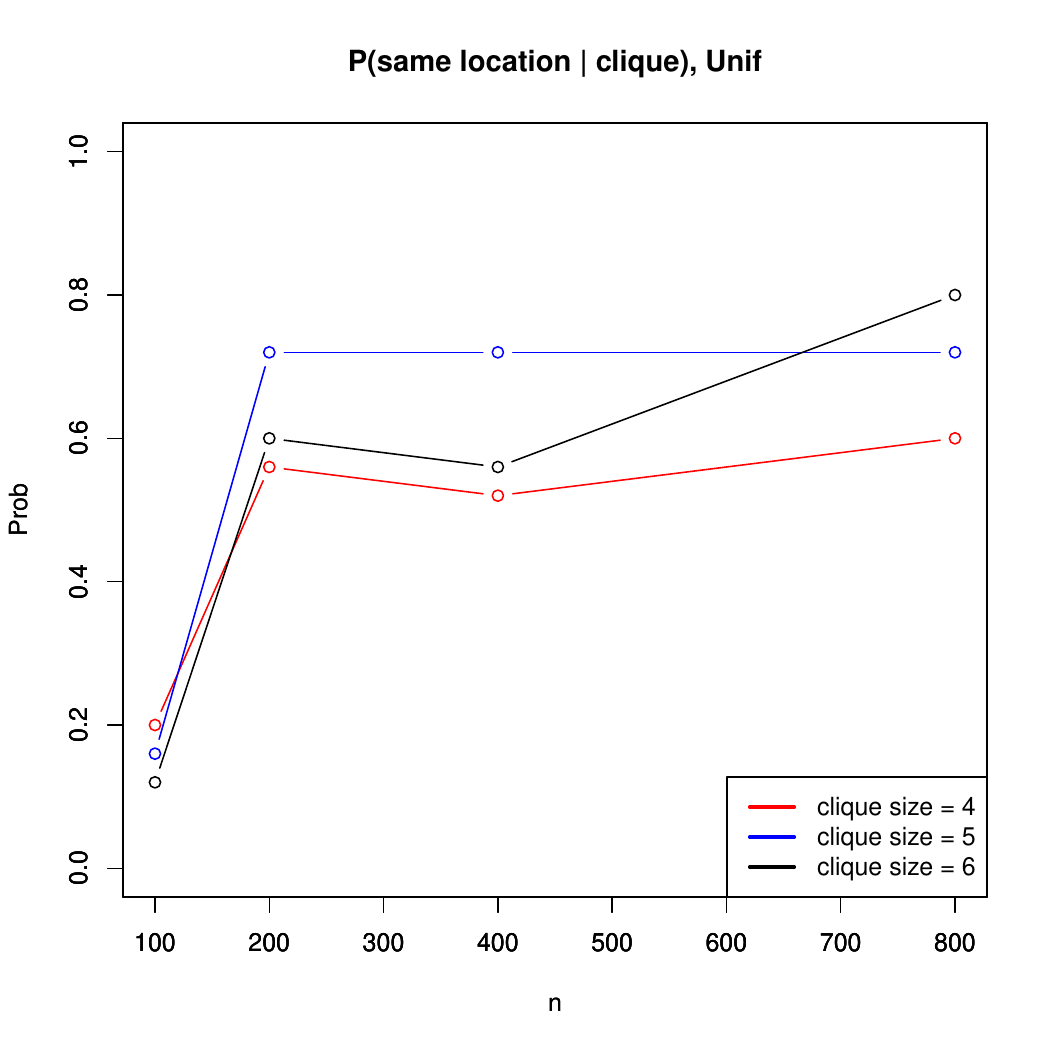} }}%
    \caption{\footnotesize{ We generate 25 networks on $n$ nodes for $n \in \{100, 200, 400, 800\}$. Using $C = \log(n)$, we generate $C$ communities in a lattice model (A), a GMM (B), or a uniform model (C). We check how many times out of 25 nodes in an arbitrary clique are at the same location. We plot the corresponding probabilities above for clique sizes 4, 5, 6.}}%
    \label{fig: p_same_loc}%
\end{figure}

We also verify in Figure \ref{fig: p_same_loc} that as the size of a clique goes up, the probability that node locations in the latent space are close to each increases

\setcounter{table}{0}
\setcounter{figure}{0}
\label{sec: CM_geometry}
 \section{Testing geometry using the Cayley-Menger determinant}
 
For $K := 4$ points with pairwise distances $D = \{d_{ij}\}$, define the \emph{Cayley-Menger} determinant  to be the determinant of the matrix

\begin{equation*}
    CM := \begin{pmatrix}
    0 & 1 & 1 & 1 & 1 \\
    1 & 0 & d_{12}^2 & d_{13}^2 & d_{14}^2 \\
    1 & d_{21}^2 & 0 & d_{23}^2  & d_{24}^2 \\
    1 & d_{31}^2 & d_{32}^2 & 0 & d_{34}^2 \\
    1 & d_{41}^2 & d_{42}^2 & d_{43}^2 & 0
    \end{pmatrix} \;.
\end{equation*}
When these points are in $\mathbb{R}^2$, then the determinant of the matrix $CM$ is 0. The goal of this section is to derive an estimator of geometry that uses this idea. Similar results hold for points from a spherical space or hyperbolic space, with slightly modified matrices.

We now propose a test of geometry using this method. To ensure that this is a reasonable comparison to our previous results, we generate a stochastic block model with locations $z_1, \dotsc, z_K$, and assign edges based on $P(G_{ij} = 1 | z_i, z_j) = \exp(-d(z_i, z_j))$. These locations correspond to $K$ cliques that we might find in a network. We then compute the determinant of $\widehat{CM}$, which is defined as above except that we use $\hat d_{ij}^2$ in place of $d_{ij}^2$. Since the distribution of the random variable under $H_0: CM = 0$ is hard to derive, let us do a parametric bootstrap. For $b = 1, \dotsc, B$, we repeat the following steps:
\begin{enumerate}
    \item For each $i < j$, draw with replacement $\ell^2$ edges from the set of edges that exist between locations $z_i$ and $z_j$. This is the re-sampling step.
    \item Compute the average number of re-sampled edges between $i$ and $j$ for any $i < j$. Call this number $ p^\star_{ij}$ and set $d^\star_{ij} = -\log(p^\star_{ij})$. 
    \item Compute $CM_b^\star = \text{det}(CM^\star)$, where $CM^\star$ is the matrix above with $d_{ij}$ now replaced by $d_{ij}^\star$.
\end{enumerate}
If 0 falls outside of the interval
\begin{equation*}
    (2\text{det}(\widehat{CM})- q_{1 - \alpha/2}\{\text{det}(CM)^\star\},  2\text{det}(\widehat{CM})- q_{\alpha/2}\{\text{det}(CM)^\star\})
\end{equation*}
then we reject $H_0$, where $q_{\alpha/2}\{\text{det}(CM)^\star\}$ is the $\alpha/2$ quantile of the empirical distribution $\{\text{det}(CM)^\star\}$. 

\begin{figure}
    \centering
    \includegraphics[scale = 0.5]{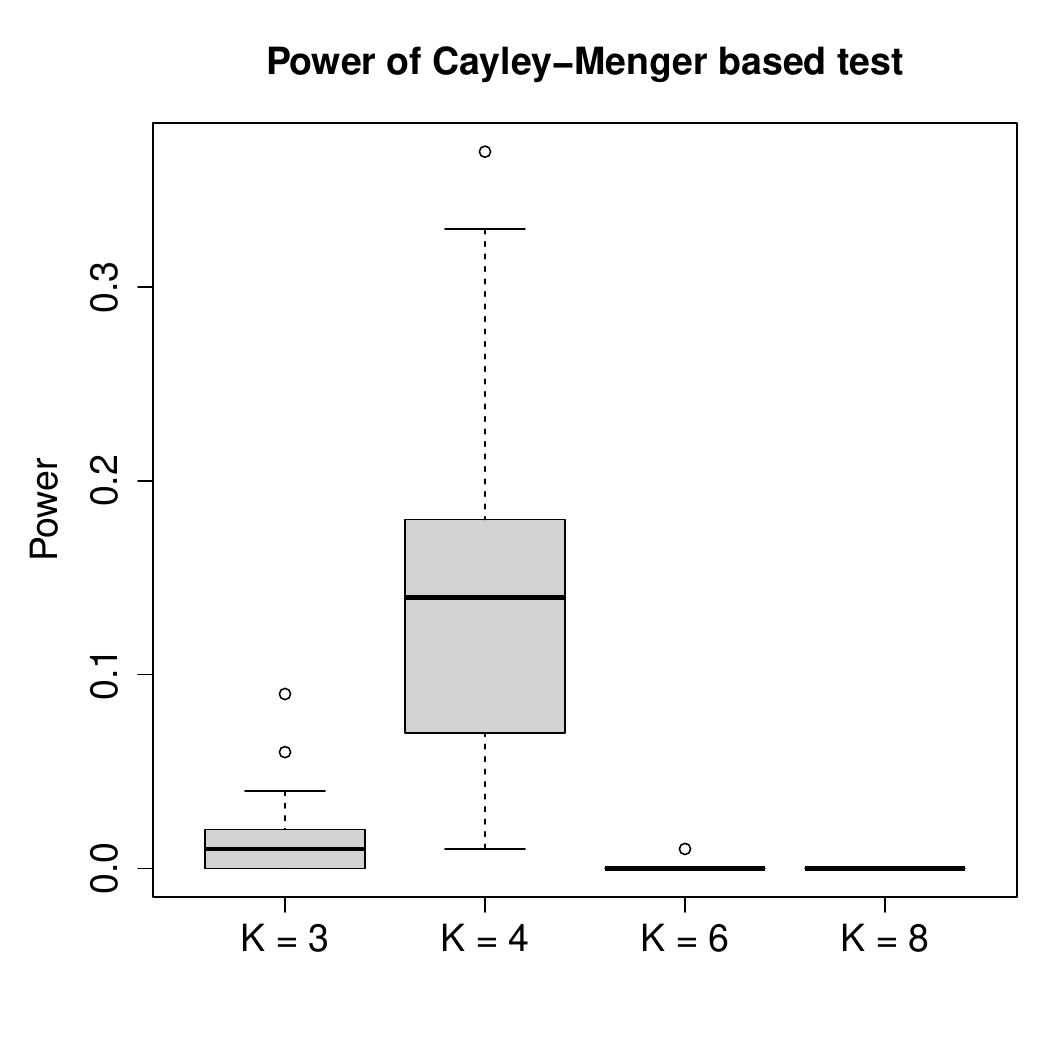}
    \caption{Power of the Cayley-Menger based test of the Euclidean hypothesis. On the $x$-axis we plot the number of points $K$ that were used. On the $y$-axis we plot the average number of rejections (out of 100) for 25 sets of $K$ locations drawn from the sphere $\mathbf{S}^2(1/2).$}
    \label{fig: CM_Sphere_Power}
\end{figure}

We plot the power of this method using spherical data in Figure \ref{fig: CM_Sphere_Power}. We see that when we use $K = p + 2$, then the method works its best, but it still performs relatively poorly. So this seems to suggest that only at this value of $K$ does it work well.

\end{document}